\documentclass{aa}  
\usepackage{aalongtable,lscape}
\usepackage{graphicx}
\usepackage{txfonts}
\usepackage[figuresright]{rotating}
\begin{document}
\title{HRC-I/{\em Chandra} X-ray observations towards $\sigma$~Orionis} 
\titlerunning{HRC-I/{\em Chandra} X-ray observations towards $\sigma$~Orionis}
%
%\subtitle{}
%
\author{J. A. Caballero\inst{1,2}
    	\and
	J. F. Albacete-Colombo\inst{3}
    	\and
	J. L\'opez-Santiago\inst{2}}
\offprints{Jos\'e Antonio Caballero, \email{caballero@cab.inta-csic.es}}
\institute{Centro de Astrobiolog\'{\i}a (CSIC-INTA), Carretera de Ajalvir km~4,
28850 Torrej\'on de Ardoz, Madrid, Spain
\and
Departamento de Astrof\'{\i}sica y Ciencias de la Atm\'osfera,
Facultad de F\'{\i}sica, Universidad Complutense de Madrid, 28040
Madrid, Spain
\and
Centro Universitario Regional Zona Atl\'antica, Universidad Nacional del
Comahue, Monse\~nor Esandi y Ayacucho, 8500 Viedma, R\'{\i}o Negro, Argentina}
\date{Received 25 April 2010 / Accepted 11 Jun 2010}

% \abstract{}{}{}{}{} 
% 5 {} token are mandatory
\abstract
{} 
{We investigated the X-ray emission from young stars and brown dwarfs in the
$\sigma$~Orionis cluster ($\tau \sim$ 3\,Ma, $d \sim$ 385\,pc) and its relation
to mass, presence of circumstellar discs, and separation to the cluster centre
by taking advantage of the superb spatial resolution of the {\em Chandra X-ray
Observatory}.} 
{We used public HRC-I/{\em Chandra} data from a 97.6\,ks pointing towards
the cluster centre and complemented them with X-ray data from IPC/{\em
Einstein}, HRI/{\em ROSAT}, EPIC/{\em XMM-Newton}, and ACIS-S/{\em
Chandra} together with optical and infrared photometry and spectroscopy from the
literature and public catalogues. 
On our HRC-I/{\em Chandra} data, we measured count rates, estimated X-ray
fluxes, and searched for short-term variability.
We also looked for long-term variability by comparing with previous X-ray
observations.}
{Among the 107 detected X-ray sources, there were 70 cluster stars with known
signposts of youth, two young brown dwarfs, 12 cluster member candidates, four
field dwarfs, and two galaxies with optical-infrared counterpart. 
The remaining sources had extragalactic nature.
Based on a robust Poisson-$\chi^2$ analysis, nine cluster stars displayed flares
or rotational modulation during the HRC-I observations, while other eight
stars and one brown dwarf showed X-ray flux variations between the HRC-I and
IPC, HRI, and EPIC epochs. 
We constructed a cluster X-ray luminosity function from O9.5 (about
18\,$M_\odot$) to M6.5 (about 0.06\,$M_\odot$).  
We found: ($i$) a tendency of early-type stars in multiple systems or with
spectroscopic peculiarities to display X-ray emission, ($ii$) that the two
detected brown dwarfs and the least-massive star are among the $\sigma$~Orionis
objects with the highest $L_X/L_J$ ratios, and ($iii$) that a large fraction of
known classical T~Tauri stars in the cluster are absent in this and other X-ray
surveys.
Finally, from a spatial distribution analysis, we quantified the impact of
the sensitivity degradation towards the HRC-I borders on the detection of faint
X-ray sources, and concluded that dozens X-ray $\sigma$~Orionis stars and brown
dwarfs are still to be detected.}  
{}
\maketitle
%
%________________________________________________________________

\section{Introduction}
\label{section.introduction}

%______________________________________________ Figure 
\begin{figure*}
\centering
\includegraphics[width=0.49\textwidth]{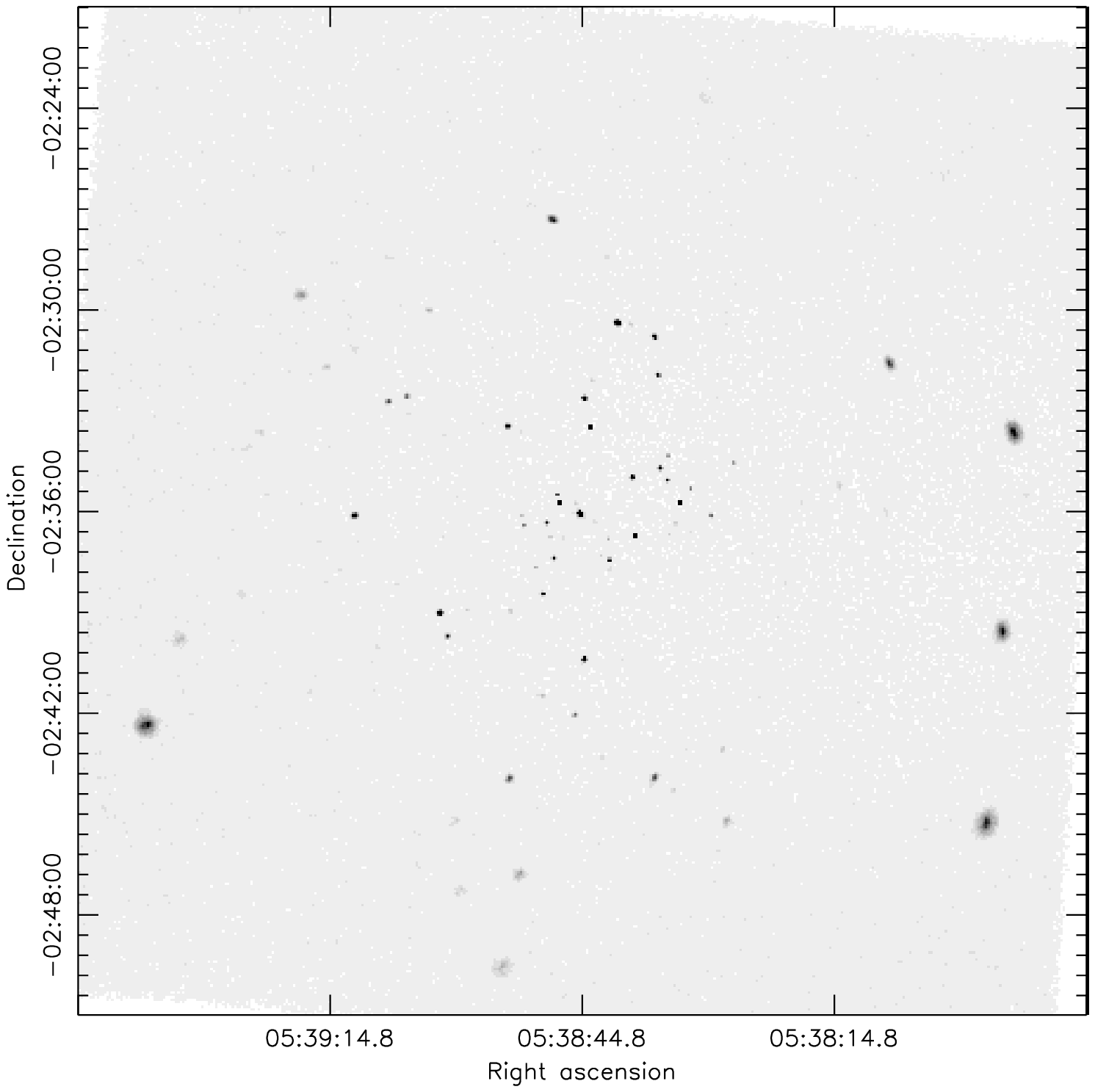}
\includegraphics[width=0.49\textwidth]{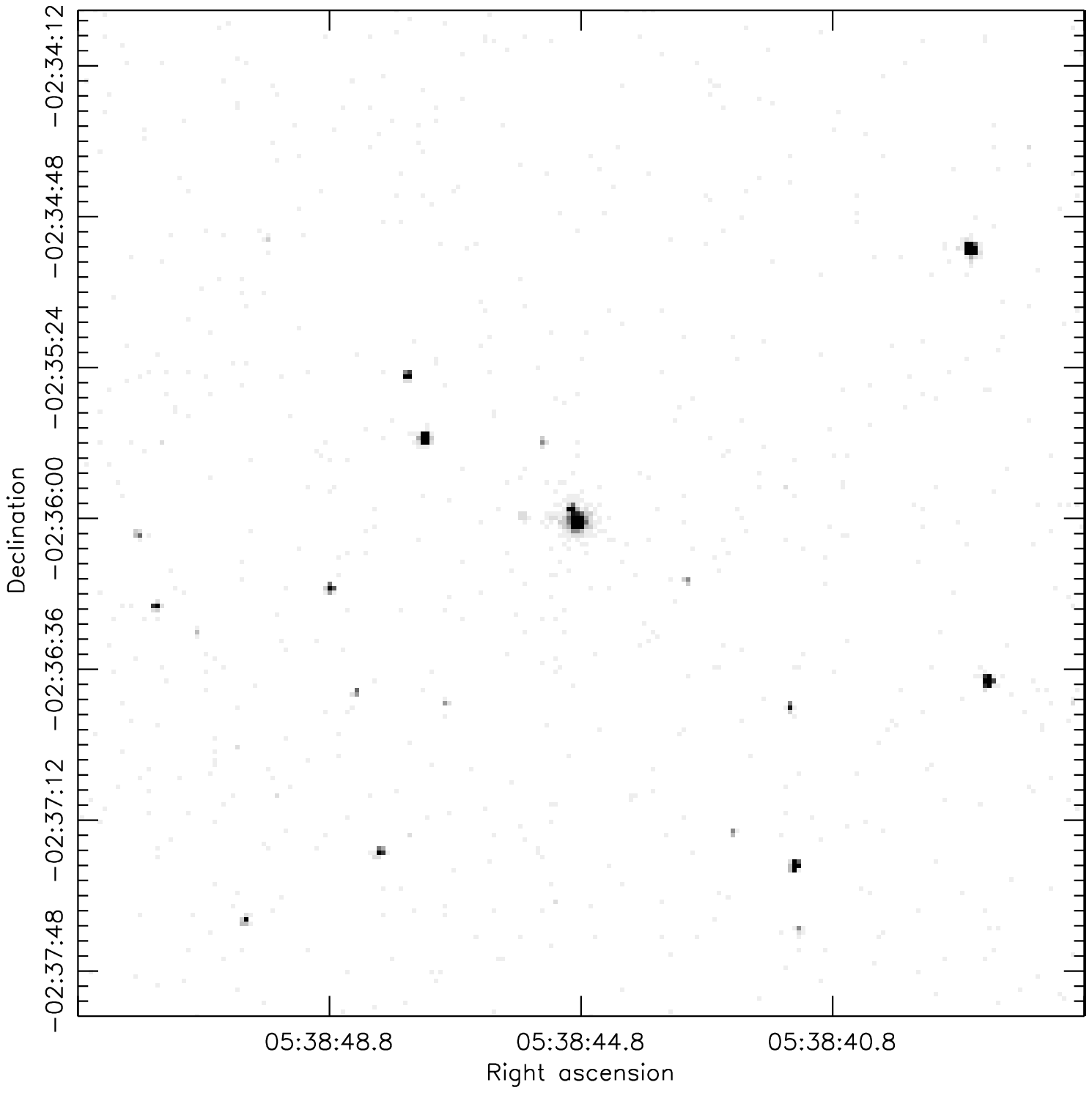}
\caption{HRC-I/{\em Chandra} images centred on $\sigma$~Ori~AB.
Approximate sizes are 30$\times$30\,arcmin$^2$ ({\em left}; note the borders of
the field of view in the corners) and 4$\times$4\,arcmin$^2$ ({\em right};
see also the Fig.~4 in Caballero 2007b).
North is up, east is left.} 
\label{xfig_hrc-i}
% jls.idl
\end{figure*}
% It comes from \section{Analysis and results}
%

The Trapezium-like system \object{$\sigma$~Ori}, the fourth brightest ``star''
in the Orion Belt, illuminates the \object{Horsehead Nebula} and injects energy
into its homonymous cluster, \object{$\sigma$~Orionis} (Garrison 1967; Wolk
1996; B\'ejar et~al. 1999).
Its age ($\tau \sim$ 3\,Ma -- Zapatero Osorio et~al. 2002; Sherry et~al. 2004),
relative closeness ($d \sim$ 385\,pc -- Caballero 2008b; Mayne \& Naylor 2008),
low extinction (0.04\,mag $< E(B-V) <$ 0.09\,mag -- B\'ejar et~al. 2004;
Sherry et~al. 2008), and high spatial density (Caballero 2008a) make the cluster
an ideal site to look for and characterise substellar objects (Zapatero Osorio
et~al. 2000; B\'ejar et~al. 2001; Caballero et~al. 2007; Bihain et~al. 2009).  
The cluster is also investigated, for example, to study circumstellar discs
based on optical spectroscopy (Kenyon et~al. 2005; Sacco et~al. 2008; Gatti
et~al. 2008) or mid-infrared photometry (Oliveira et~al. 2006; Caballero
2007a; Zapatero Osorio et~al. 2008; Luhman et~al. 2008) and young X-ray emitter
stars (Sanz-Forcada et~al. 2004; Franciosini et~al. 2006; Skinner
et~al. 2008; L\'opez-Santiago \& Caballero 2008 and references therein).

X-ray observations in young open clusters, such as $\sigma$~Orionis, provide
information on winds of early-type stars, high-temperature coronae of late-type
stars, absorption by circumstellar discs, magnetic activity associated to fast
rotation, the cluster X-ray luminosity function, and, in general, the evolution
of young (pre-)main-sequence stars. 
Except for the {\em ROSAT} variability analysis in Caballero et~al. (2009),
the latest X-ray studies in $\sigma$~Orionis have been carried out using
instruments onboard the {\em XMM-Newton} and {\em Chandra} space missions.
In this work, we analyse in detail observations of a large portion of the
cluster accomplished with the {\em Chandra} High Resolution Camera (HRC).
The lower sentivity of HRC with respect to EPIC/{\em XMM-Newton} (European
Photon Imaging Cameras) used by Franciosini et~al. (2006) was compensated by the
better spatial resolution and a longer exposure time, of almost 100\,ks.  
Besides, the HRC observations in $\sigma$~Orionis were more sensitive and
covered a larger field of view than those performed with ACIS/{\em Chandra}
(Advanced CCD Imaging Spectrometer) by Skinner et~al. 
(2008)\footnote{Skinner et~al. (2008) also used the High Energy Transmission 
Grating, HETG, for the brightest sources.}.
HRC observations provide, however, no spectral information. 

Some preliminary results based on the HRC/{\em Chandra} dataset, which is
publicly available from the {\em Chandra} Data Archive\footnote{\tt
http://cxc.harvard.edu/cda/} since 2003, have been advanced by Adams et~al.
(2002, 2003, 2004, 2005) and Caballero (2005, 2007b).
Here, we detect X-ray sources on the deep HRC image, cross-identify them with
optical, near-infrared, and previously-known X-ray sources, classify them into
young and field stars and galaxies using state-of-the-art spectro-, astro-, and
photometric data, compare with previous X-ray observations, and study the X-ray
luminosity funcion in the cluster, the frequency of X-ray emitters, and its
relation to spatial location, disc occurrence, and stellar mass.

\section{Analysis and results}
\label{section.analysis}

\subsection{Data retrieval}
\label{section.dataretrieval}

HRC, held in the {\em Chandra} focal plane array together with ACIS, is a double
CsI-coated microchannel plate detector similar to the High Resolution Imaging
photon-counting detectors onboard the {\em Einstein Observatory} and {\em
ROSAT}. 
However, HRC has substantially increased capability compared with them in X-ray
quantum efficiency (in the energy range 0.08--10.0\,keV), detector size
(90$\times$90\,mm$^2$ or 16\,Mpx, which translates into a field of view of
31$\times$31\,arcmin$^2$), internal background rate, and, specially, spatial
resolution (down to 0.016\,arcsec). 

Using the web version of ChaSeR at the {\em Chandra} Data Archive, we searched
and retrieved the package of primary data products associated to the
observations with identification number 2560 (sequence number 200168, principal
investigator S.~Wolk). 
Observations were carried out on 2002 Nov 21--22 and took a total exposure time
of 97.6\,ks.
The field of view was approximately centred on $\sigma$~Ori~D (Mayrit~13084), a
B2V star located at 13\,arcsec to the massive binary (possibly triple) star
$\sigma$~Ori~AB at the bottom of the gravitational well in the centre of the
$\sigma$~Orionis cluster.
%RA= 84.6904318796
%DEC= -2.5981948102.

\subsection{Reduction}
\label{section.reduction}

Data reduction, starting with the level-1 event list provided by the  
processing pipeline at the {\em Chandra} X-ray Center, was performed using
the {\em Chandra} Interactive Analysis of Observations software
CIAO~3.4\footnote{\tt http://cxc.harvard.edu/ciao3.4/} and the
{\em Chandra} Calibration Database CALDB~3.4.1\footnote{\tt
http://cxc.harvard.edu/caldb3/}.  
We produced a level-2 event file using the CIAO task {\tt hrc\_process\_events}.
The data were filtered to remove events that did not have a good event ``grade''
or that had one or more of the ``status bits'' set to unity (see the
definitions of ``grade'' and ``status bits'' at the {\em Chandra}/CIAO 
dictionary\footnote{\tt http://chandra.ledas.ac.uk/ciao/dictionary/}). 
Intervals of solar background flaring were searched for, but none were found
(see, however, Section~\ref{section.xraylightcurves}).
As a result, we assumed a constant background and did not applied
time filtering.
An exposure map, needed by the source detection algorithm and to re-normalise
source count rates, was calculated with the CIAO tool {\tt mkexpmap} assuming a
monochromatic spectrum ($k_{\rm B}T$ = 1.0\,keV). 
See further details in Albacete-Colombo et~al. (2008), where an alike reduction
process was performed.
%Tambien podes guiarte con lo que escribi en mis papers de CygOB2 y Trumpler 16.

\subsection{Source detection}
\label{section.sourcedetection}

Source detection was accomplished with the Palermo Wavelet Detection code 
PWDetect\footnote{\tt http://www.astropa.unipa.it/progetti\_ricerca/PWDetect/}
version 1.3.2 (Damiani et~al. 1997a) on the level-2 event list restricted to the
0.5--10\,keV energy band and specifically compiled to run for a maximun of
7\,10$^6$ events. 
PWDetect analyses the data at different spatial scales, from 0.25 to 16\,arcsec,
allowing the detection of both point-like and moderately extended sources and
the efficient resolution of close sources pairs. 
The most important input parameter required by the code is the final threshold
significance for detection, $S_{\rm min}$ (in equivalent Gaussian $\sigma$s),
which depends on the background level, detector, and desired number of spurious
detections per field due to Poisson noise, as determined from extensive
simulations of source-free fields (cf. Damiani et~al. 1997a).
We determined the total number of background counts detected during the
entire exposure over the full HRC-I detector at 4.5\,10$^6$ photons with a
proprietary IDL script.
This background level translated into a final detection threshold of $S_{\rm
min}$ = 5.1$\sigma$ if we impose only {\em one spurious} detection in the field
of view. 

A total of 109 HRC-I sources with $S >$ 5.1$\sigma$ were found with PWDetect.
We visually inspected each X-ray source and identified two ``double
detections'', corresponding to the stars Mayrit~3020~AB (No.~25) and
Mayrit~156353 (No.~11).  
In detail, for each optical counterpart, PWDetect revealed two X-ray
sources, one bright and one faint and slightly decentred, separated by a few
tens of arcsecond. 
This separation is smaller than the sizes of the point spread functions of the
X-ray sources.
The double detections may arise because of bad adopted background estimate near
bright X-ray sources (Damiani et~al. 1997a, 1997b). 
We discarded the faint X-ray sources in the two cases\footnote{We thank
I.~Pillitteri for helpful guidance in this subject.} and kept the remaining 107
sources as reliable X-ray detections.
Their coordinates, significances of detection ($S$), angular separations to the
centre of field of view (offaxis), count rate, and associated uncertainties are
listed in Table~\ref{table.xraydetections}. 
The sources are sorted by decreasing significance of detection.

%______________________________________________ Figure 
\begin{figure}
\centering
\includegraphics[width=0.49\textwidth]{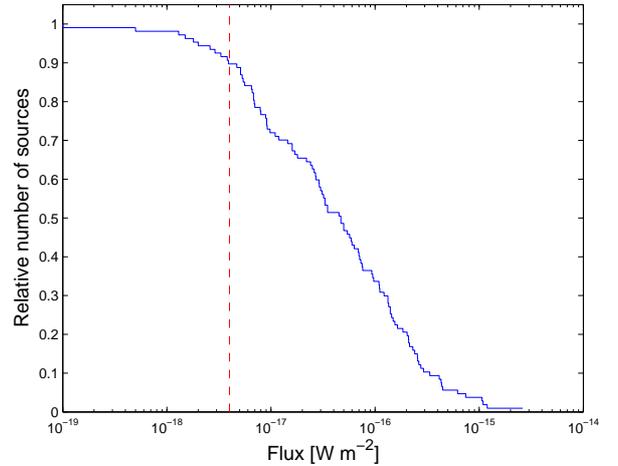}
\caption{Relative cumulative number of the HRC-I/{\em Chandra} X-ray
sources as a function of apparent flux. 
The vertical [red] dashed line at 0.4\,10$^{-17}$\,W\,m$^{-2}$ indicates the
approximate completeness flux of our survey.}  
\label{xfig_relN_flux}
% sox13.m
\end{figure}

In addition, we estimated the apparent X-ray flux\footnote{Throughout this
work, we use the word `flux' for denoting the quantity $\lambda F_\lambda$.
For transforming between the Syst\`eme international d'unit\'es and the
centimetre-gram-second system, use the conversion factor
10$^{-14}$\,erg\,cm$^{-2}$\,s$^{-1}$ $\equiv$ 10$^{-17}$\,W\,m$^{-2}$.
Using $d$ = 385\,pc to the $\sigma$~Orionis cluster, a flux ${\mathcal F}$
= 10$^{-17}$\,W\,m$^{-2}$ translates into a {\em cgs} luminosity $\log{L_X}
\approx$ 29.25.} for each source. 
We integrated the counts over a circular area three times wider than the one
used by PWDetect, which is in turn smaller than the size of the local point
spread function.  
More than 97\,\% of the photons of a source fall within the circular area.
A mean background level was subtracted after integrating the counts over an area
of the same radius (but free of X-ray emission) in the vicinity of each source.
Finally, for the conversion betweeen counts and energy, we used the factor
$\overline{E_\gamma}$ = 1.2\,keV (mean energy per X-ray photon), which is
representative of late-type young stars in $\sigma$~Orionis.
This value was obtained by determining a weighted mean of the coronal
temperatures of the stars in Table~3 in L\'opez-Santiago \& Caballero (2008). 
The completeness flux limit, which marks an inflection point in the cumulative
number of X-ray sources as a function of apparent flux, was
0.4\,10$^{-17}$\,W\,m$^{-2}$ (Fig.~\ref{xfig_relN_flux}).
The actual completness limit varies with the offaxis separation
(Section~\ref{section.spatial}).

\subsection{Cross-identification}
\label{subsection.cross}

%__________________________________________________ 
   \begin{table*}
      \caption[]{X-ray stars not tabulated in the Mayrit catalogue 
      (Caballero~2008c)$^{a}$.}  
         \label{table.nonmayrit}
     $$ 
         \begin{tabular}{lcl cc cccc l}
            \hline
            \hline
            \noalign{\smallskip}
No.	&	& Name			& $\alpha$ 	& $\delta$	& $i$			& $J$ 			& $H$			& $K_{\rm s}$		& References$^c$	\\  
	&	&			& (J2000) 	& (J2000) 	& [mag]			& [mag]			& [mag]			& [mag]			& 			\\  
            \noalign{\smallskip}
            \hline
            \noalign{\smallskip}
25	& *	& Mayrit 3020 AB	& 05 38 44.84 	& --02 35 57.1 	& ....			&   10.4$\pm$0.2	&  10.70$\pm$0.07	& 10.480$\pm$0.010	& vLO03, Ca05, Bo09	\\ % Class~II		      \\  
31	& *	& [W96] 4771--1056	& 05 39 00.52 	& --02 39 39.0 	& 12.371$\pm$0.06	& 11.665$\pm$0.028	& 11.221$\pm$0.024	& 11.110$\pm$0.022	& Wo96, Sk08		\\ % No seq., Li~{\sc i}?     \\  
39	& *	& Mayrit 168291 AB	& 05 38 34.31 	& --02 35 00.0 	& 12.272$\pm$0.03	& 11.216$\pm$0.031	& 10.565$\pm$0.033	& 10.354$\pm$0.030	& He07, Ga08, Sk08	\\ % Li~{\sc i}  	      \\  
46	&	& Mayrit 1093033	& 05 39 24.56 	& --02 20 44.1 	& 12.727$\pm$0.16	& 11.371$\pm$0.026	& 10.778$\pm$0.024	& 10.554$\pm$0.023	& He07			\\ % Candidate, no new	      \\  
47	& *	& Mayrit 68229		& 05 38 41.35 	& --02 36 44.4 	& 14.404$\pm$0.03	& 12.988$\pm$0.026	& 12.330$\pm$0.024	& 12.084$\pm$0.025	& Wo96, Ca07b, Sk08	\\ % Candidate  	      \\    
48	&	& Mayrit 172264		& 05 38 33.35 	& --02 36 17.6 	& 13.393$\pm$0.02	& 12.052$\pm$0.027	& 12.295$\pm$0.023	& 11.107$\pm$0.027	& Sh04, Sk08 		\\ % Candidate  	      \\  
51	&	& [SWW2004] 166 	& 05 38 53.06 	& --02 38 53.6 	& 12.717$\pm$0.02	& 11.625$\pm$0.026	& 11.034$\pm$0.026	& 10.828$\pm$0.025	& Sh04, Ol06, Sk08 	\\ % NM  	      	      \\  
57	& *	& Mayrit 492211 	& 05 38 27.74 	& --02 43 00.9 	& 13.636$\pm$0.03	& 11.189$\pm$0.030	& 11.447$\pm$0.024	& 10.287$\pm$0.024	& Sh04, Sa08		\\ % Li~{\sc i}  	      \\  
58	& *	& Mayrit 21023 		& 05 38 45.31	& --02 35 41.3 	& ....			&  13.41$\pm$0.09	&  12.98$\pm$0.06	&  12.73$\pm$0.09	& Ca07b, Bo09		\\ % Candidate  	      \\  
85	&	& Mayrit 270196 	& 05 38 39.72 	& --02 40 19.7 	& 15.489$\pm$0.05	& 13.746$\pm$0.031	& 13.099$\pm$0.026	& 12.883$\pm$0.028	& Sk08			\\ % Candidate  	      \\  
95$^b$	&	& Mayrit 605079 	& 05 39 24.35 	& --02 34 01.3 	& 14.501$\pm$0.19	& 12.978$\pm$0.030	& 12.272$\pm$0.026	& 12.058$\pm$0.021	& Sh04, Sa07, Sa08	\\ % Possible		      \\  
98	&	& Mayrit 1178039 	& 05 39 33.78 	& --02 20 39.8 	& 13.783$\pm$0.15	& 12.367$\pm$0.026	& 11.598$\pm$0.023	& 11.429$\pm$0.023	& Sh04, Ol06	 	\\ % Candidate  	      \\  
99	&	& Mayrit 957055 	& 05 39 37.29 	& --02 26 56.7 	& 13.000$\pm$0.03	& 11.698$\pm$0.026	& 10.974$\pm$0.024	& 10.773$\pm$0.021	& Sh04			\\ % Candidate  	      \\  
           \noalign{\smallskip}
            \hline
         \end{tabular}
     $$ 
\begin{list}{}{}
\item[$^{a}$] Stars marked with an asterisk, `*', are commented in
Section~\ref{section.notes.table1}.
\item[$^{b}$] See Section~\ref{section.clustermembers} for a discussion on
Mayrit~605079. 
\item[$^{c}$] Reference abbreviations --
Wo96: Wolk (1996);
vLO03: van~Loon \& Oliveira (2003);
Sh04: Sherry et~al. (2004);
Ca05: Caballero (2005);
Ol06: Oliveira et~al. (2006);
Ca07b: Caballero (2007b);
Sa07: Sacco et~al. (2007);
He07: Hern\'andez et~al. (2007);
Ga08: Gatti et~al. (2008);
Sa08: Sacco et~al. (2008);
Sk08: Skinner et~al. (2008);
Bo09: Bouy et~al. (2009).
\end{list}
   \end{table*}

We cross-matched the 107 X-ray sources in Table~\ref{table.xraydetections} with
optical and near-infrared catalogues.
First, we searched for their optical/near-infrared counterparts in the Mayrit
catalogue of young stars and brown dwarfs in the $\sigma$~Orionis cluster
(Caballero 2008c). 
He tabulated coordinates, $iJHK_{\rm s}$ magnitudes (from the DENIS and 2MASS
catalogues -- Epchtein et~al. 1997; Skrutskie et~al. 2006), and youth features
of a large number of confirmed and candidate cluster members.
He also tabulated foreground field dwarfs and background galaxies.
Of the 107 X-ray sources in our work, 77 were in the Mayrit catalogue.
Secondly, we found the optical/near-infrared counterparts of other 13 X-ray
sources not tabulated in the Mayrit catalogue, listed in
Table~\ref{table.nonmayrit}.  
Caballero (2008c) did not record them because they had no 2MASS counterpart
(Nos.~25 and~58) or known youth features at that time and were located bluewards
of his conservative selection criterion in the $i$ vs. $i-K_{\rm s}$ diagram
(the remaining 11 stars).
However, most of the 11 ``blue'' X-ray stars are ``red'' enough to have been
considered in previous photometric searches in the cluster (see references in
footnote to Table~\ref{table.nonmayrit}). 

%______________________________________________ Figure 
\begin{figure}
\centering
\includegraphics[width=0.49\textwidth]{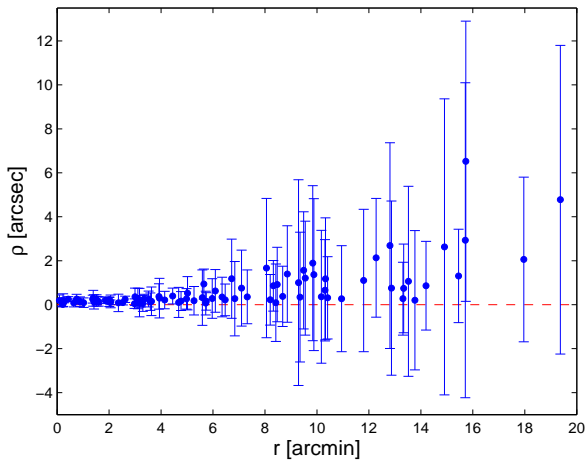}
\caption{Separation between the 90 correlated HRC-I sources and their 2MASS 
counterparts as a function of separation to the cluster centre 
($\rho$ vs. $r$ diagram).
The horizontal (red) dashed line marks $\rho$ = 0\,arcsec (all the data points
are located above this line).} 
\label{xfig_rho_r}
% sox03.m
\end{figure}

In total, we found the optical/near-infrared counterparts of 90 X-ray sources.
The separations between the coordinates of the 2MASS and our X-ray sources is
plotted against the separation to the centre of the field of view in
Fig.~\ref{xfig_rho_r}. 
None of them separates from zero by more than 1$\sigma$ (accounting for the
errors in the determination of the photo-centroids of the HRC-I and 2MASS
sources).
Average separations are $\Delta \alpha$ = 0.2$\pm$1.0\,arcsec and $\Delta
\delta$ =  0.0$\pm$0.7\,arcsec.
Square-mean-roots in the innermost 3\,arcmin, where the HRC-I point spread
functions are sharper, get below 0.1\,arcsec.

The remaining 17 non-cross-matched X-ray sources and their closest 2MASS sources
are listed in Table~\ref{table.noncounterpart}. % ($107 - 90$)
Following L\'opez-Santiago \& Caballero (2008), we also looked for the optical
photographic counterparts in the USNO-B1 catalogue (Monet et~al. 2003).
We had no success with the cross-matching.
In all cases, the separations between the coordinates of the HRC-I and 2MASS
sources are larger than 2$\sigma$ and get larger than 6$\sigma$ in 13 cases.
These 13 HRC-I sources must have counterparts fainter than the USNO-B1, DENIS,
and 2MASS limiting magnitudes at $B_J \sim$ 21.0\,mag, $R_F \sim$ 20.0\,mag, $i
\sim$ 18.0\,mag, $J \sim$ 17.1\,mag, $H \sim$ 16.4\,mag, and $K_{\rm s} \sim$
14.3\,mag, respectively. 
We are not confident about the non-cross-matching of the other four X-ray
sources, which are separated to their closest 2MASS sources by less than
3$\sigma$. 
In two cases, Nos.~62 and~96, nearby galaxies undetected by USNO-B1,
DENIS, or 2MASS are visible in public images (see footnotes to
Table~\ref{table.noncounterpart}). 
Finally, in the other two cases, Nos.~97 and~107, the errors in coordinates of
X-ray sources could be underestimated and the 2MASS sources, which are cluster
member candidates (Burningham et~al. 2005; Caballero 2007b), may be the actual
optical counterparts (note the small angular separation of No.~97).

%__________________________________________________ 
   \begin{table}
      \caption[]{The closest 2MASS sources to X-ray galaxy candidates without
      optical/near-infrared counterpart listed in 
      Table~\ref{table.xraydetections}$^{a}$.}  
         \label{table.noncounterpart}
     $$ 
         \begin{tabular}{lc cc c l}
            \hline
            \hline
            \noalign{\smallskip}
No.	&	& $\alpha$ 	& $\delta$	& $\rho$	& Name			\\  
	&	& (J2000) 	& (J2000) 	& [arcsec]	&			\\  
            \noalign{\smallskip}
            \hline
            \noalign{\smallskip}
24	&	& 05 38 35.10   & --02 34 55.9  & 18.1  	& ...			\\ %  100
56	&	& 05 38 39.65   & --02 30 21.0  & 13.9  	& [SWW2004] 79  	\\ %   10.4
62	& *	& 05 38 14.22   & --02 35 07.3  & 6.31  	& [W96] rJ053814--0235	\\ %    2.47
68	&	& 05 38 59.65   & --02 38 15.6  & 41.2  	& ...			\\ %   52.2
76	&	& 05 38 53.37   & --02 33 22.9  & 24.4  	& Mayrit 203039 	\\ %   67.6
83	&	& 05 38 59.22   & --02 33 31.6  & 28.2  	& ....  		\\ %   41.4
91	&	& 05 38 41.37   & --02 28 31.8  & 12.3  	& ...			\\ %    6.83
93	& *	& 05 38 44.70   & --02 43 22.3  & 38.9  	& ...			\\ %   15.9
96	& *	& 05 39 06.64   & --02 38 08.1  & 3.22  	& ...			\\ %    2.30
97	& *	& 05 38 43.86   & --02 37 06.8  & 0.766 	& Mayrit 68191  	\\ %    3.06
100	&	& 05 38 24.49   & --02 29 22.8  & 27.9  	& ...			\\ %   10.5
101	&	& 05 38 50.42   & --02 36 43.1  & 11.5  	& ...			\\ %   35.9
102	&	& 05 38 42.00   & --02 39 23.2  & 15.7  	& ...			\\ %  112
104	&	& 05 38 36.77   & --02 42 54.6  & 13.7  	& ...			\\ %    9.25
105	&	& 05 38 34.79   & --02 34 15.8  & 16.2  	& Mayrit 182305 	\\ %   90.2
106	&	& 05 39 10.21   & --02 37 09.8  & 25.9  	& ...			\\ %   18.0
107	& *	& 05 38 28.25   & --02 32 27.4  & 4.16  	& [BNL2005] 1.02 156	\\ %    2.75
           \noalign{\smallskip}
            \hline
         \end{tabular}
     $$ 
\begin{list}{}{}
\item[$^{a}$] Sources marked with an asterisk, `*', are commented in
Section~\ref{section.notes.table2}.
\end{list}
   \end{table}

\subsection{Source classification}
\label{section.classification}

%______________________________________________ Figure 
\begin{figure}
\centering
\includegraphics[width=0.49\textwidth]{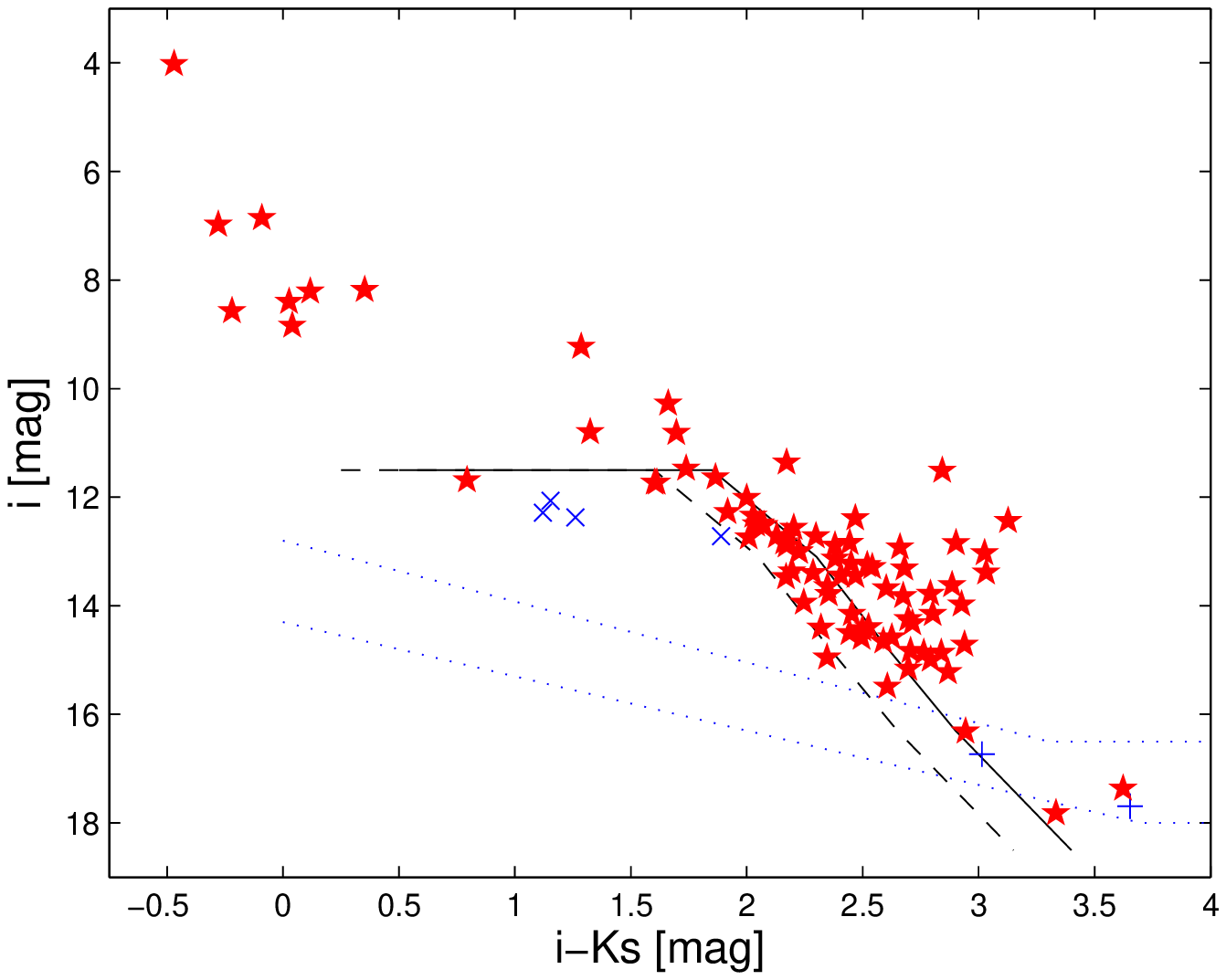}
\includegraphics[width=0.49\textwidth]{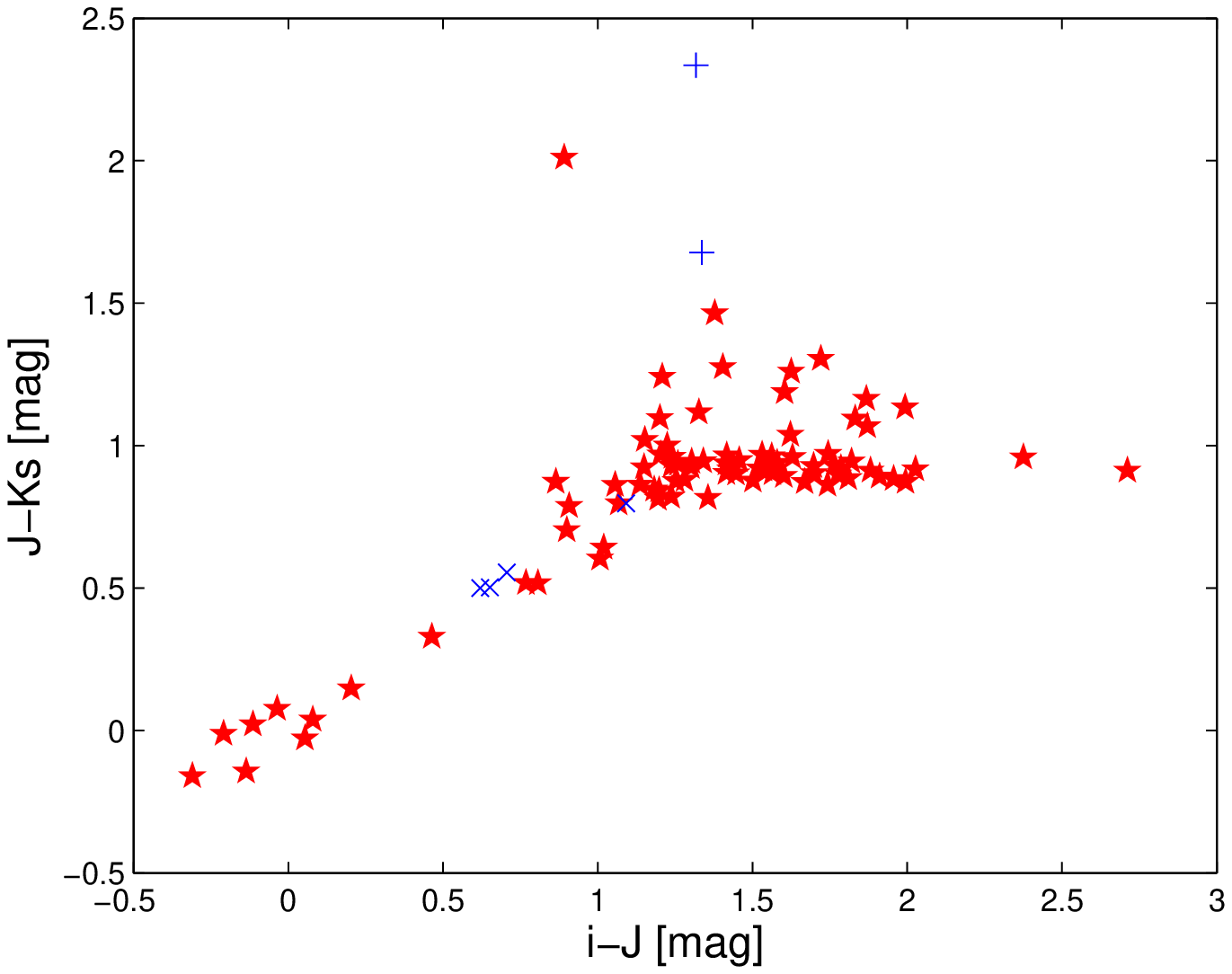}
\caption{Colour-magnitude and colour-colour diagrams.
The different symbols represent:
cluster star and brown dwarf members and candidates (--red-- filled stars);
field stars (--blue-- crosses), and galaxies (--blue-- pluses).
In the $i$ vs. $i-K_{\rm s}$ diagram in the top, the dotted (blue) lines are the
approximate completeness and detection limits of the combined DENIS-2MASS
cross-correlation. 
The solid (black) line is the criterion for selecting cluster stars and brown
dwarfs without known features of youth in $\sigma$~Orionis used by Caballero
(2008c).
The dashed (black) line is the criterion shifted bluewards by 0.25\,mag.
The reddest sources in the $J-K_{\rm s}$ vs. $i-J$ diagram in the bottom, with
colours $J-K_{\rm s} >$ 1.5\,mag, are the galaxies UCM0536--0239 and 2E~1456 and
the T~Tauri star Mayrit~609206 (V505~Ori).}  
\label{xfig_iJKs}
% sox05.m
\end{figure}

On the one hand, we have classified the 90 HRC-I sources with near-infrared
counterpart into 84 young cluster members and candidates, four X-ray field
stars, and two X-ray galaxies (Table~\ref{table.xraycounterparts}). 
Details on this classification are given next.
On the other hand, the 13 HRC-I sources without optical or near-infrared
counterparts at separations larger than 6$\sigma$ are galaxies (possibly active
galactic nuclei; L\'opez-Santiago \& Caballero 2008). 
The remaining four sources without (or with questionable) counterpart seem to be
two galaxies as well (Nos.~62 and~96; see above) and two cluster member
candidates (Nos.~97 and~107). 
Given the reasonable uncertainty in the actual nature of the last four sources,
we cautiously discarded them for next steps of the analysis.
Colour-magnitude and colour-colour diagrams in Fig.~\ref{xfig_iJKs} illustrate
the source classification.

\subsubsection{Cluster members and candidates}
\label{section.clustermembers}

Of the 84 young cluster members and candidates, 72 (86\,\%) have
uncontrovertible features of youth: 
OB spectral type, intense Li~{\sc i}~$\lambda$6707.8\,{\AA} resonant doublet in
absorption, mid-infrared flux excess due to a circumstellar disc, strong (broad,
asymmetrical) H$\alpha$ emission due to accretion, and/or weak alkali absorption
lines due to low gravity (Caballero 2008c and references therein;
Gonz\'alez-Hern\'andez et~al. 2008; Sacco et~al. 2008). 
Two of them are fainter than the star-brown boundary at $J \approx$ 14.5\,mag
(Caballero et~al. 2007) and are, therefore, {\em bona~fide} X-ray ``young brown
dwarfs'' (Section~\ref{section.browndwarfs}).
The other 70 cluster members are classified in
Table~\ref{table.xraycounterparts} as ``young stars''. 

There remain 12 stars that follow the photometric sequence defined by the
confirmed cluster stars in Fig.~\ref{xfig_iJKs} and that we classify as ``young
star candidates''. 
All of them have been classified in the
same way in other photometric (Wolk 1996; Sherry et~al. 2004; Scholz \&
Eisl\"offel 2004; Caballero 2007b; Hern\'andez et~al. 2007; Bouy et~al. 2009)
and X-ray (Franciosini et~al. 2006; Skinner et~al. 2008) searches in the
cluster. 
Of the young star candidates, there is spectroscopic information only for one.
\object{Mayrit~605079} (No.~95, [SWW2004]~127), a photometric member candidate
in Sherry et~al. (2004), was spectroscopically followed up by Sacco et~al. (2007,
2008).
They measured a radial velocity consistent with cluster membership, a faint
H$\alpha$ (chromospheric) emission, and a peculiar under-abundance of lithium.
They derived nuclear and isochronal ages about 10\,Ma older than expected for
$\sigma$~Orionis stars.
Mayrit~605079 might belong to a differentiated young stellar population in the
Orion Belt (Jeffries et~al. 2006; Caballero 2007a; Maxted et~al. 2008) or be
instead an active field M-dwarf interloper with CN contamination around
the Li~{\sc i} line (Caballero~2010). 
% (70+2) + 12

\subsubsection{Field stars}
\label{subsection.fieldstars}

Caballero (2006) took high-resolution spectra of the two stars associated to the
HRC-I sources Nos.~42 and~69, and found no trace of Li~{\sc i} in absorption
(except for H$\alpha$ when it is in emission, the Li~{\sc i} line is the most
obvious spectroscopic feature in young $\sigma$~Orionis stars of the same
magnitude as Nos.~42 and~69). 
The two of them were classified as non-cluster members by Caballero (2008c).

The star associated to the HRC-I source No.~51 was a photometric cluster member
candidate in Sherry et~al. (2004), but it has no lithium absorption, radial
velocity, and H$\alpha$ emission consistent with membership in $\sigma$~Orionis
according to Sacco et~al. (2008).

A fourth star, associated to the HRC-I source No.~31, was discovered and
spectroscopically investigated by Wolk (1996).
Its X-ray emission has been measured with {\em ROSAT} (Wolk 1996), {\em
XMM-Newton} (Franciosini et~al. 2006), and {\em Chandra} (Skinner et~al. 2008).
Given its location in the colour-magnitude diagram in Fig.~\ref{xfig_iJKs},
close to the confirmed field stars investigated by Caballero (2006) and its
unclear spectroscopic information (see footnote to Table~\ref{table.nonmayrit}),
we classify it as a ``possible field star''.

\subsubsection{Galaxies}
\label{subsection.galaxies}

There are two galaxies among the 90 HRC-I sources with 2MASS counterpart.
One is the very bright X-ray galaxy \object{2E~1456} (No.~9), which is extended
in optical and near-infrared images.
Besides, it has blue colours in the optical and red ones in the near infrared
(Caballero 2008c), an X-ray spectral energy distribution typical of an active
galactic nucleus (L\'opez-Santiago \& Caballero 2008), and irregular X-ray
variability (Caballero et~al. 2009). 
Bright X-ray galaxies towards the $\sigma$~Orionis cluster are not uncommon (see
also 2E~1448 in L\'opez-Santiago \& Caballero 2008, which is out of the HRC-I
field of view).
The other cross-matched galaxy is \object{UCM0536--0239} (No.~64).
It is a Type~1 obscured quasi-stellar object at a spectroscopic redshift
$z_{sp}$ = 0.2362$\pm$0.0005 (Caballero et~al. 2008 and references therein).
The two galaxies have peculiar colours if compared to stars without thick discs
(Fig.~\ref{xfig_iJKs}).

\subsection{X-ray light curves}
\label{section.xraylightcurves}

%______________________________________________ Figure 
\begin{figure}
\centering
\includegraphics[width=0.45\textwidth]{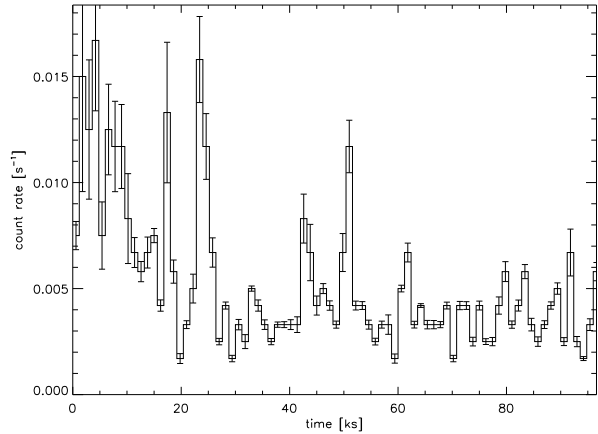}
\caption{A median HRC-I background light curve.
Note the high, decreasing, background level during the beginning of the
observation.} 
\label{xfig_background}
% jls
\end{figure}
%

%______________________________________________ Figure 
\begin{figure}
\centering
\includegraphics[width=0.49\textwidth]{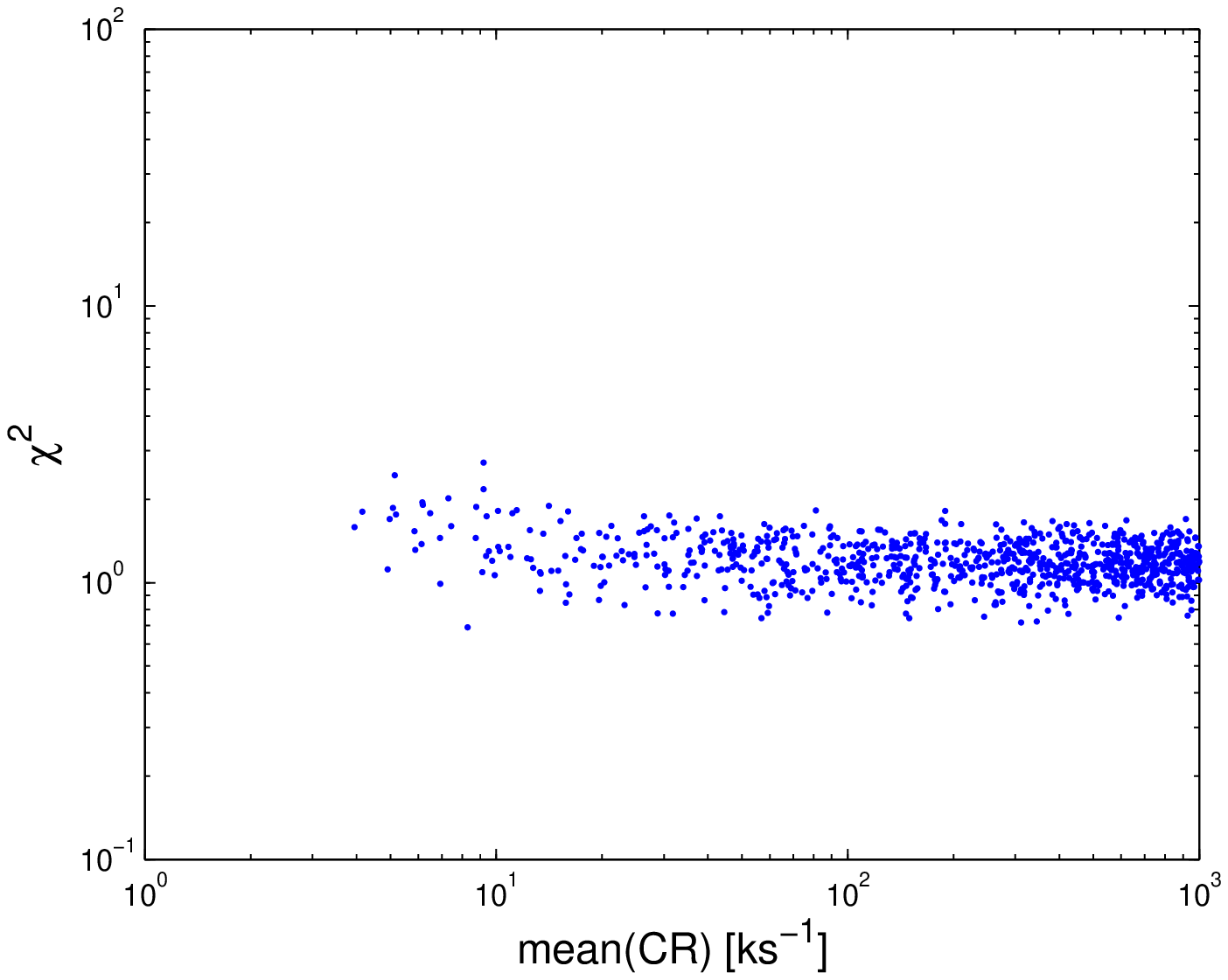}
\includegraphics[width=0.49\textwidth]{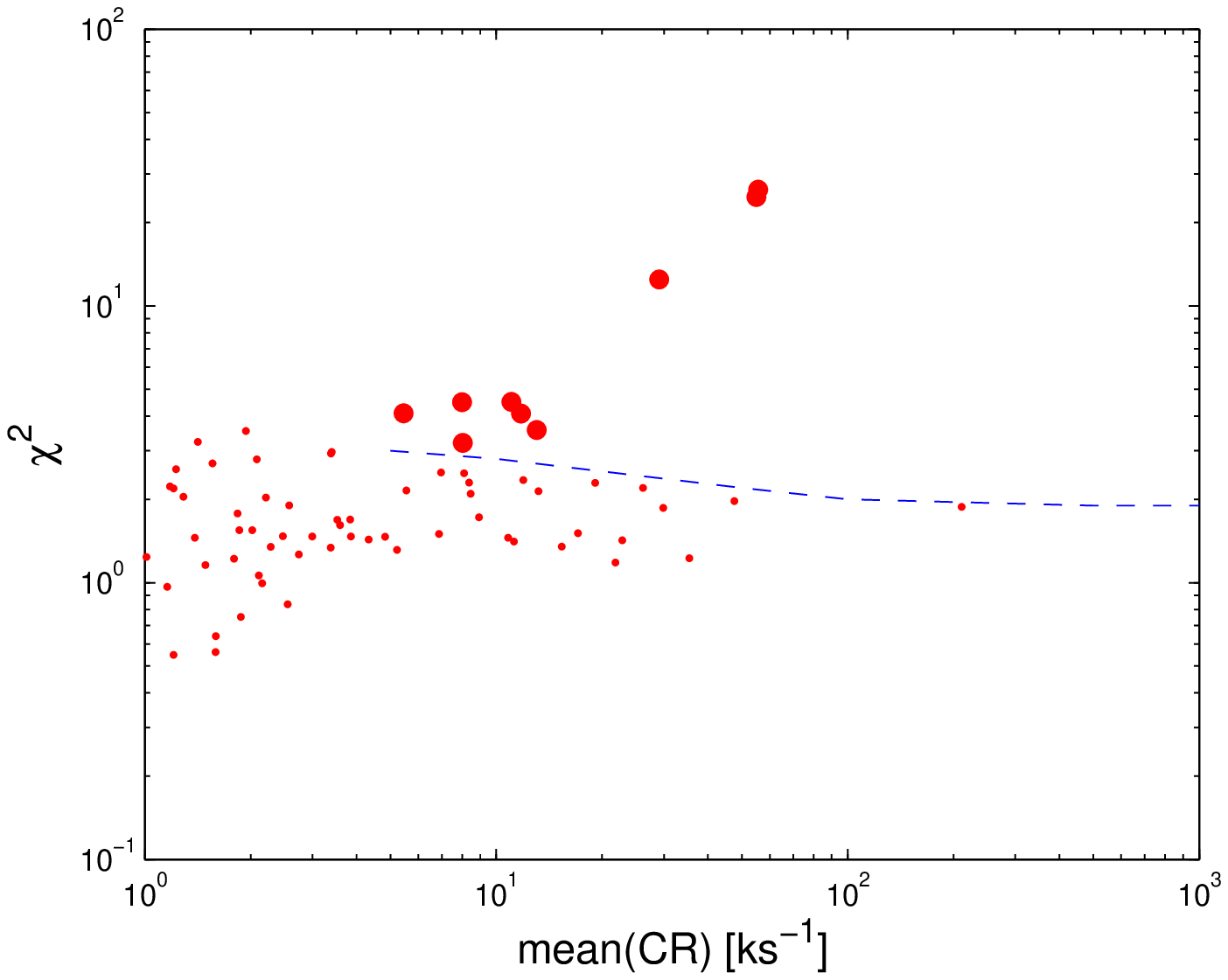}
\caption{{\em Top:} $\chi^2$ as a function of the mean count rate ($\chi^2_j$
vs. $\overline{CR_j}$ diagram) for 10$^3$ of the 10$^5$ X-ray simulated series.
{\em Bottom:} same as top window, but for the 107 X-ray real series. X-ray
sources above the dashed line have probabilities larger than 99.5\,\% of being
actual variables.
Light curves with mean count rates lower than 5\,ks$^{-1}$ were not used in the
statistical analysis.
Compare this figure with the Fig.~6 in Caballero et~al. (2009).} 
\label{xfig_chi2_meancr}
% ./Multicurve4/mc05b.m
% ./Multicurve4/mc01.m
\end{figure}

We built 107 X-ray light curves to look for flares and rotational modulation in
young stars. 
For each X-ray source, we integrated the numbers of HRC-I counts in two
circular areas of the same radius, one centred on the source itself and the
other one in a region free of X-ray sources for subtracting the background
level.
The integration radii varied between 7 and 30\,arcsec depending on the offaxis
distance (i.e., the size of the point spread function)
The bin size was fixed to 1200\,s. 
We discarded the first 5\,ks of each light curve because they were
affected by a relatively high background (this effect was only appreciable in
the faintest sources; Fig.~\ref{xfig_background}).

Next, we followed the same Poisson-$\chi^2$ analysis as in Caballero et~al.
(2009) on the 107 X-ray light curves to indentify variable sources
(Fig.~\ref{xfig_chi2_meancr}). 
This analysis provides similar results as applying Kolmogorov-Smirnov tests or
carrying out a visual inspection of the light curves.
We used the parameters $A$ = 76, $B$ = 0.40\,ks$^{-2}$, and $s$ = 2 in the
sigmoid relation between the number of events and the mean count rate, and the
expression $\delta {\rm CR_i} = 0.91287 {\rm CR_i}^{1/2}$ in the relation
between the individual count rates and their errors. 
In the case of the {\em Chandra} data, the above relations had much lower
uncertainties than for the {\em ROSAT} data in Caballero et~al. (2009).

\begin{figure*}
\centering
\includegraphics[width=0.32\textwidth]{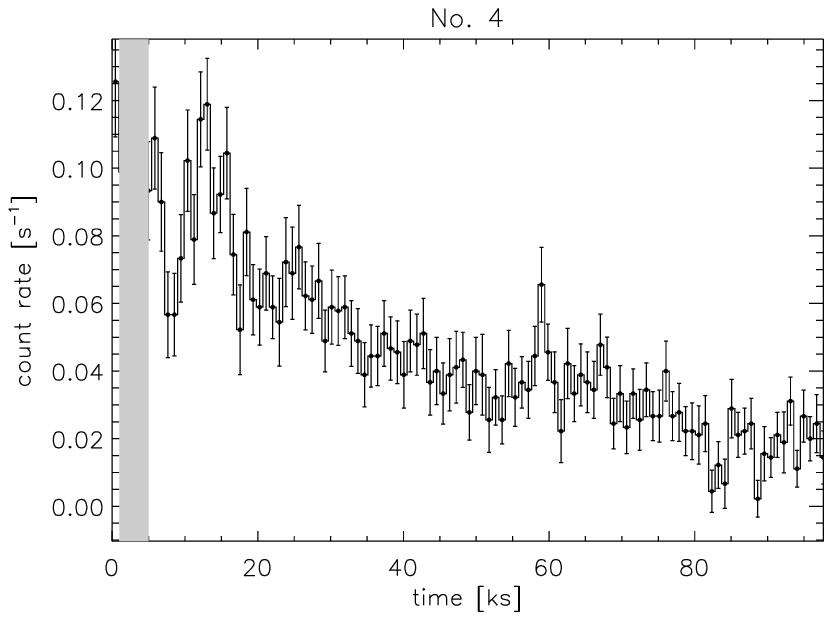}
\includegraphics[width=0.32\textwidth]{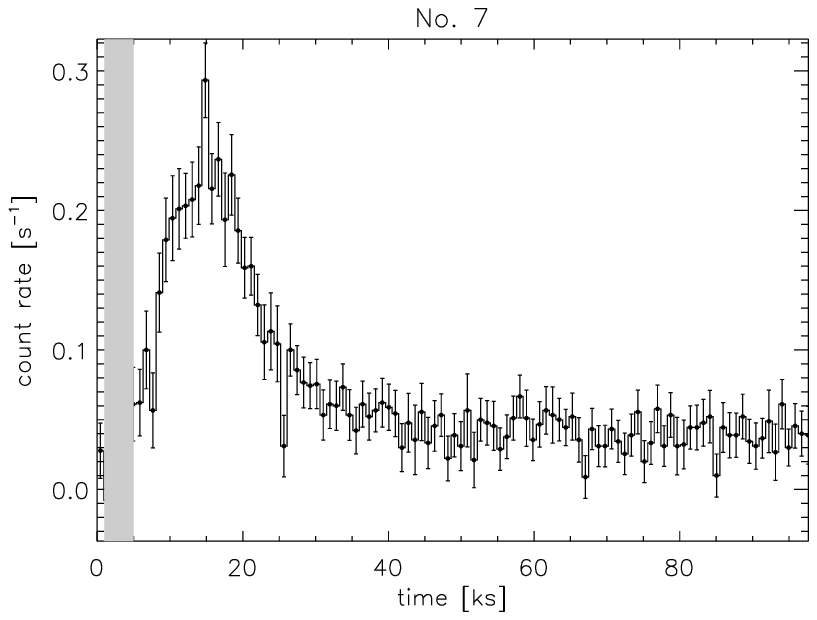}
\includegraphics[width=0.32\textwidth]{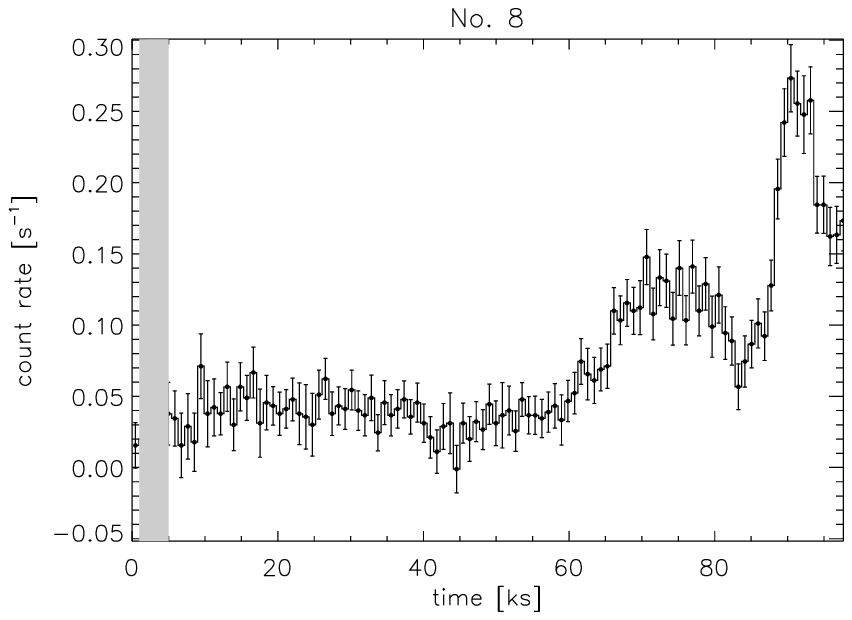}
\includegraphics[width=0.32\textwidth]{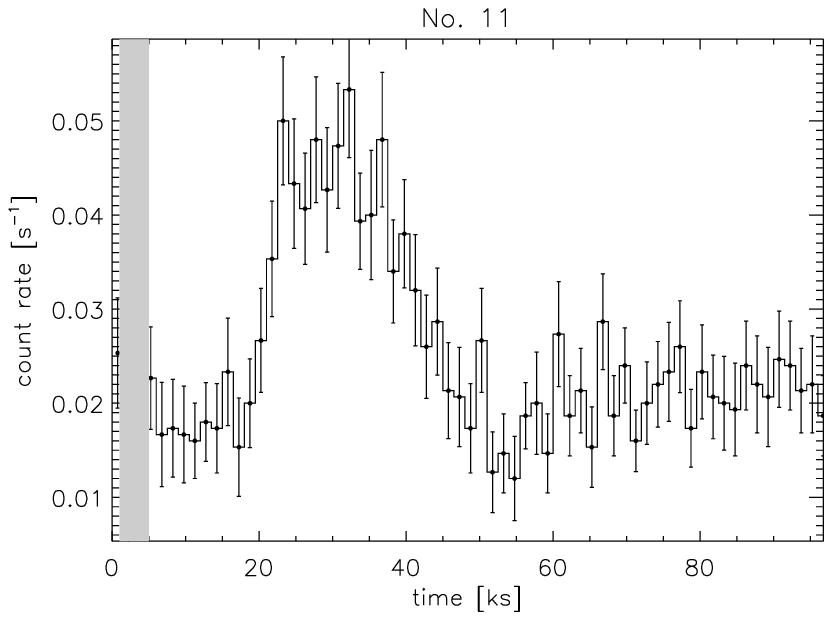}
\includegraphics[width=0.32\textwidth]{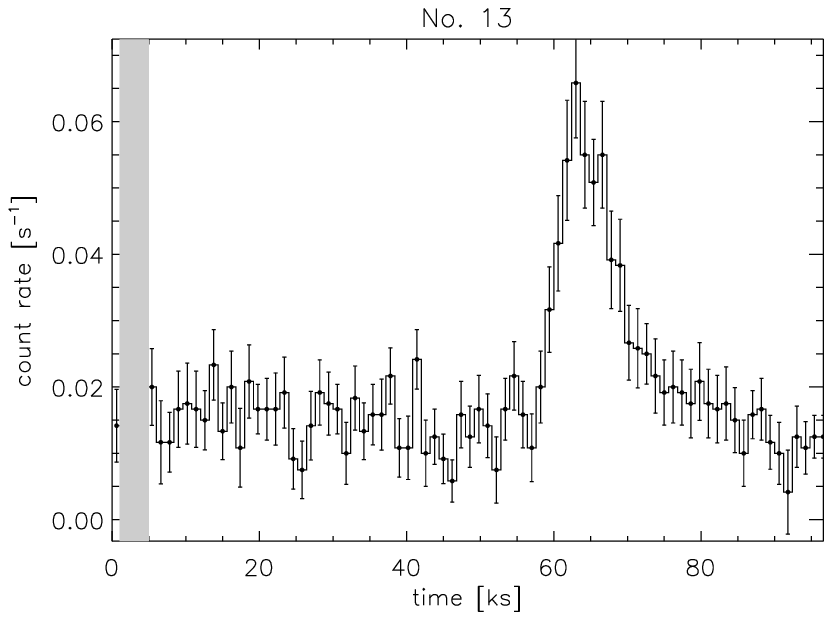}
\includegraphics[width=0.32\textwidth]{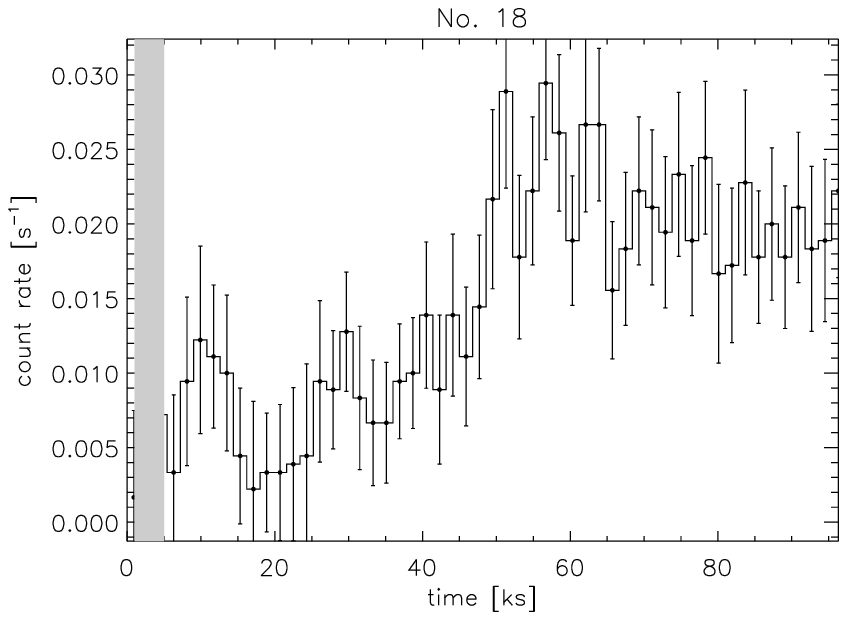}
\includegraphics[width=0.32\textwidth]{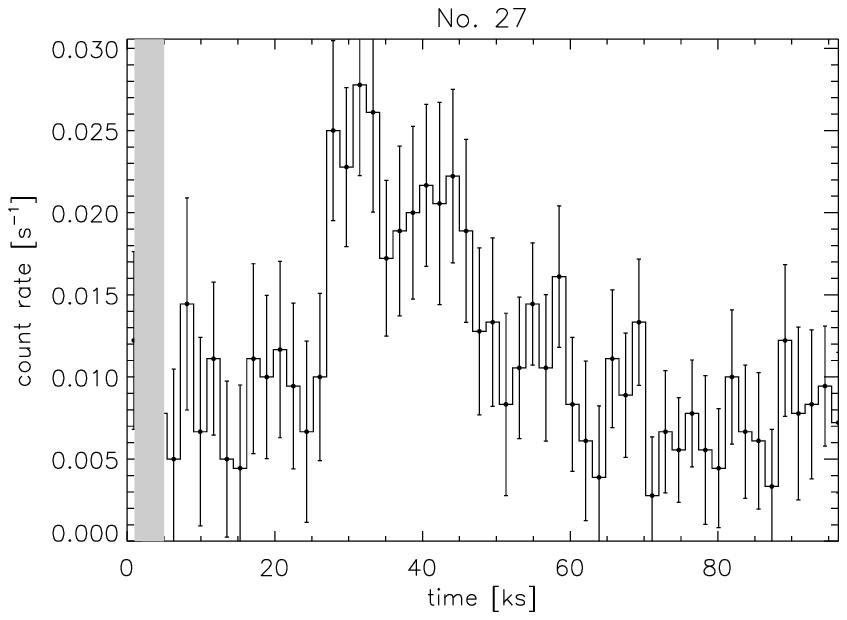}
\includegraphics[width=0.32\textwidth]{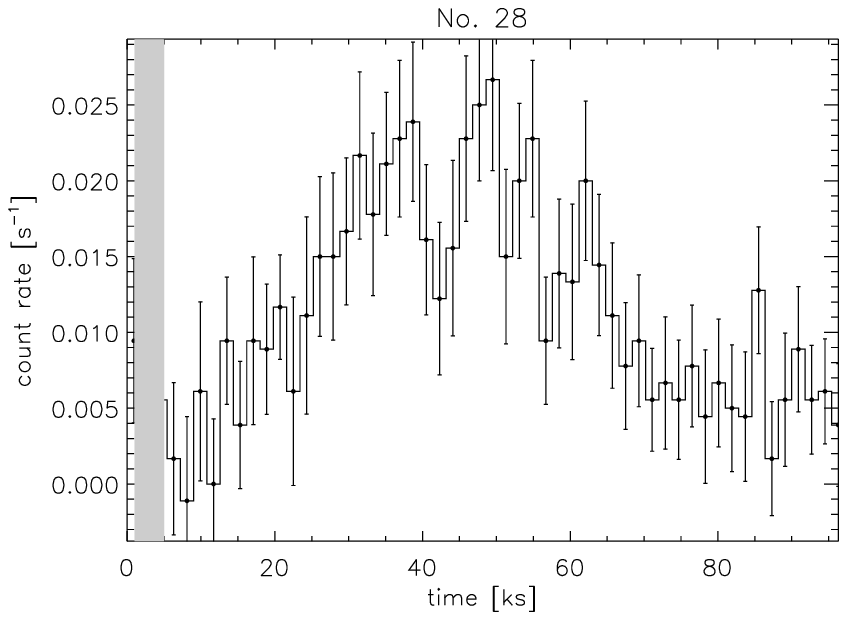}
\includegraphics[width=0.32\textwidth]{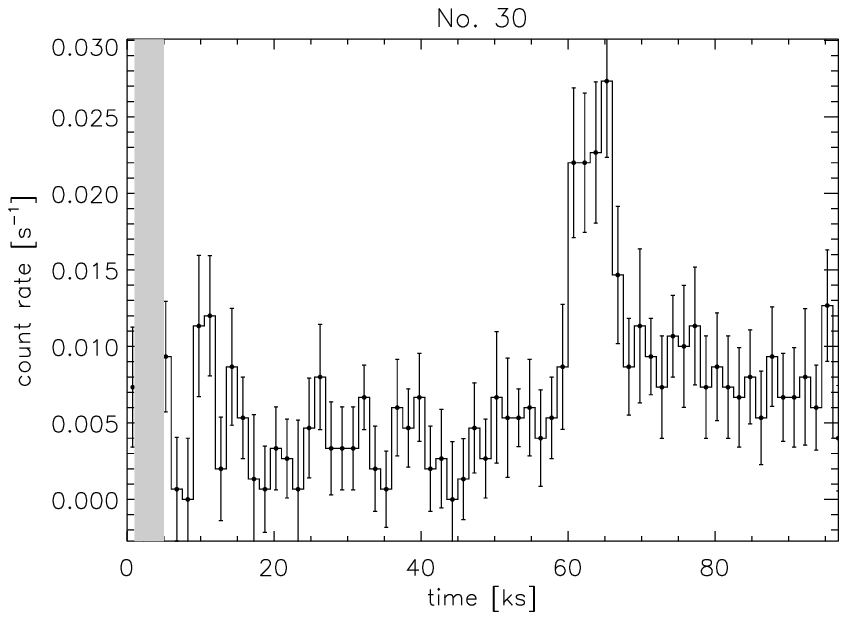}
\caption{HRC-I/{\em Chandra} light curves of the nine X-ray variable stars in
Table~\ref{table.variablestars}. 
The grey areas between 1 and 5\,ks indicate portions of all the light curves
affected by high background.} 
\label{xfig_n0}
\end{figure*}

Nine X-ray sources had probabilities of variability larger than a conservative
value of $p_{\rm var}$ = 99.5\,\% (Table~\ref{table.variablestars} and
Fig.~\ref{xfig_n0}).
The nine of them are $\sigma$~Orionis stars with signposts of youth.
Three stars (Nos.~7, 8, and~13) displayed apparent flares with
peak-to-quiescence ratios of about six and durations longer than 20\,ks. 
Besides, we detected in star No.~4 the long-lasting decay of a flare with an
expected peak-to-quiescence ratio larger than six. 
Other three stars (Nos.~11, 27, 30) also displayed flares during the
observations. 
On the contrary to the other two stars, the flare observed in the star No.~30
was relatively faint and short (it showed a ``spike'' flare following the
nomenclature by Wolk et~al. 2005). 

%______________________________________________ Figure 
%
% jls.idl
%

%__________________________________________________ 
   \begin{table}
      \caption[]{Sources with a probability of X-ray variability in the HRC-I
      data larger than $p_{\rm var}$ = 99.5\,\%.}    
         \label{table.variablestars}
     $$ 
         \begin{tabular}{l l cc l}
            \hline
            \hline
            \noalign{\smallskip}
No.	& Name				& $\overline{CR}$	& $\chi^2$	& Variability 		\\ % & \#    & JF	  Factor
	& 				& [ks$^{-1}$]		&		& type			\\ % &       & (--1)   
            \noalign{\smallskip}
            \hline
            \noalign{\smallskip}
4	& \object{Mayrit 348349} 	& 29.1			& 12.5		& Flare decay  		\\ % & 38    & 50	  >5.5       Haro 5-13
7	& \object{Mayrit 789281} 	& 54.9			& 24.8		& Flare  		\\ % & 2     & 1	  5.9	     2E 1454
8	& \object{Mayrit 863116} AB	& 55.6			& 26.3		& Flare with structure	\\ % & 107   & 139	  6.0	     RX J0539.6-0242 AB
11	& \object{Mayrit 156353} 	& 13.0			&  3.6		& Flare  		\\ % & 49    & 64	  3.3	     [SWW2004] J053843.449-023325.33
13	& \object{Mayrit 180277} 	& 11.8			&  4.1		& Flare  		\\ % & 20    & 27	  4.5	     [W96] rJ053832-0235b
18	& \object{Mayrit 403090}	& 11.0			&  4.5		& Rot. modulation?	\\ % & 96    & 124		     [W96] 4771-1038
27	& \object{Mayrit 489165}	&  8.0			&  3.2		& Flare  		\\ % & 77    & 99	  ~5	     [SWW2004] J053853.162-024353.05
28	& \object{Mayrit 489196} 	&  8.0			&  4.5		& Rot. modulation?	\\ % & 31    & 40	  3.4	     TY Ori
30	& \object{Mayrit 397060} 	&  5.4			&  4.1		& Flare  		\\ % & 92    & 120	  3.8	     V507 Ori
           \noalign{\smallskip}
            \hline
         \end{tabular}
     $$ 
   \end{table}
%

%______________________________________________ Figure 
\begin{figure*}
\centering
\includegraphics[width=0.33\textwidth]{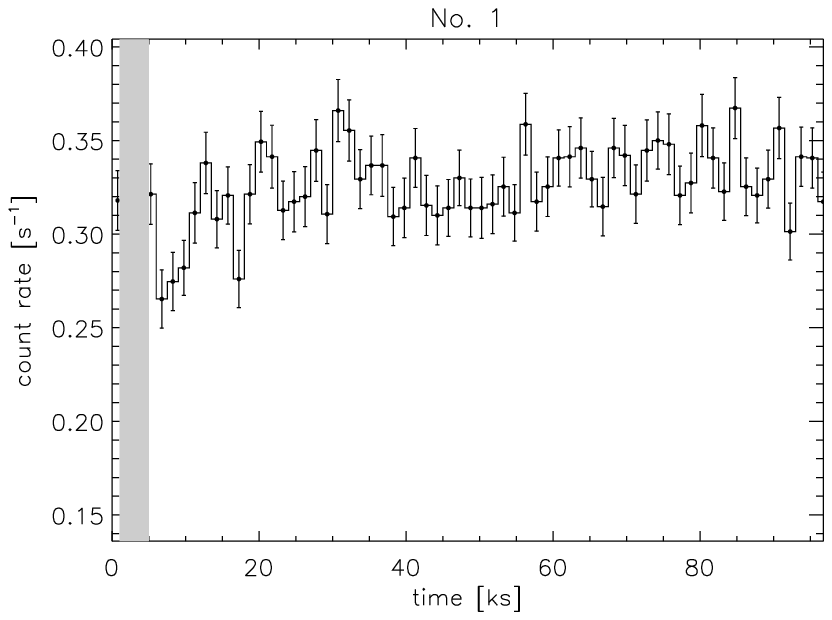}
\includegraphics[width=0.33\textwidth]{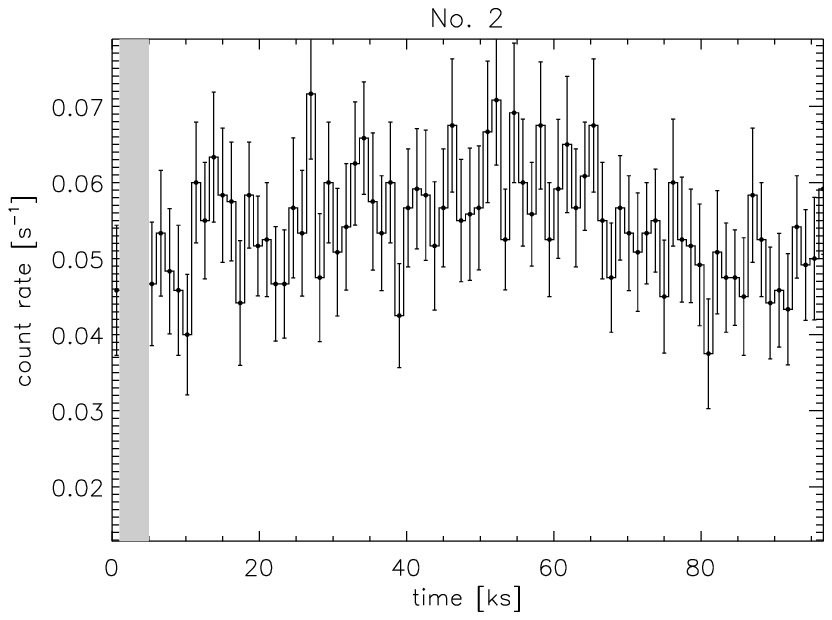}
\includegraphics[width=0.33\textwidth]{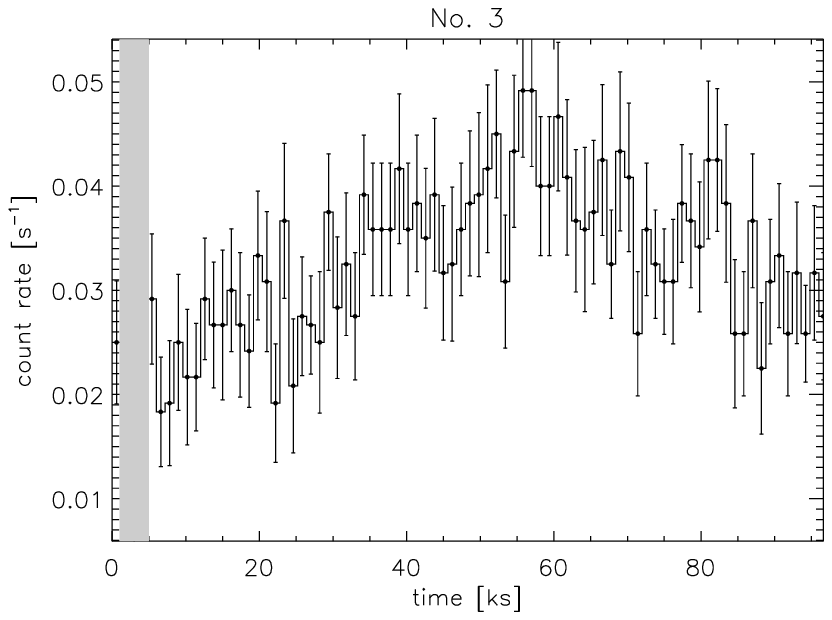}
\caption{Same as Fig.~\ref{xfig_n0}, but for three brightest X-ray stars:
Mayrit~AB ($\sigma$~Ori~AB, No.~1), Mayrit~114305~AB ([W96]~4771--1147~AB,
No.~2), and Mayrit~42062~AB ($\sigma$~Ori~E, No.~3).}  
\label{xfig_n001to3}
\end{figure*}
%
% jls.idl
%

The two remaining stars, Nos.~18 and~28, showed variations not clearly  
attributable to ``usual'' flares. 
The light curve of the source No.~28 is similar to that observed for
$\sigma$~Ori~E, a star with rotationally-modulated X-ray emission (see below).
The case of the source No.~18 is more complex.  
The count-rate enhancement suffered at about 40\,ks from the beginning of the
observation could be related to a persistent flare, although occultation of part
of the corona by a companion or of an active region by stellar rotation should
not be discarded.
Nevertheless, since HRC-I does not provide spectral energy information, we could
not perform an analysis of the time-resolved spectra to corroborate the
hypothesis of rotational modulation in the light curves of the stars Nos.~18
and~28.  

To date, there have been few incontestable cases of X-ray rotational modulation
in the $\sigma$~Orionis cluster (e.g., Franciosini et~al. 2006).
The most documented case is that of the bright B2Vpe star $\sigma$~Ori~E
(Mayrit~42062~AB, No.~3), which was found to have an X-ray emission modulated
with a period consistent with the stellar rotation, $P \sim$ 1.19\,d
($P \sim$ 103\,ks; Townsend et~al. 2010 and references therein), by Skinner
et~al. (2008). 
Our Poisson-$\chi^2$ analysis gave $\sigma$~Ori~E a low probability of
variability.
However, Caballero et~al. (2009) noticed that the methodology was sensitive to
flaring activity, but not to low-amplitude modulation.
We visually inspected the X-ray light curve of $\sigma$~Ori~E and detected a
modulation with a sinusoidal-like variation of the HRC-I count rate between 20
and 50\,ks$^{-1}$ and an estimated period slightly longer than the duration of
the observations ($>$97.6\,ks), which is also consistent with the rotational
period.  
On the contrary to Groote \& Schmitt (2004), Sanz-Forcada et~al. (2004), and 
Caballero et~al. (2009), who reported strong X-ray flares in the light curves of
$\sigma$~Ori~E, we found any (the flares are originated in its low-mass
companion; Caballero et~al. 2009). 
The light curve of $\sigma$~Ori~E is displayed in the right panel of
Fig.~\ref{xfig_n001to3} in comparison with the two brightest X-ray sources in
our HRC-I observations. 
The supposed stable light curve of $\sigma$~Ori~AB (Mayrit~AB, No.~1), whose
X-ray emission is likely originated in a strong wind (in particular for
$\sigma$~Ori~AB: Sanz-Forcada et~al. 2004; Skinner et~al. 2008 -- in general for
OB stars: Lucy \& White 1980; Owocki \&  Cohen 1999; Kudritzki \& Puls 2000
G\"udel \& Naz\'e 2009), had a $\chi^2$ value slightly below the limit $p_{\rm
var}$ that we adopted for variability, just as it occurred during the
ACIS-S observations by Skinner et~al. (2008).  
The light curve of the classical T~Tauri star Mayrit~114305~AB
([W96]~4771--1147~AB, No.~2) had a lower $\chi^2$ value of about 1.2, but 
showed a hint of rotational modulation.

\subsection{Beyond the completeness}
\label{section.beyond}

%__________________________________________________ 
   \begin{table*}
      \caption[]{Previously-known sources in the 10-spurious search and not in 
      Table~\ref{table.xraydetections}.}   
         \label{table.beyond}
     $$ 
         \begin{tabular}{l cc ccc c ll l}
            \hline
            \hline
            \noalign{\smallskip}
Name			& $\alpha$ 	& $\delta$ 	& $\Delta \alpha$, $\Delta \delta$	& $S$		& Offaxis	& CR	    	& NX  	& CXO	& Remarks		\\
			& (J2000)	& (J2000)	& [arcsec] 				& ($\sigma$)	& [arcmin] 	& [ks$^{-1}$]   & (Fr06)& (Sk08)&			\\
            \noalign{\smallskip}
            \hline
            \noalign{\smallskip}
Mayrit~734047		& 05 39 20.44 	& --02 27 36.8  & 5.54					& 4.92		& 12.0		& 2.1$\pm$0.6	& 159	& ...	& Li~{\sc i}, H$\alpha$ \\   % 132 84.835403 -2.461000  12426.3 20133.5  0.00211 0.00061	   084.835180	   -02.460224			12.149  0.027	11.425  0.026	11.168  0.023	3.4590  % Mayrit 734047      S Ori J053920.5-022737, [SWW2004] 100   Li I, Halpha, X
Mayrit~468096	     	& 05 39 15.83 	& --02 36 50.7  & 2.09					& 4.83		& 7.60		& 0.6$\pm$0.2	& 151	& ...	& Li~{\sc i}, Class II	\\   % 128 84.816254 -2.614055  12949.6 15950.9  0.00063 0.00020	   084.815958	   -02.614090			13.251  0.027	12.537  0.027	12.219  0.030	3.1500  % Mayrit 468096      [FPS2006] NX 151, [HHM2007] 967 II, X
Mayrit~441103	     	& 05 39 13.47   & --02 37 39.1  & 1.58					& 4.62		& 7.13		& 0.42$\pm$0.14 & 148	& ...	& ...			\\   % 125 84.805702 -2.627647  13237.8 15579.5  0.00042 0.00014	   084.806121	   -02.627531			13.409  0.027	12.771  0.023	12.497  0.027	3.0000  % Mayrit 441103      [FPS2006] NX 148	     X
{[FPS2006]~NX~120}	& 05 38 59.51   & --02 35 28.6  & 0.83				& 4.67		& 3.47		& 0.23$\pm$0.08 & 120	& 40	& Galaxy		\\   % 106 84.747955 -2.591248  14814.2 16574.3  0.00023 0.00008	   0	       0			       	0       0       0       0       0       0       2.8750  % ??? (close to 2MASS J05385930-0235282, [FPS2006] NX 120)
Mayrit~270181	     	& 05 38 44.49   & --02 40 30.5  & 0.90					& 4.71		& 4.60		& 0.22$\pm$0.09 & ...	& ...	& Li~{\sc i}, low~$g$	\\   % 69  84.685295 -2.674706  16524.7 14293.6  0.00022 0.00009	   084.685389	   -02.675140		       	13.365  0.034   12.724  0.033   12.497  0.035   2.4444  % Mayrit 270181       S Ori J053844.4-024030, [KJN2005] 5     Li I, low g
           \noalign{\smallskip}
            \hline
         \end{tabular}
     $$ 
   \end{table*}

We performed a new search of X-ray sources in our HRC-I data by imposing a
less-restrictive identification criterion.
In Section~\ref{section.sourcedetection}, we established only one spurious X-ray
source among the 107 (actually 109) detections.
In this case, we eased the identification of very faint sources close to the
noise limit by setting to ten the maximum of spurious X-ray sources with
PWDetect.
The corresponding background level translated into a final threshold
of significance of detection $S_{\rm min}$ = 4.6$\sigma$ (it was $S_{\rm min}$ =
5.1$\sigma$ for a maximum of one spurious X-ray source). 
The less-restrictive choice resulted in the detection of 142 sources (i.e., we
gained about 24 new reliable sources by accepting nine extra spurious
detections). 
However, the gain was not considerable because of the large contamination by
extragalactic sources at low X-ray count rates.

Of the 33 newly identified sources, we list five in Table~\ref{table.beyond}.
Four of them were identified in the X-ray observations by Franciosini et~al.
(2006). 
One of the four sources was also identified by Skinner et~al. (2008), which
supports our X-ray detections beyond the completeness.
There is optical/near-infrared counterpart for all HRC-I sources in
Table~\ref{table.beyond} except for [FPS2006]~NX~120 ([SSC2008]~40), which
is probably a galaxy (Franciosini et~al. 2006)\footnote{At less than
6\,arcsec from [FPS2006]~NX~120 lie 2MASS J05385930--0235282, a fore- or
background source based on $iJHK_{\rm s}$ colours, and {[BZR99]~S\,Ori~72}, a
young L/T-transition cluster member candidate or active galactic nucleus (Bihain
et~al. 2009).}.
%Besides, it {\em apparently} falls close to the mid-infrared source
%\object{$\sigma$~Ori~IRS~3B}, whose coordinates were incorrectly provided by
%Caballero et~al. (2008). 
%The correct coordinates of $\sigma$~Ori~IRS~3B are: 05~38~48.7 --02~35~30 J2000.
The four cross-matched X-ray sources are $\sigma$~Orionis cluster members and
candidates with faint X-ray emission (Caballero 2008c).
Of them, only Mayrit~441103 has no known feature of youth.
We followed the criterion in L\'opez-Santiago \& Caballero (2008) to discard the
remaining 29 X-ray sources without 2MASS counterpart (including [SSC2008]~40) as
stellar/substellar candidates, and classified them as objects of extragalactic
nature.

\section{Discussion}
\label{section.discussion}

\subsection{Short-term X-ray variability: HRC-I light curves}
\label{section.shortterm}

The nine X-ray variable sources in Table~\ref{table.variablestars} are young
stars in the $\sigma$~Orionis cluster. 
This makes a minimum frequency of X-ray variability of 11\,\% (9/84; it
increases to 12\,\% if we take into account $\sigma$~Ori~E).
The reader should compare this value with the ones of 36 and 39\,\% reported by
Franciosini et~al. (2006) and Caballero et~al. (2009), respectively, in the same
cluster, but using different sampling and datasets (in practice, 43\,ks of
continuous observations with {\em XMM-Newton} --Franciosini et~al. 2006-- and
one {\em ROSAT} visit per day during 34 days --Caballero et~al. 2009--). 
Although Skinner et~al. (2008) did not provide a frequency, we estimated a rough
value at 25\,\% from their data (see below).
We ascribed the low frequency derived by us to our conservative variability
criterion, rather than to the different completeness depths of the surveys.
Our value of 11\,\% is a lower limit to the X-ray frequency because there are
probable variable young stars that did not pass our filter.
For example, stars Nos.~16 (Mayrit~97212) and~17 (Mayrit~157155), which were not
listed in Table~\ref{table.variablestars}, displayed hints of rotational
modulation and flaring activity, respectively, after a visual inspection.

We also compared our derived flare rate with other measurements in the
literature.
With seven flares detected during our observation among 84 young stars and
candidates, we derived $1/1180$ flares per star per kilosecond.
This value decreased to less than about $1/1070$ when we discarded the
early-type (OB) cluster stars (Section~\ref{section.earlytype}). 
Both corrected and uncorrected values are consistent with previous
determinations of flare rates, although we did not consider the completeness
for flare detection.
For example, with different instruments, sensitivities, flare definitions and
energies, data biases, extragalactic contaminations, and stellar spectral-type
intervals, Wolk et~al. (2005), Albacete-Colombo et~al. (2007), and Stelzer
et~al. (2007) reported flare rates of $1/1150$, $1/610$, and $1/1320$ flares per
star per kilosecond, respectively, in star-forming regions slightly younger than
$\sigma$~Orionis ({Orion Nebula Cluster}, {Cyg~OB2}, and
{Taurus}; $\tau \sim$ 1--2\,Ma).  

Several stars in Table~\ref{table.variablestars} have been previously reported
to display X-ray variability.
By applying Kolmogorov-Smirnov tests on the unbinned photon arrival times,
Franciosini et~al. (2006) found that roughly a half of the (weak-line and
classical) T~Tauri stars in $\sigma$~Orionis were variable at the 99\,\%
confidence level. 
Eight cluster members with signposts of youth and two candidate members showed
clear flares during their {\em XMM-Newton} observations. 
Of them, we were able to detect the X-ray emission in the HRC-I image of five
stars, of which only one displayed variability during our observations (No.~28,
Mayrit~489196, [FPS2006]~NX~61), but of rotational-modulation type.
However, the two X-ray light curves obtained with {\em XMM-Newton} and {\em
Chandra} resemble each other, so we may face the same variability type (e.g., a
low-amplitude, long-lasting flaring activity). 
Franciosini et~al. (2006) also reported five young stars showing significant
variability not clearly attributable to flares.
We detected the five of them and found that one, No.~11 (Mayrit~156353,
[FPS2006]~NX~76), displayed a flare during our observations.
In contrast, during the entire {\em XMM-Newton} observations, the star showed a
steady decay by a factor of $\sim$2, which we attribute to the decay of a
long-lasting flare. 
The frequencies of X-ray rotational modulation reported by us and Franciosini
et~al. (2006) are consistent with the approximate interval 1--3\,\%.

The list of ten variables in Skinner et~al. (2008) included Mayrit~42062~AB
($\sigma$~Ori~E; see above), the unseen galaxy associated to No.~24 (a slow
low-amplitude variable X-ray with unusual hardness and without
optical/near-infrared counterpart), and some young stars with slow decline
(No.~5, Mayrit~203039) or increase (No.~25, Mayrit~3020~AB) in count rate. 
Besides, X-ray flares were visible in Mayrit~105249 (No.~12; variable in
Franciosini et~al. 2006) and, possibly, Mayrit~92149~AB (No.~29).
If we do not take into account $\sigma$~Ori~E, there are no stars in common in
the lists of variable X-ray sources in Skinner et~al. (2008) and our work. 

Besides, two of the five most variable stars in the study by Caballero et~al.
(2009; Section~\ref{section.hrirosat}) appear also in
Table~\ref{table.variablestars}. 
They are Mayrit~863116~AB (No.~8) and Mayrit~156353 (No.~11).
Interestingly, the HRC-I light curve of the bright star Mayrit~863116~AB showed
a flare with structure.
The double hump may be originated in a series of two flares of different shape
or in only one flare that was occulted by stellar rotation, a companion, or a
disc (Mayrit~863116~AB seems to be a spectroscopic binary with a warm
circumstellar disc; Caballero et~al. 2009). 
	
There are only a few stars that have been repeatedly found to display the same
X-ray variability type, such as the bright early-type $\sigma$~Ori~E and
T~Tauri Mayrit~863116~AB stars.  
To sum up, some young X-ray stars that displayed variability at other epochs did
not do it during our observations, and vice~versa.
This result was expected from the relatively low flare rate measured above of
one flare per star every one or two weeks.
As a result, the variability frequencies given above depend on several
factors including the sensitivity, length, and  energy bandpass of the
observation and can only be taken as lower limits.

\subsection{Long-term X-ray variability: comparison to previous X-ray surveys in
$\sigma$~Orionis} 
\label{section.longterm}

%______________________________________________ Figure 
\begin{figure*}
\centering
\includegraphics[width=0.49\textwidth]{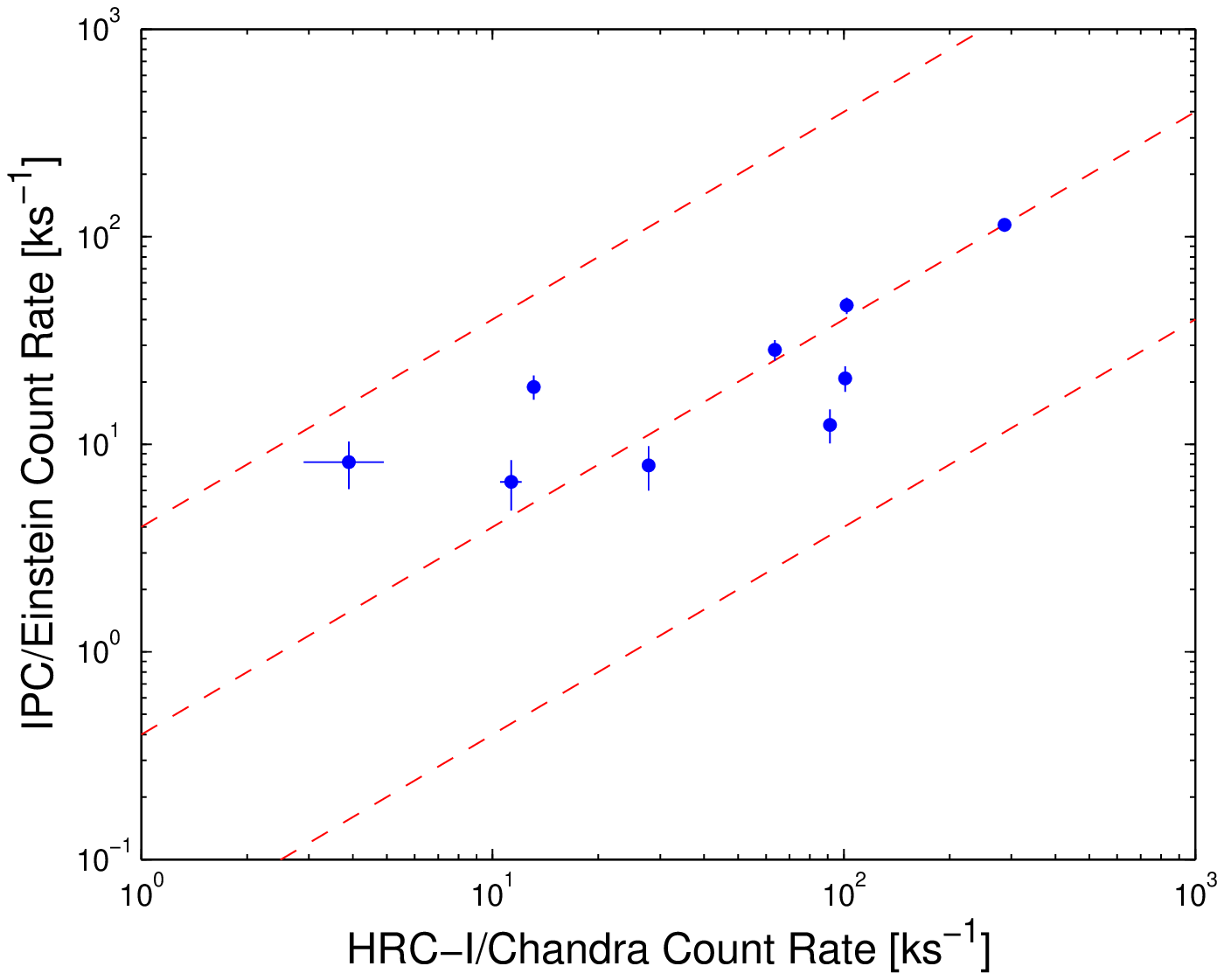}
\includegraphics[width=0.49\textwidth]{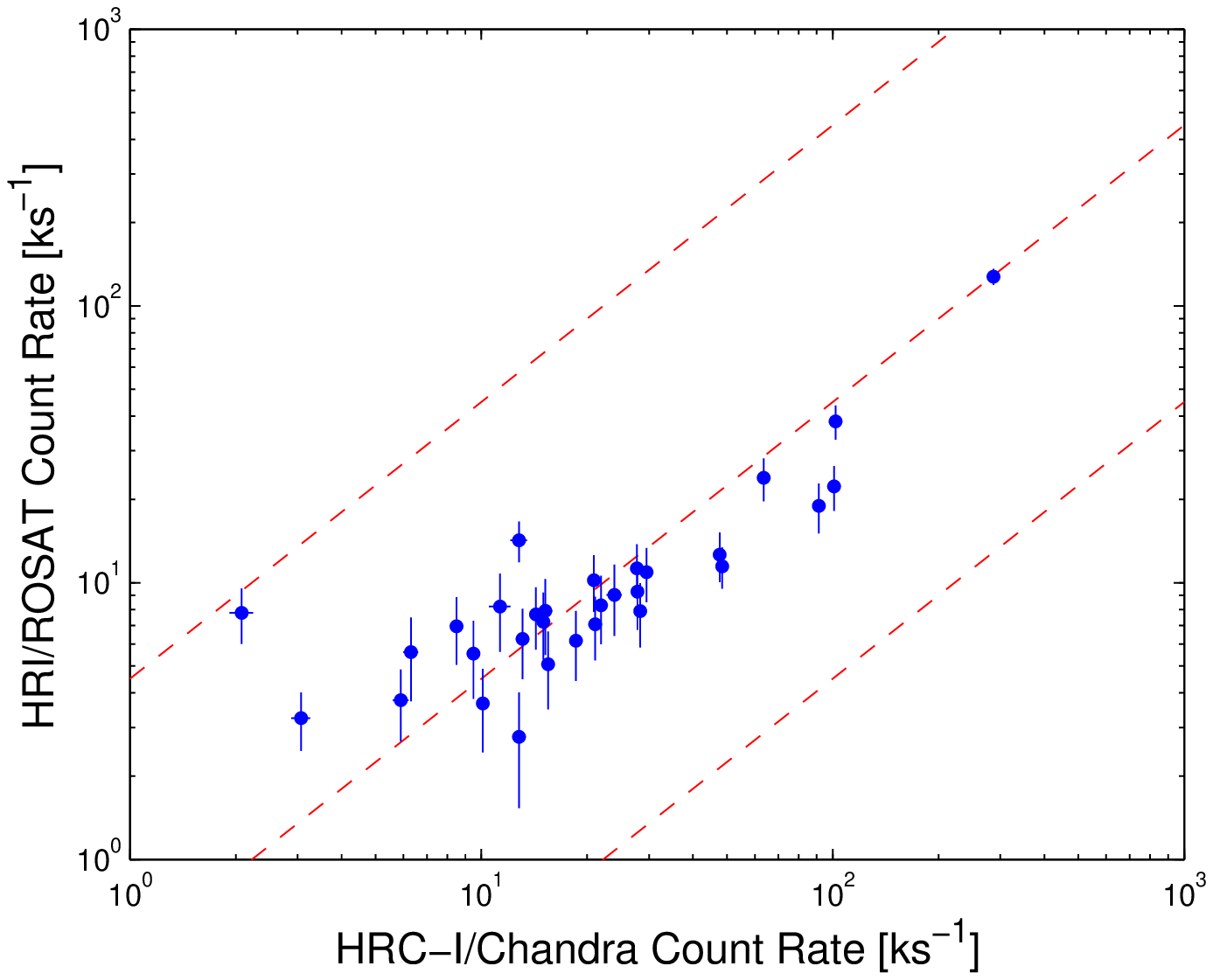}
\includegraphics[width=0.49\textwidth]{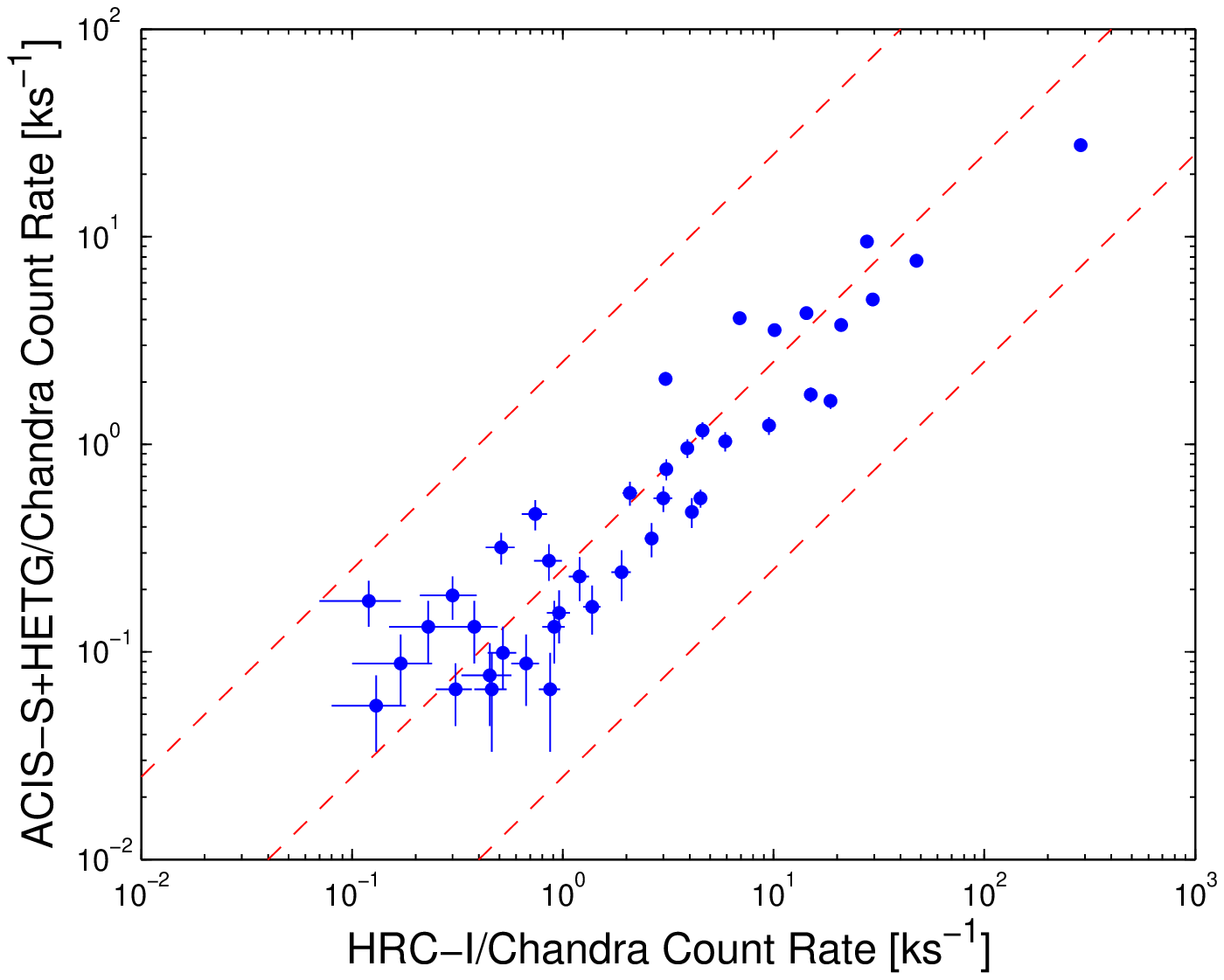}
\includegraphics[width=0.49\textwidth]{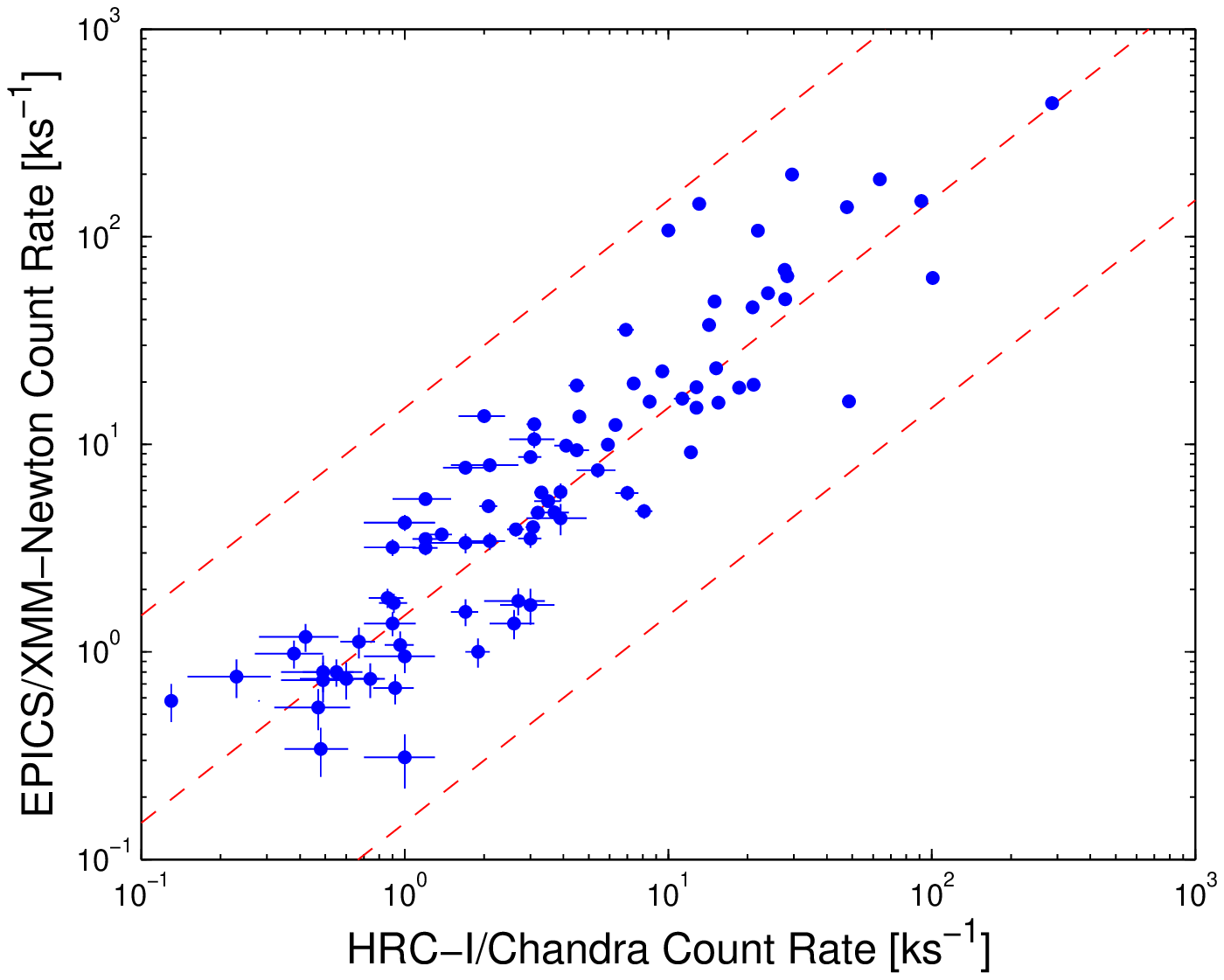}
\caption{Count rates of IPC/{\em Einstein} ({\em top left}), HRI/{\em ROSAT}
({\em top right}), ACIS-S+HETG/{\em Chandra} ({\em bottom left}), and EPIC/{\em
XMM-Newton} ({bottom right}) as a function of count rates of
HRC-I/{\em Chandra}.
The dashed lines indicate IPC--, HRI--, ACIS-S+HETG--, and EPIC--HRC-I
count-rate ratios of 4.00, 0.40, and 0.04, 4.50, 0.45, and 0.045, 2.50, 0.25,
and 0.025, and 15.0, 1.50, and 0.15 from top to bottom, respectively.
The OB-type binary star $\sigma$~Ori~AB has {\em not} been used as a
reference in the ACIS-S+HETG--HRC-I comparison.} 
\label{xfig_othermission_crhrc-i}
% xfig_cripc_crhrc-i: 		sox08.m
% xfig_crhri_crhrc-i: 		sox10.m
% xfig_cracis-i_crhrc-i: 	sox11.m
% xfig_crepics_crhrc-i:		sox12.m
\end{figure*}
%

%__________________________________________________ 
   \begin{table}
      \caption[]{Energy bands, spatial resolutions, and field of view of some
      X-ray instruments onboard space missions$^{a}$.}   
         \label{table.spacemissions}
     $$ 
         \begin{tabular}{ll ccc}
            \hline
            \hline
            \noalign{\smallskip}
Space 		& Instrument	& Energy	& Resolution	& FoV	\\
mission		& 		& [keV]		& [arcsec]	& [arcmin]\\
            \noalign{\smallskip}
            \hline
            \noalign{\smallskip}
{\em Einstein} 	& HRI		& 0.2--3.0	& 4		& 25	\\
	 	& IPC		& 0.3--3.5	& 60		& 75	\\
{\em ROSAT} 	& HRI		& 0.1--2.4	& 5		& 20\,$\times$\,20\\
	 	& PSPC		& 0.1--2.4	& 15		& 114	\\
{\em Chandra} 	& HRC-I		& 0.08--10	& 0.4		& 31\,$\times$\,31\\
	 	& ACIS-S	& 0.2--10	& 1.2		& 16\,$\times$\,16\\
{\em XMM-Newton}& PN		& 0.2--15	& 6		& 30 	\\
		& MOS		& 0.2--12	& 6		& 30	\\
          \noalign{\smallskip}
            \hline
         \end{tabular}
     $$ 
\begin{list}{}{}
\item[$^{a}$] See an exhaustive compilation of parameters of X-ray detectors at 
{\tt http://space.mit.edu/$\sim$jonathan/xray\_detect.html}.
\end{list}
   \end{table}

All the large space missions able to observe low- to mid-energy X-rays, i.e.
{\em Einstein Observatory} (HEAO-2), {\em ROSAT} (R\"ontgensatellit), {\em
XMM-Newton}, and {\em Chandra}, have observed the $\sigma$~Orionis region in
detail (Section~\ref{section.comparison}).
Besides, the Advanced Satellite for Cosmology and Astrophysics ({\em ASCA})
observed nearby areas close to the Horsehead Nebula and {Alnitak}
($\zeta$~Ori).
In principle, the different pointing centres and exposure times of the
observations, the singular apertures, fields of view, spatial resolutions and,
specially, detector responses of the instrument/telescope systems
(Table~\ref{table.spacemissions}), and the ``colours'' and intrinsic variability
of the X-ray sources avoid a direct comparison between previous results and
ours.
In spite of these differences, we expected to find a correlation between
count rates measured by HRC-I and the other used instruments and to identify
X-ray sources that deviate from the general trends.
See, e.g., the {\em Einstein}-{\em ROSAT} comparison in the Pleiades by Stauffer
et~al. (1994).

%__________________________________________________ 
   \begin{table}
      \caption[]{Long-term X-ray variable stars.}   
         \label{table.variable}
     $$ 
         \begin{tabular}{ll c l}
            \hline
            \hline
            \noalign{\smallskip}
No.	& Name				& Variability 	& Instrument 	\\
	&				& factor$^{b}$	& 		\\
            \noalign{\smallskip}
            \hline
            \noalign{\smallskip}
3	& \object{Mayrit 42062} AB	& 4.5		& EPIC	\\
4	& Mayrit 348349			& 0.22		& EPIC	\\
16	& \object{Mayrit 97212}		& 7.2		& EPIC	\\
20	& \object{Mayrit 344337} AB	& 7.3		& EPIC	\\
37	& \object{Mayrit 102101} AB	& 8.3		& HRI	\\ % 1995~Mar~04
80	& \object{Mayrit 497054}	& 4.6		& EPIC	\\
84$^{a}$& \object{Mayrit 433123}	& 0.21		& EPIC	\\
97	& No.~97			& 5.9		& ACIS-S\\ 
99	& \object{Mayrit 957055}	& 5.2		& IPC	\\ % 1980 Mar 02
...	& \object{Mayrit 631045}$^{f}$	& $\gtrsim$4.8	& EPIC	\\
...	& \object{Mayrit 662301}$^{f}$	& $\gtrsim$6.3	& EPIC	\\
...	& \object{Mayrit 841079}$^{f}$	& $\gtrsim$4.4	& EPIC	\\
         \noalign{\smallskip}
            \hline
         \end{tabular}
     $$ 
\begin{list}{}{}
\item[$^{a}$] See Section~\ref{section.browndwarfs} for a discussion on
the brown dwarf No.~84/Mayrit~433123. 
\item[$^{b}$] Quotient of the measured count-rate ratio CR$_{1}$/CR$_{2}$ and 
the average count-rate ratio (CR$_{1}$/CR$_{2}$)$_{0}$, where 1 denotes the 
space-mission instrument listed in the last column and 2 denotes 
HRC-I/{\em Chandra}.
The values of (CR$_{1}$/CR$_{2}$)$_{0}$ are 0.40 (IPC/{\em Einstein}), 0.45 
(HRI/{\em ROSAT}), 0.25 (ACIS-S/{\em Chandra}), and 1.50 
(EPIC/{\em XMM-Newton}).
\end{list}
   \end{table}

The ``long-term variability'' found in our comparison and summarised in
Table~\ref{table.variable} and Fig.~\ref{xfig_othermission_crhrc-i} may
actually be the result of observing an X-ray source with short- or mid-term
variability (in scales of hours or a few days; e.g., flares) at two separated
epochs. 
In particular, nine $\sigma$~Orionis stars and one galaxy displayed
quotients of the measured and average count-rate ratios larger than 4 or smaller
than 1/4.
Some of them showed variations of a factor 7 or more or were identified to
vary in different comparisons:
\begin{itemize}
\item No.~3/Mayrit~42062~AB underwent flaring-like activity during the EPIC
observations. 
\item No.~4/Mayrit~348349 showed an apparent flare decay during our HRC-I
observations and other strong flare during the HRI/{\em ROSAT} ones.
\item No.~16/Mayrit~97212 and No.~20/Mayrit~344337~AB showed significant
variability not clearly attributable to flares during the EPIC observations.
\item No.~37(Mayrit~102101~AB underwent a strong flare during HRI observations.
\item The stars Mayrit~631045, Mayrit~662301, and Mayrit~841079, with
designations NX~149, NX~7, and NX~174, respectively, in Franciosini et~al.
(2006; Section~\ref{section.epicxmmnewton}) displayed flares and were bright
enough during EPIC observations to be fitted to one-temperature models.
Mayrit~841079 (V603~Ori) is the source of the Herbig-Haro object \object{HH~445}
(Reipurth et~al. 1998; Andrews et~al. 2004).
\end{itemize}

\subsection{The cluster X-ray luminosity function}
\label{section.luminosityfunction}

%______________________________________________ Figure 
\begin{figure}
\centering
\includegraphics[width=0.49\textwidth]{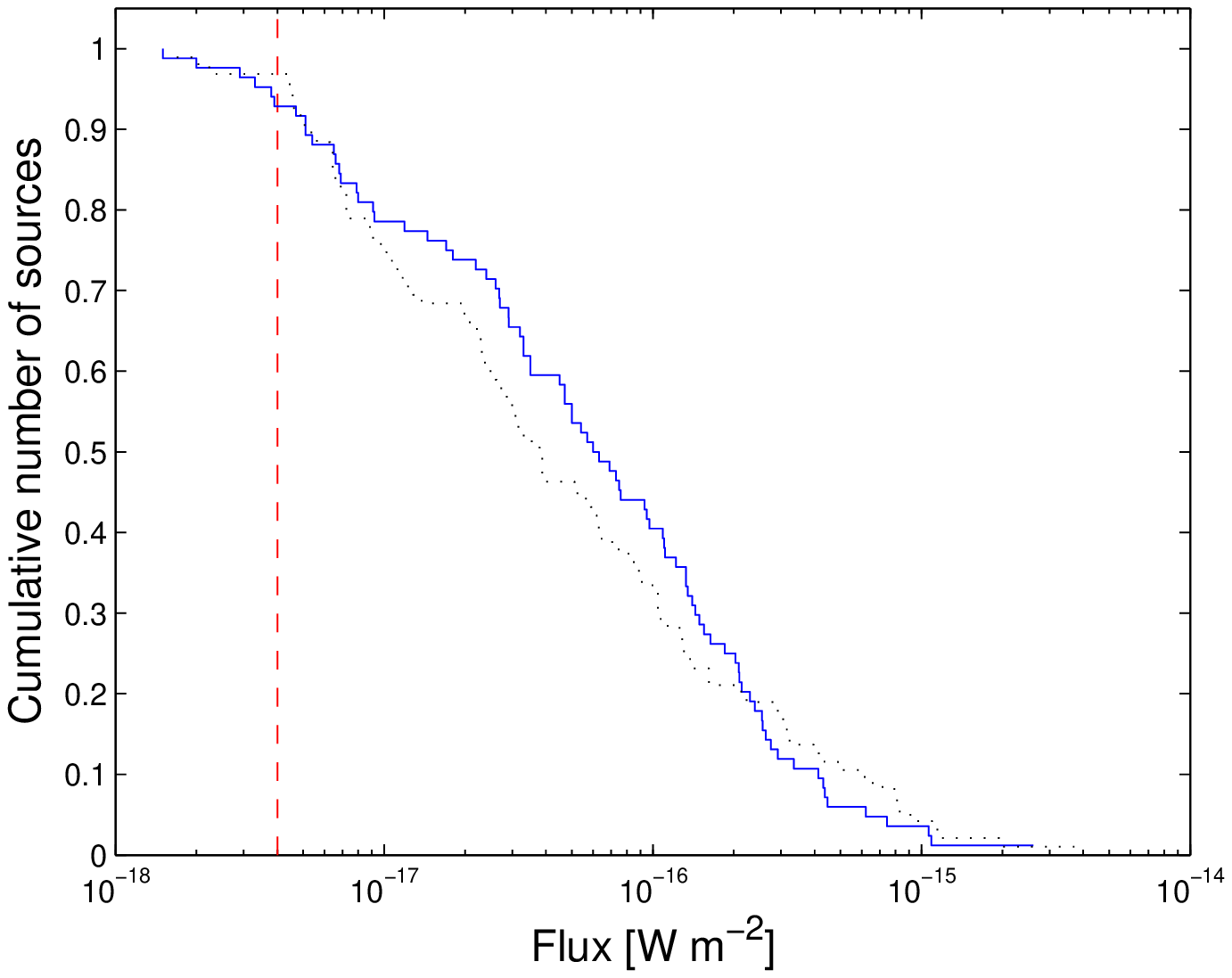}
\includegraphics[width=0.49\textwidth]{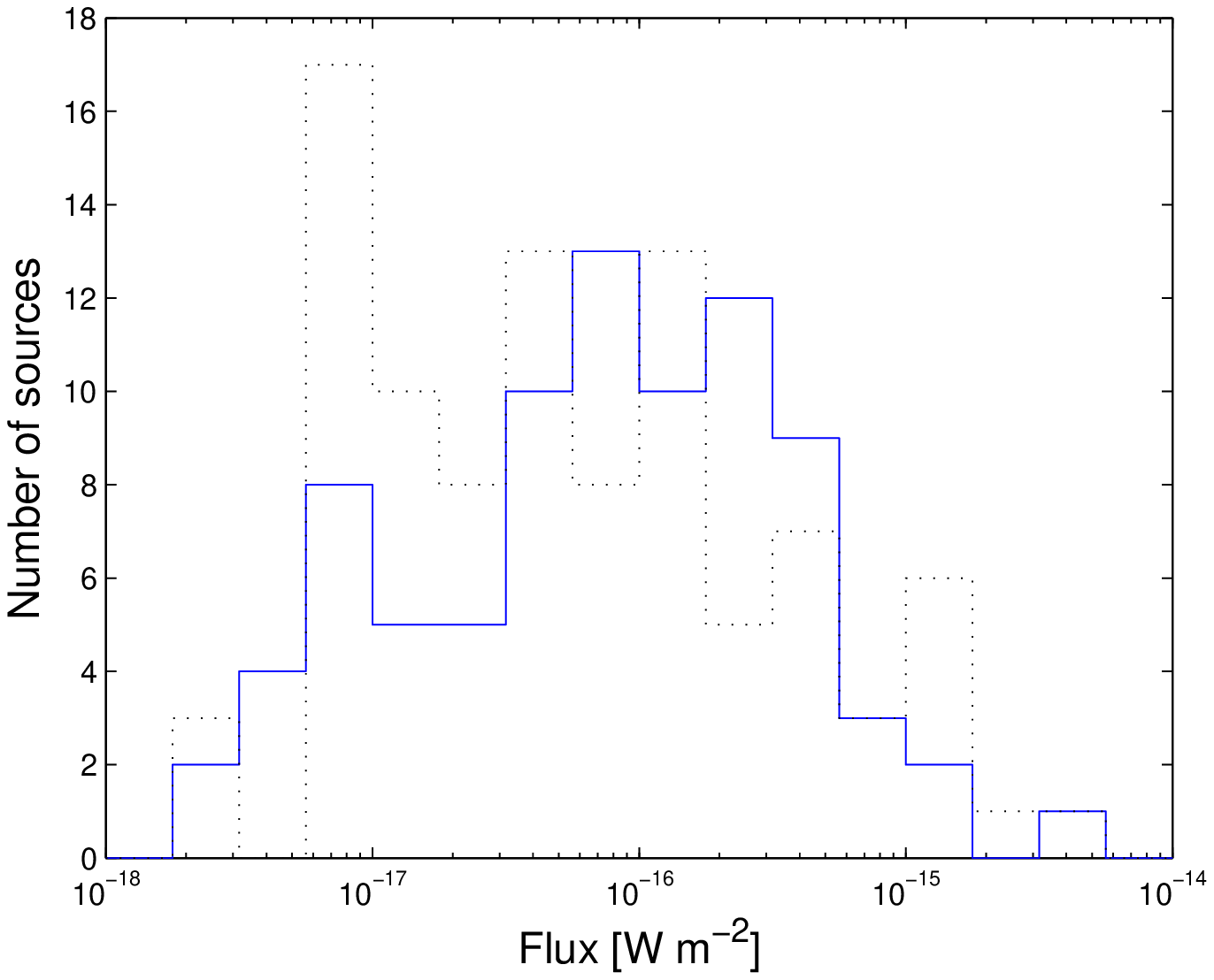}
\caption{{\em Top panel:} same as Fig.~\ref{xfig_relN_flux}, but only for young
stars, young star candidates, and possible young stars in $\sigma$~Orionis (as
classified in Table~\ref{table.xraydetections}). 
The dotted line indicates the relative cumulative number of Franciosini
et~al. (2006) EPIC X-ray sources as a function of apparent flux. 
Except for a $4 \pi d^2$ factor, the two curves delineate the cumulative
X-ray luminosity function of the cluster.
{\em Bottom panel:} same as the top panel, but in an histogram.}   
\label{xfig_xlf}
% sox14.m
\end{figure}
%

%______________________________________________ Figure 
\begin{figure}
\centering
\includegraphics[width=0.49\textwidth]{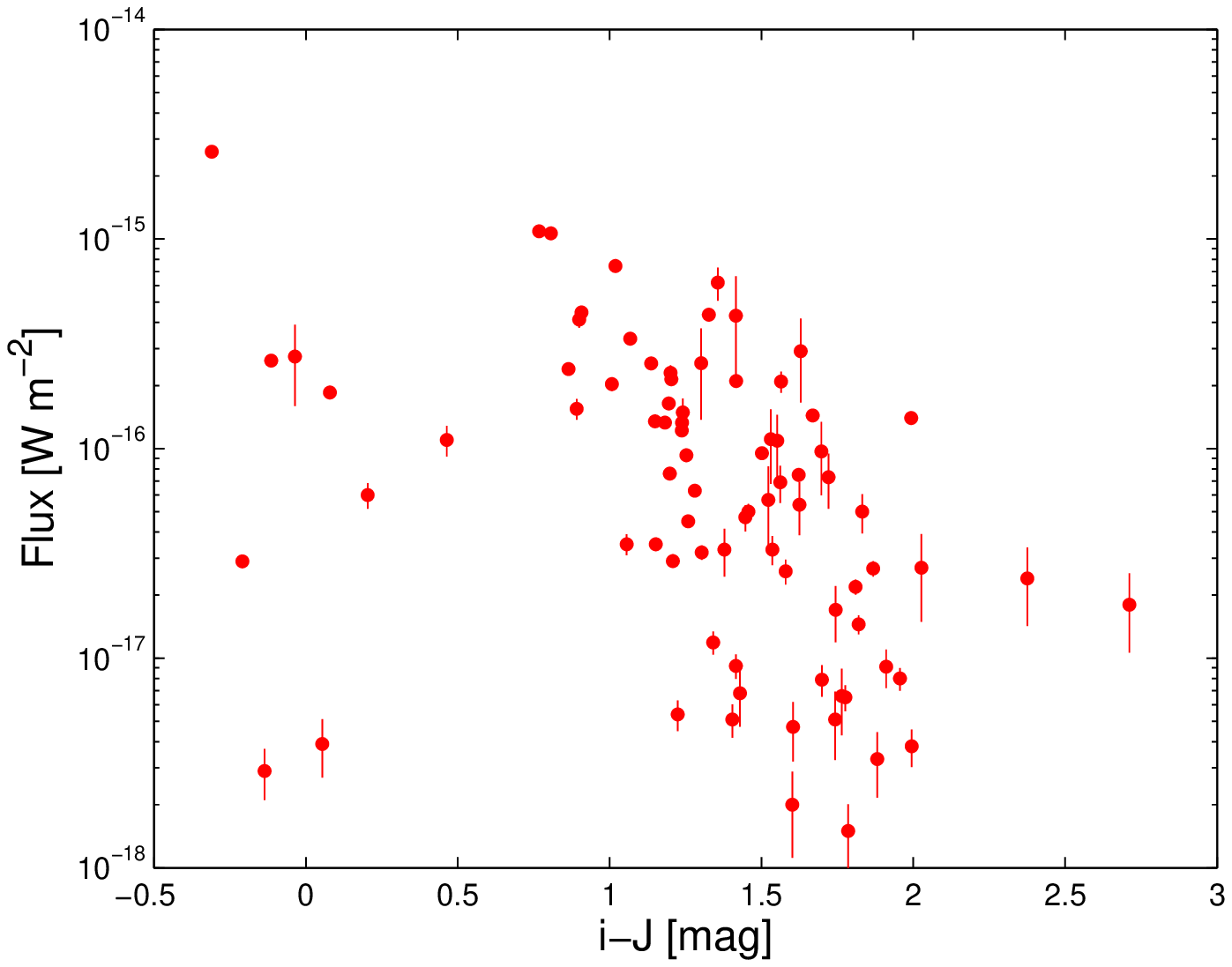}
\includegraphics[width=0.49\textwidth]{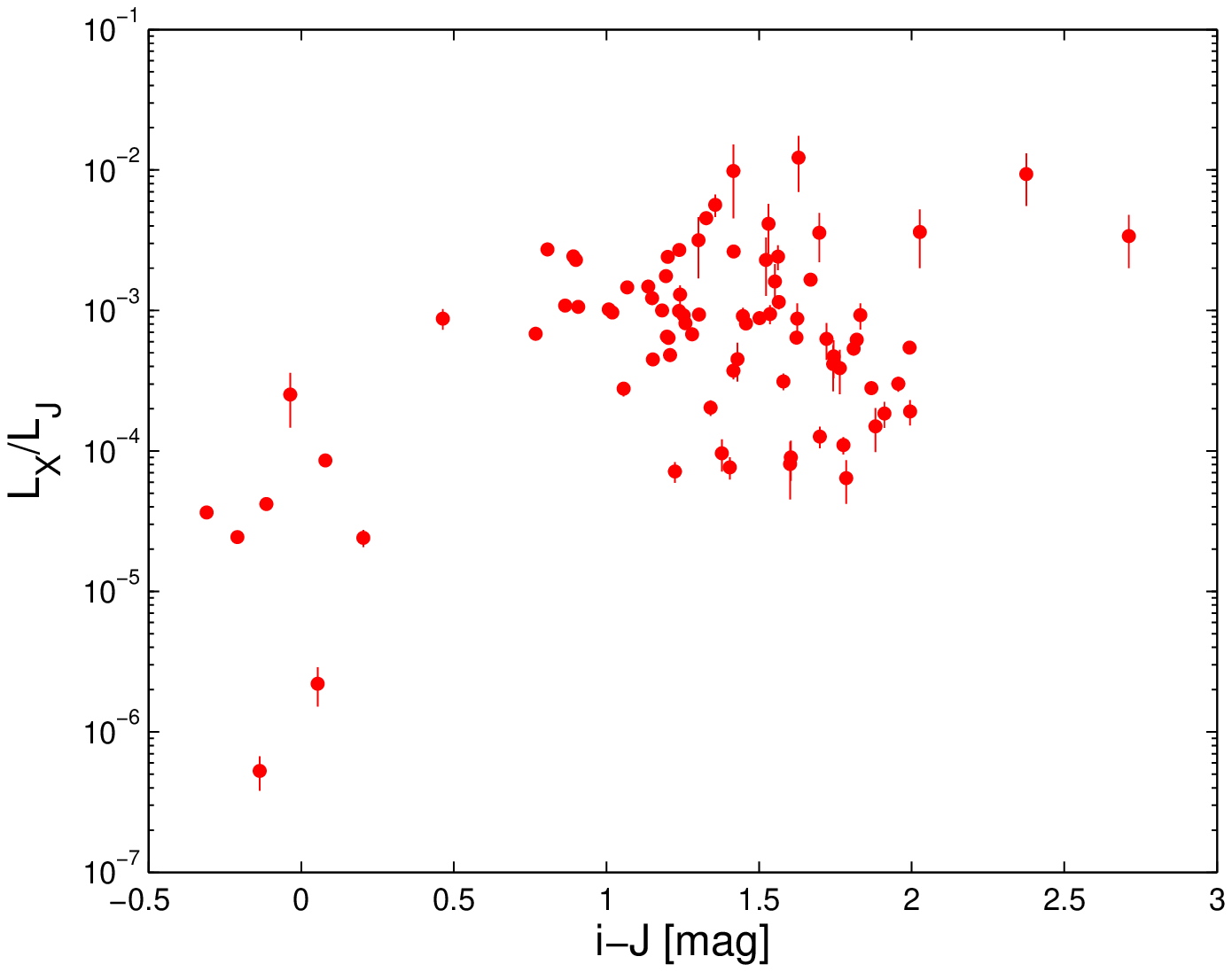}
\caption{X-ray flux ({\em top}) and X-ray-to-$J$-band lumninosity ratio ({\em
bottom}) as a function of the $i-J$ colour.
Error bars account for the uncertainty in count rate and offaxis separation.} 
\label{xfig_fluxLXLJ_i-J}
% sox14.m
\end{figure}

The X-ray luminosity functions (XLFs) of young star clusters have been
extensively studied during the last three decades. 
The {\em ROSAT} XLFs of the {Pleiades}, {Hyades}, or
{$\alpha$~Persei} ($\tau \sim$ 90--600\,Ma, $d \sim$ 45--190\,pc --
Stauffer et~al. 1994; Stern et~al. 1995; Randich et~al. 1996) represented a
cornerstone until the advent of {\em Chandra} and {\em XMM-Newton}.
By taking advantage of the improved spatial resolution of these space missions
currently under operation, clusters at longer heliocentric distances but with
much younger ages than the three of them above have been studied in detail
since, such as the {Orion Nebula Cluster}, {IC~348},
{NGC~1333}, {NGC~2264}, or {M~17} ($\tau \sim$ 1--10\,Ma,
$d \sim$ 260--1600\,pc -- Feigelson et~al. 2002; Preibisch \& Zinnecker 2002;
Getman et~al. 2002; Flaccomio et~al. 2006; Broos et~al. 2007). 
In spite of the low number of X-ray emitters investigated in $\sigma$~Orionis
with respect to the star-forming regions listed above, it has sill a number of
advantadges, e.g., nearness, very low visual extinction, and wide knowledge of
its stellar and substellar populations (Section~\ref{section.introduction}).

Franciosini et~al. (2006) already investigated the XLF of $\sigma$~Orionis.
We illustrate the classical approach with Fig.~\ref{xfig_xlf}.
The HRC-I median flux of all the cluster members and candidates, without
attending to its spectral type, is 6.2\,10$^{-17}$\,W\,m$^{-2}$.
We transformed back the X-ray luminosities tabulated by Franciosini et~al.
(2006) to fluxes (see below).
For seven $\sigma$~Orionis stars detected by them but without luminosity
determination, we used their EPIC count rates and count-rate-to-flux
conversion factor.
Except for slight differences that can be ascribed to the different spectral
sensitivity of HRC-I and EPIC and mehod of flux estimation, the Franciosini
et~al. (2006) XLF and ours are quite similar. 

Because of the long-lasting debate on the actual cluster distance and the
absence of spectral-type determination for all the $\sigma$~Orionis
members and candidates, we preferred instead the diagrams in
Fig.~\ref{xfig_fluxLXLJ_i-J} for our XLF discussion. 
Both the apparent X-ray flux (top panel) and the X-ray-to-$J$-band lumninosity
ratio (bottom panel) are independent of the actual distance, while there are
accurate $i-J$ measurements for all the X-ray stars and brown dwarfs in
$\sigma$~Orionis, mostly taken from Caballero (2008c).
The optical/near-infrared colour $i-J$ is a suitable indicator of effective
temperature (i.e., of spectral type).
The use of other colours involving bluer optical and redder near-infrared bands
(e.g., $V-J$, $i-K_{\rm s}$) is currently impractical because of no data
availability (all the faintest cluster members lack $B$-, $V$-, and $R$-band
measurements) or flux excesses at wavelengths longer than 1.2\,$\mu$m in cluster
members with circum(sub)stellar material.
The X-ray-to-$J$-band lumninosity ratio, $L_X / L_J$, is defined by:
\begin{equation}
\frac{L_X}{L_J} = \frac{4 \pi d^2 {\mathcal F}_X}{4 \pi d^2 {\mathcal F}_J},
\end{equation}
\noindent where ${\mathcal F} \equiv \lambda F_\lambda$ is the apparent flux,
in watts per square meter, and the apparent $J$-band flux ${\mathcal F}_J$ is
approximately proportional to the apparent bolometric flux ${\mathcal F}_{\rm
bol}$.  
The spectral energy distribution of late-K- and M-type stars peak at the $J$
band, which is besides the band least affected by photometric variability and
presence of discs.
The $L_X / L_J$ ratio is thus a proxy for $L_X / L_{\rm bol}$.

Diagrams showing X-ray-to-$J$-band luminosity ratio as a function of 
colour/effective temperature/spectral type, as in the bottom panel in
Fig.~\ref{xfig_fluxLXLJ_i-J}, have been shown by, e.g., Micela et~al. (1999),
Reid (2003), and Daemgen et~al. (2007).
% H\"unsch et~al. (1999), Reid et~al. (1995)
In our diagram, three different regions can be separated: massive early-type
stars (mostly OB), intermediate- and low-mass stars (GKM), and brown dwarfs
(with spectral types later than about M5.5 in $\sigma$~Orionis).

\subsubsection{Early-type stars}
\label{section.earlytype}

With HRC-I/{\em Chandra}, we identified eight $\sigma$~Orionis stars with
spectral types earlier than F0, listed in Table~\ref{table.earlytype}.
The list includes three stars in the eponymous $\sigma$~Ori Trapezium-like
system with spectral types B2 or earlier.
In Fig.~\ref{xfig_fluxLXLJ_i-J}, the eight of them have colours $i-J
\lesssim$ 0.2\,mag and display a wide range of $L_X / L_J$ ratios. 

The spectral types in Table~\ref{table.earlytype} were borrowed from the
bright-star compilation in Caballero (2007a), except for the secondaries in the
binary systems Nos.~3 and~10 (a colon, ``:'', after a spectral type denotes
uncertainty; the letters ``p'' and ``e'' indicate peculiarity and emission,
respectively). 
We estimated a K--M: spectral type for Mayrit~42062~B, the companion at $\rho
\approx$ 0.33\,arcsec to $\sigma$~Ori~E, based on its approximate $K_{\rm s}$
magnitude as evaluated by Bouy et~al. (2009).
The estimation of the late B-early A spectral type for Mayrit~306125~B, the
companion at $\rho \approx$ 0.47\,arcsec to Mayrit~306125~A (HD~37525), was
taken from Caballero et~al. (2009). 
The brightest star in the cluster, No.~1/$\sigma$~Ori~AB + ``F'', seems to be
actually a close triple systems of OB stars (Frost \& Adams 1904; Bolton 1974;
Caballero 2008a; S. Sim\'on-D\'{\i}az et~al., in~prep.).
Only two stars, No.~53/Mayrit~524060 and No.~88/Mayrit~960106, are not known to
form part of a multiple system.

Of the eight early-type stars, three (Nos. 1, 3, and~10) were bright enough in
X-rays for HRI/{\em ROSAT} to be analysed by Caballero et~al. (2009).
Other three stars (Nos. 34, 53, and~74) were detected with EPIC/{\em
XMM-Newton} by Franciosini et~al. (2006).
In practice, they could not resolve the X-ray emission coming from the system
HD~294272 (No.~34/Mayrit~189303 and No.~74/Mayrit~182303).
The pair was first resolved in X-rays by Caballero (2007a) using our HRC-I/{\em
Chandra} dataset.
Of the other two stars, No.~88/Mayrit~960106 was detected with PSPC/{\em ROSAT}
by White et~al. (2000) but escaped other X-ray surveys. 
The presence of the last star, No.~70/Mayrit~13084 ($\sigma$~Ori~D), in the
current HRC-I data was already noticed by Sanz-Forcada et~al. (2004), Caballero
(2007b), and Skinner et~al. (2008), but it has never been analysed.
The B2V star was not detected either with HRI-PSPC/{\em ROSAT}, EPIC/{\em
XMM-Newton}, or ACIS-S/{\em Chandra}.

The early-type stars with the lowest $L_X / L_J$ ratios were No.~70/Mayrit~13084
and No.~74/Mayrit~182303, which justified previous undetections, while the star
with the highest $L_X / L_J$ ratio was No.~88/Mayrit~960106. 
This is the B9-type giant V1147~Ori, an $\alpha^2$~CVn-type variable with
peculiar silicon abundance (Joncas \& Borra 1981; North 1984; Catalano \& Renson
1998). 
Its undetection in previous surveys with HRI/{\em ROSAT}, EPIC/{\em
XMM-Newton}, and ACIS-S/{\em Chandra} may reside simply in its location in
$\sigma$~Orionis, at about 16\,arcmin to the east of the cluster centre.

Only a few $\sigma$~Orionis stars more massive than 2.5\,$M_\odot$ (Caballero
2007a) have not been detected with HRC-I/{\em Chandra}.
They are {Mayrit~208324} (HD~294271, B5V), 
{Mayrit~1116300}\footnote{L\'opez-Santiago \& Caballero (2008) provided a
restrictive upper limit of the EPIC/{\em XMM-Newton} apparent flux of
Mayrit~1116300.} (HD~37333, A1Va -- but see Naylor 2009), and 
{Mayrit~11238} ($\sigma$~Ori~C, A2V). 
The star {HD~37699}, a young B5V star with an envelope at 25.8\,arcmin to
the cluster centre, seems to be associated to the stellar population near the
Horsehead Nebula (Caballero \& Dinis 2008).

In summary, with HRC-I/{\em Chandra} we detected all the $\sigma$~Orionis stars
more massive than 5\,$M_\odot$ ($\sigma$~Ori~AB, D, E) and roughly two thirds of
the stars with masses in the interval 2.5 to 5\,$M_\odot$.
Stars in multiple systems or with spectral peculiarities tend to be among the
stars with detected X-ray emission.

%__________________________________________________ 
   \begin{table}
      \caption[]{Early-type stars in $\sigma$~Orionis detected with HRC-I/{\em
      Chandra}.}   
         \label{table.earlytype}
     $$ 
         \begin{tabular}{lll l}
            \hline
            \hline
            \noalign{\smallskip}
No.	& Name				& Alternative		& Spectral 	\\
	&				& name   		& type		\\
            \noalign{\smallskip}
            \hline
            \noalign{\smallskip}
1	& Mayrit AB			& $\sigma$~Ori~AB + ``F''& O9.5V + B0.5V + ? \\
3	& Mayrit 42062 AB		& $\sigma$~Ori~E	& B2Vpe + K--M:	\\
10	& \object{Mayrit 306125} AB	& HD 37525 AB		& B5Vp + B--A: 	\\
34	& \object{Mayrit 189303}	& HD 294272 B		& B8V	 	\\
53	& \object{Mayrit 524060}	& HD 37564		& A8V:	 	\\
70	& \object{Mayrit 13084}		& $\sigma$~Ori~D	& B2V	 	\\
74	& \object{Mayrit 182305}	& HD 294272 A		& B9.5III	\\
88	& \object{Mayrit 960106}	& V1147 Ori		& B9IIIp	\\
         \noalign{\smallskip}
            \hline
         \end{tabular}
     $$ 
   \end{table}

\subsubsection{Brown dwarfs} 
\label{section.browndwarfs}

%__________________________________________________ 
   \begin{table*}
      \caption[]{Intermediate- and low-mass X-ray stars in $\sigma$~Orionis with
      colours $J-K_{\rm s} >$ 1.15\,mag$^{a}$.}   
         \label{table.ctts}
     $$ 
         \begin{tabular}{lll c ccccc}
            \hline
            \hline
            \noalign{\smallskip}
No.	& Name			& Alternative		& $J-K_{\rm s}$	& Sp.	& pEW(Li {\sc i})       & pEW(H$\alpha$)	& SED	& Phot. 	\\
	&			& name   		& [mag]		& type	& [m\AA]	        & [\AA]			& class	& variable	\\
            \noalign{\smallskip}
            \hline
            \noalign{\smallskip}
29	& Mayrit~92149~AB	& [W96] rJ053847--0237	& 1.24$\pm$0.06	& M1.0:	& 481$\pm$8	        & --20.9$\pm$1.2	& II	& no		\\ % M1.0:, [Li I (Sa08), Halpha, II, X], Wo96,Fr06,He07
36	& Mayrit~203283		& [W96] rJ053831--0235	& 1.16$\pm$0.04	& M0.0:	& 479$\pm$6	        & --10.2$\pm$0.9	& II	& no		\\ % M0.0:, [Li I (Sa08), Halpha, II, X], Wo96,ZO02a,Fr06,He07
45	& Mayrit~609206		& V505 Ori		& 2.01$\pm$0.04	& K7.0	& 431$\pm$11	        & --25.1$\pm$0.7	& II	& yes		\\ % K7.0, [Li I (Sa08), Halpha, Em, mIR, X], HM53,Wi89,ZO02a,OvL04
61	& Mayrit~30241		& [HHM2007] 687		& 1.28$\pm$0.05	& ...	& ...			& ...			& II	& no		\\ % [X, Ks excess], Ca07b 
72	& Mayrit~521199		& TX Ori		& 1.46$\pm$0.04	& K4	& ...			& --16.6		& II	& yes		\\ % K4 [Halpha, Ca II, mIR, X, Si], HM53,HR72,OvL04,He07
75	& Mayrit~622103		& BG Ori		& 1.30$\pm$0.04	& M0.5:	& 480$\pm$7	        &   --40$\pm$3		& II	& yes		\\ % M0.5:, [Li I (Sa08), Halpha, II, X], HM53,Ca06,Fr06,He07 
79	& Mayrit~203260		& Haro 5--11		& 1.19$\pm$0.04	& M2.0:	& 342$\pm$2	        &  --198$\pm$12		& II	& no		\\ % M2.0:, [Li I (Sa08), Halpha, II], Wi89,WB04,Ca06,He07 
80	& Mayrit~497054		& V509 Ori		& 1.26$\pm$0.03	& M0.5:	& 263$\pm$4	        & --25.8$\pm$0.8	& II	& yes		\\ % M0.5:, [Li I (Sa08), Halpha, II, X]. Wi89,Ca06,Fr06,He07
         \noalign{\smallskip}
            \hline
         \end{tabular}
     $$ 
\begin{list}{}{}
\item[$^{a}$] Spectral types and Li~{\sc i} and H$\alpha$ pseudo-equivalent
widths are from Zapatero Osorio et~al. (2002) and Sacco et~al. (2008).
The colon after the spectral type denotes a estimation based on photometry.
\end{list}
   \end{table*}

Two red cluster members with high $L_X / L_J$ ratios stand out in the upper
right corner of the bottom panel in Fig.~\ref{xfig_fluxLXLJ_i-J}, with
colours $i-J \sim$ 2.4--2.7\,mag.
They are two of the only three X-ray brown dwarfs detected in $\sigma$~Orionis
with EPIC/{\em XMM-Newton} by Franciosini et~al. (2006):
No.~84/Mayrit~433123 (S\,Ori~25 -- B\'ejar et~al. 1999; Muzerolle et~al.
2003; Barrado y Navascu\'es et~al. 2003; Caballero et~al. 2004, 2007) and
No.~82/\object{Mayrit~396273} (S\,Ori~J053818.2--023539 -- B\'ejar et~al. 2004;
Kenyon et~al. 2005; Maxted et~al. 2008).
The third X-ray cluster brown dwarf, unidentified in our dataset, is
\object{Mayrit~487350} ([SE2004]~70, NX~67), which underwent a flare during the EPIC
observations and is located at a relatively short projected physical separation
to the planetary-mass object {\em candidate} {S\,Ori~68} (Scholz \&
Eisl\"offel 2004; Caballero et~al. 2006). 

For Mayrit~396273, L\'opez-Santiago \& Caballero (2008) imposed a maximum
X-ray flux of 2.9\,10$^{-17}$\,W\,m$^{-2}$ from their EPIC/{\em XMM-Newton}
observations to the west of $\sigma$~Orionis, consistent with the flux reported
here (1.0$\pm$0.3\,10$^{-17}$\,W\,m$^{-2}$) and the flux estimated from the
Franciosini et~al. (2006) count rate ($\sim$0.6\,10$^{-17}$\,W\,m$^{-2}$).
The brown dwarf may have a high X-ray quiescent level or underwent flares
during both Franciosini et~al. (2006) and our observations.
Mayrit~396273 has the highest $L_X/L_J$ ratio in $\sigma$~Orionis after the two
young star candidates No.~94/Mayrit~887313 and No.~98/Mayrit~1178039 (which are
located at large offaxis separations).

The other brown dwarf, Mayrit~433123, is a photometric variable,
emission-line, accreting, substellar object of only about 0.058\,$M_\odot$, well
below the hydrogen burning mass limit (Caballero et~al. 2007).
From the long-term X-ray variability analysis in Section~\ref{section.longterm},
Mayrit~433123 was about five times brighter at the HRC-I/{\em Chandra} epoch
than at the EPIC/{\em XMM-Newton} one, which indicates that the brown dwarf
could flare during our observations.

Unfortunately, we could not perform a spectral analysis of the two
substellar objects and the low statistics prevented us to achieve conclusions on
the origin of the X-ray emission from their light curves. 
One of the scenarios that could explain the X-ray emission in brown dwarfs
is accretion from a circumsubstellar disc, since the high electrical
resistivities in the neutral atmospheres of ultracool dwarfs are expected
to prevent significant dynamo action (Mohanty et~al. 2002; Stelzer et~al.
2010).  
In fact, Mayrit~433123, with M6.5 spectral type and pEW(H$\alpha$) $\approx$
--44\,\AA, satisfies the empirical criterion to classify accreting T~Tauri
stars and substellar analogues using low-resolution optical spectroscopy of
Barrado y Navascu\'es \& Mart\'{\i}n (2003).
Besides, it seems to be rotationally locked to an imperceptible disc inclined $i
\approx$ 46\,deg with respect to us (Caballero et~al. 2004, 2007; Luhman et~al.
2008).  
However, if a brown dwarf is young enough, it could still retain  a (non
self-sustained) priomordial field. 
Furthermore, Stelzer et~al. (2006) found that accreting brown dwarfs have lower
X-ray luminosity than non-accreting ones and suggested that substellar activity
is subject to the same mechanisms that suppress X-ray emission in
pre-main-squence stars during the T~Tauri phase.
The object statistics (two or three X-ray brown dwarfs) is still too poor to
conclude whether X-rays from brown dwarfs originate via the same processes as
from low-mass stars.

Using the same HRC-I/{\em Chandra} dataset, but with a coarse
identification process, Caballero (2007b) listed two additional faint X-ray
sources that were not identified by us, even during the 10-spurious search
(Section~\ref{section.beyond}).
They could be related to the young very low-mass star {Mayrit~50279}
(Sacco et~al. 2008) and the X-ray source {[FPS2006]~NX~77}.
Caballero (2007b) associated the latter to an infrared source with $J \sim$
19.0\,mag and $J-K_{\rm s} \sim$ 1.8\,mag (tentatively called Mayrit~72345).
If it belonged to $\sigma$~Orionis, it would be an L-type, planetary-mass object
with an estimated mass of 7\,$M_{\rm Jup}$.
Bouy et~al. (2009) agreed with this classification.
However, it would have an extraordinary luminosity ratio larger than
$L_X/L_{\rm bol} \sim 10^{-1}$ and, thus, we consider it instead an active
background galaxy candidate with very red infrared colours.

\subsubsection{Intermediate- and low-mass stars}
\label{section.ctts}

There are a few remarkable X-ray stars among the remaining cluster members and
candidates that are neither early-type stars nor young brown dwarfs.
One of them is No.~63/\object{Mayrit~591158} ([W96]~4771--0026), which has a
relatively blue colour $i-K_{\rm s} \approx$ 0.46\,mag and lies in the $L_X/L_J$
vs. $i-J$ diagram halfway between OB and active KM $\sigma$~Orionis stars.
Mayrit~591158 has cosmic lithium abundance, an effective temperature of about
6000\,K, a high rotational velocity of $v \sin{i}$ = 60$\pm$5\,km\,s$^{-1}$, a
partially-filled H$\alpha$ absorption line, and [S~{\sc ii}] and [N~{\sc ii}]
lines in emission (Caballero 2006; Gonz\'alez-Hern\'andez et~al. 2008). 
This star is significantly warmer than the other six X-ray stars in the diagram
with colours 0.5\,mag $\lesssim i-J \lesssim$ 1.0\,mag, all of which have strong
lithium absorption lines and spectral type (or effective temperature)
determinations between late-G--K0 and K7. 
As a result, Mayrit~591158 is the only X-ray emitter in $\sigma$~Orionis with a
spectral type between F and mid-G\footnote{Furthermore, Mayrit~591158 and
Mayrit~524060 (A8V:) are the only X-ray emitters in $\sigma$~Orionis with
spectral types between early-A and mid-G.}. 
This fact is probably associated to the high reported rotational velocity, which
may favour an enhancement of the magnetic activity.
%Other blue: 2 (K0), 5 (K5), 7 (late-G or early-K, 5235\,K), 8 (G5--K0), 35 (K5:,
%4520\,K), 45 (K7e).

Other remarkable X-ray source is the young low-mass star candidate
No.~103/\object{Mayrit~578123} ([FPS2006]~NX~153), which is the third faintest
X-ray source in our sample and has a high $L_X/L_J$ ratio.
We estimated a mass of about 0.08--0.09\,$M_\odot$ from its $J$-band magnitude
as in Caballero et~al. (2007).
There is no spectroscopy available of Mayrit~578123 to confirm its membership in
$\sigma$~Orionis.

It has been widely discussed in the literature whether classical
(accreting) T~Tauri stars have a lower frequency and intensity of X-ray emission
than weak-line (non-accreting) T~Tauri stars (e.g., Feigelson et~al. 1993;
Neuh\"auser et~al. 1995; Preibisch \& Zinnecker 2002; Telleschi et~al. 2007
-- see also Stelzer et~al. 2006 for a discussion on X-ray emission from
T~Tauri-like brown dwarfs). 
In the $\sigma$~Orionis cluster, Franciosini et~al. (2006), Caballero (2007b),
and L\'opez-Santiago \& Caballero (2008) confirmed the real deficiency in
classical T~Tauri stars in the XLF.
Some hypothesis have been presented to explain this deficiency, such as cooling
of active regions by accretion or absorption of X-rays by dust in a
circumstellar disc.
In the second picture, the geometry of the star-disc system with respect to us
plays a crucial r\^ole (i.e., edge-on discs occult the central object while
front-on ones do not).
Since the inclination angles of circumstellar discs are randomly distributed, we
expect no relation between the strength of both the X-ray emission and
near-infrared flux excess. 

%______________________________________________ Figure 
\begin{figure}
\centering
\includegraphics[width=0.49\textwidth]{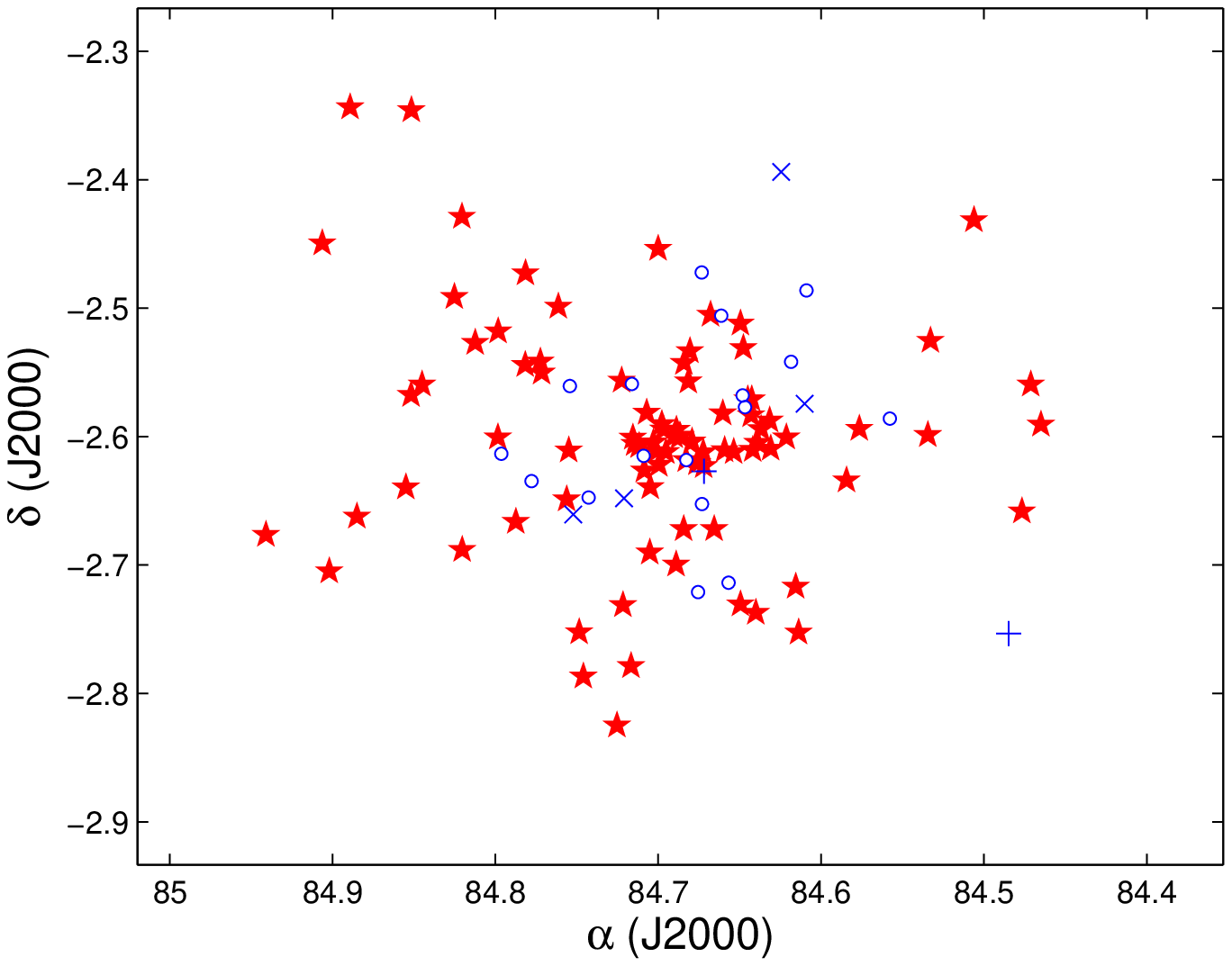}
\includegraphics[width=0.49\textwidth]{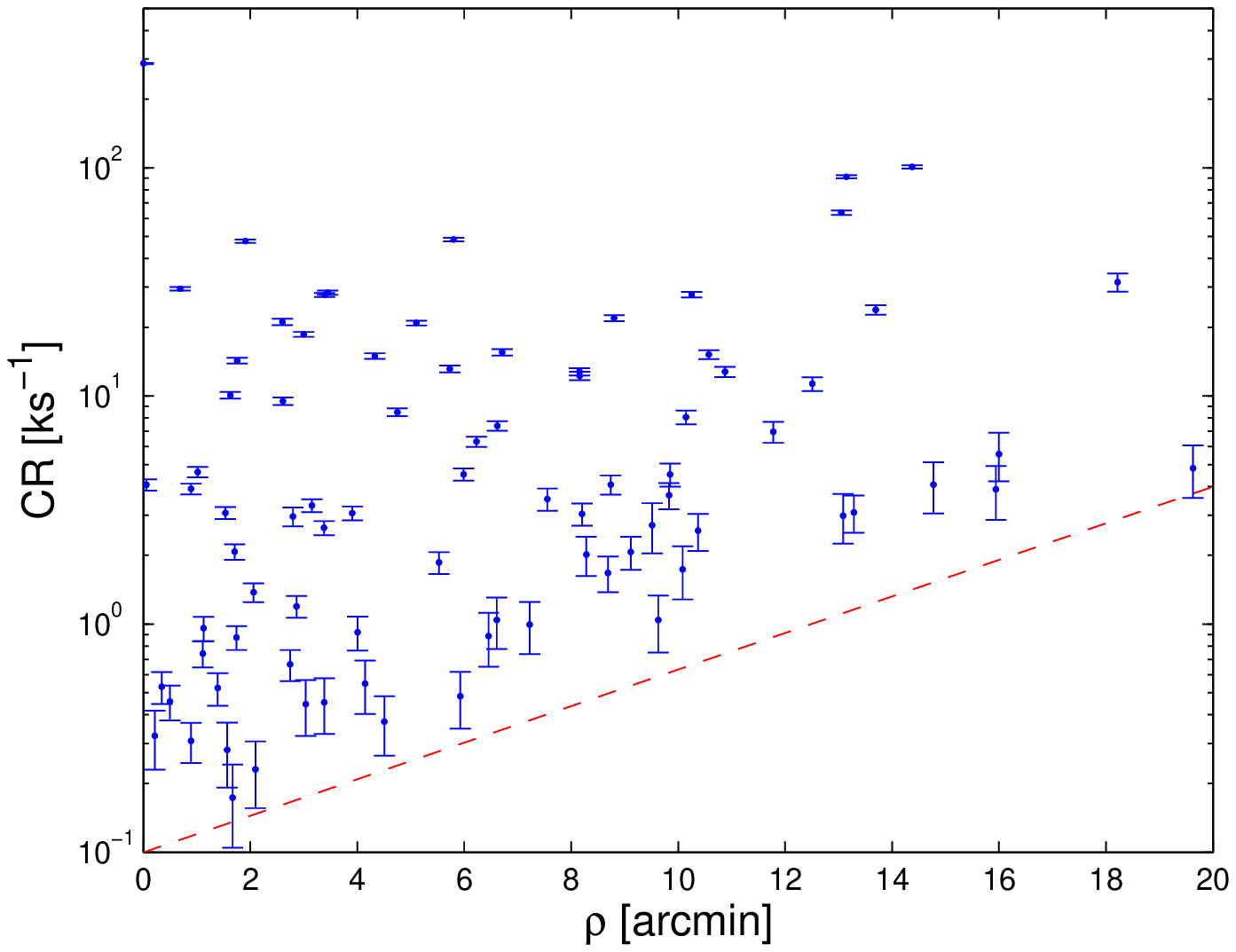}
\includegraphics[width=0.49\textwidth]{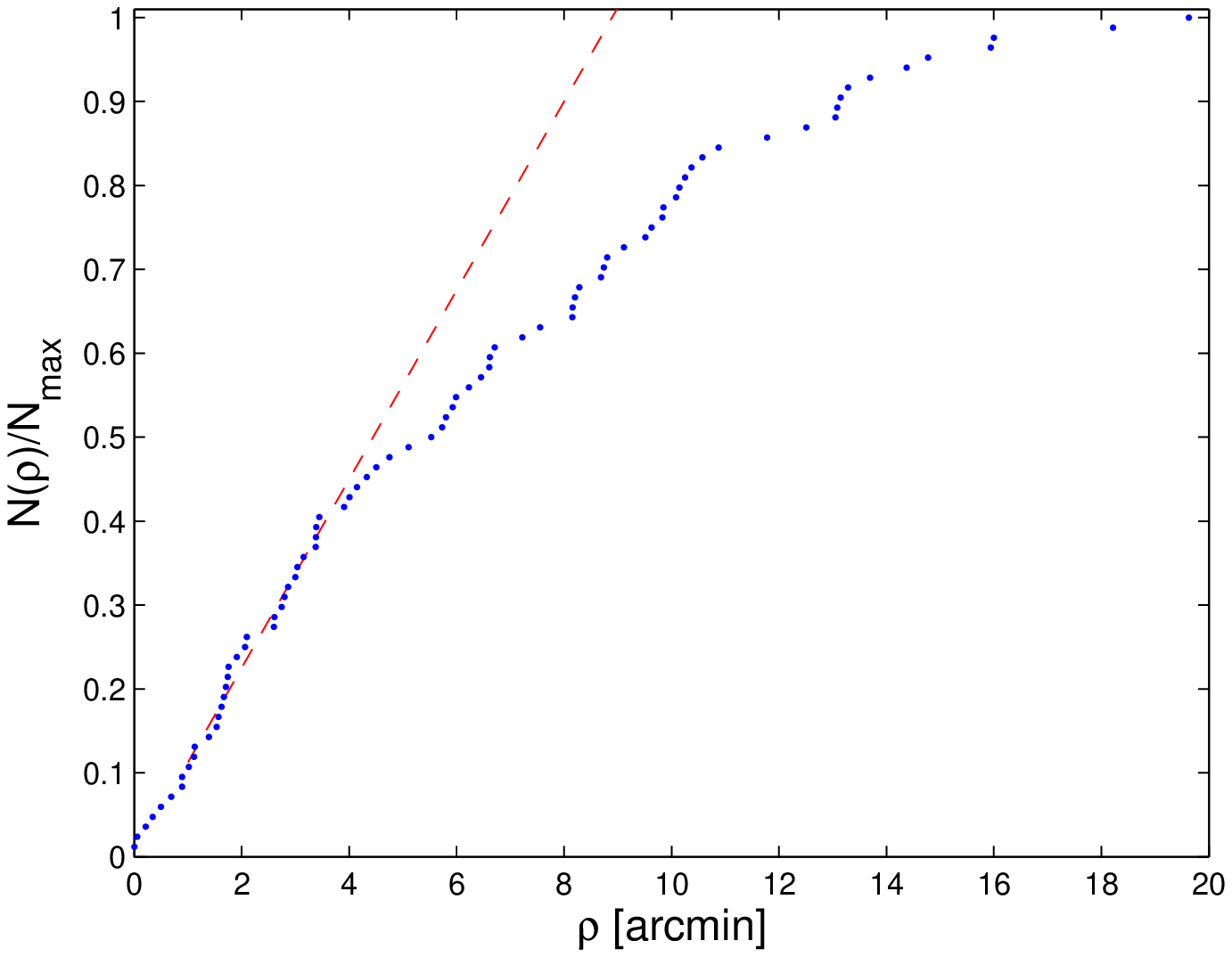}
\caption{{\em Top panel:} spatial location diagram.
The different symbols represent:
cluster star and brown dwarf members and candidates (--red-- filled stars),
field stars (--blue-- crosses), galaxies with optical/near-infrared counterpart
(--blue-- pluses), and galaxies without counterpart (--blue-- open circles). 
Size is 40 $\times$ 40\,arcmin$^2$, with centre on $\sigma$~Ori~AB.
{\em Middle panel:} count rate of $\sigma$~Orionis stars as a function of the
angular separation to the cluster centre.
The dashed line sketches the approximate lower limit for detection of the
HRC-I/{\em Chandra} observations.
{\em Bottom panel:} relative cumulative number of X-ray $\sigma$~Orionis
star and brown dwarf members and candidates as a function of angular separation
to the cluster centre, $\rho$. 
The dashed line indicates the expected values if the X-ray stars followed a
volume-density law proportional to $\rho^{-2}$.} 
\label{xfig_delta_alpha}
% sox06.m, sox09.m
\end{figure}

Following this discussion, we investigated the reddest KM-type X-ray stars in
$\sigma$~Orionis, which we expected to be classical T~Tauri stars with discs.
The eight X-ray stars with colours $J-K_{\rm s} >$ 1.15\,mag listed in
Table~\ref{table.ctts} have spectral energy distributions (from the
optical to 8.0--24\,$\mu$m) typical of discs harbours according to Hern\'andez
et~al. (2007). 
Except for No.~61/Mayrit~30241, which misses spectroscopy, all the stars satisfy
the H$\alpha$-accretion criterion of Barrado y Navascu\'es \& Mart\'{\i}n
(2003).
Of them, only two stars, No.~72/\object{Mayrit~521199} (TX~Ori) and, specially,
No.~45/\object{Mayrit~609206} (V505~Ori, with $J-K_{\rm s}$ =
2.01$\pm$0.04\,mag) have colours redder than 1.4\,mag, while in $\sigma$~Orionis
there are about a dozen KM-type stars redder than this value (Caballero 2008c).
For example, none of the stellar sources of the four Herbig-Haro objects in
$\sigma$~Orionis (Reipurth et~al. 1998), which also have very red $J-K_{\rm s}$
colours, were detected with HRC-I (but the source of HH~445 was detected by
Franciosini et~al. 2006 -- Section~\ref{section.epicxmmnewton}). 
Likewise, only six of the about thirty KM-type $\sigma$~Orionis stars redder
than $J-K_{\rm s}$ = 1.2\,mag were detected with HRC-I.
A detailed analysis of the frequency of X-ray emitters as a function of mass,
disc presence, and degree of accretion is to be done, but the values above hint
at a lower frequency and intensity of X-ray emission of classical (accreting)
T~Tauri stars in $\sigma$~Orionis than weak-line (non-accreting) T~Tauri stars.

\subsection{Spatial distribution of X-ray sources}
\label{section.spatial}

As a final analysis of the HRC-I data, we investigated the spatial distribution
of X-ray stars in $\sigma$~Orionis.
From the top panel in Fig.~\ref{xfig_delta_alpha}, the cluster stars are
concentrated towards the centre, defined by the eponymous $\sigma$~Ori~AB
system, which coincides with the centre of the field of view with a small error
of 13\,arcsec (Section~\ref{section.dataretrieval}). 
The apparent concentration of galaxies without optical/near-infrared counterpart
and field stars in the innermost 10\,arcmin is due to the combined effect of
their faintness and the decreasing sensitivity of the HRC-I detector at large
offaxis separations. 
Only relatively bright X-ray fore- and background sources, such as the field
star No.~69/[W96]~rJ053829--0223 or, specially, the galaxy 2E~1456, could be
detected at more than 10\,arcmin to the pointing centre.
The middle panel in Fig.~\ref{xfig_delta_alpha} illustrates the effect of
the degradation of the sensitivity towards the HRC-I borders: while roughly all
the X-ray sources with count rates $CR >$ 0.1\,ks$^{-1}$ were detected in the
central area, the lower limit for detection increased up to about 1\,ks$^{-1}$
at 10\,arcmin and about 4\,ks$^{-1}$ at 20\,arcmin.

According to Caballero (2008a), the radial distribution of $\sigma$~Orionis
stars (without attending to their X-ray emission) follows a power law
proportional to the angular separation to the cluster centre, $\rho^{+1}$, valid
only for $\rho \lesssim$ 20\,arcmin.
This distribution corresponds to a volume density proportional to $\rho^{-2}$,
which is expected from the collapse of an isothermal spherical molecular cloud.
From the bottom panel in Fig.~\ref{xfig_delta_alpha}, the X-ray stars in
$\sigma$~Orionis follow the power law $\rho^{+1}$ only in the innermost
4\,arcmin.
Apart from the limited field of view of the detector, at large offaxis
separations, the degradation of the sensitivity towards the HRC-I borders gets
important and many X-ray $\sigma$~Orionis stars were missed during the
observations. 
We estimated that about 30 and more than 100 young stars and brown dwarfs
were missed in the 4--10 and 10--20\,arcmin annuli, respectively. 
The sensitivity degradation must be taken into account when frequencies of X-ray
emitters are computed.

\section{Summary}
\label{section.summary}

We carried out a detailed analysis of the X-ray emission of young stars in the
$\sigma$~Orionis cluster ($\tau \sim$ 3\,Ma, $d \sim$ 385\,pc).
We analysed public HRC-I/{\em Chandra} observations obtained in November 2002.
The wide field of view, long exposure time of 97.6\,ks, and the superb spatial
resolution of HRC-I/{\em Chandra} allowed us to detect 107 X-ray sources, many
of which had not been identified in previous searches with IPC/{\em Einstein},
HRI/{\em ROSAT}, ACIS-S/{\em Chandra}, or EPIC/{\em XMM-Newton}.
After cross-matching with optical and near-infrared catalogues, we classified
the X-ray sources into 84 young cluster members and candidates, four active
field stars, and 19 galaxies, of which only two have known optical and
near-infrared counterparts. 
Among the cluster members and candidates, two are {\em bona fide} brown dwarfs
with signposts of youth.

A robust Poisson-$\chi^2$ analysis to search for X-ray variability showed that
at least seven young stars displayed flares during the HRC-I observations, while
two (or three, if we include the B2Vpe star No.~2/Mayrit~42062~AB --
$\sigma$~Ori~E) may display rotational modulation.
Some of the observed flares were intense, with peak-to-quiescence ratios of
about six and durations longer than 20\,ks (and longer than our observations in
one~case).

We compared the count rates and variability status of our HRC-I sources with the
results of previous observations with {\em Einstein}, {\em ROSAT}, {\em
Chandra}, and {\em XMM-Newton}, and found that eleven stars displayed
significant X-ray flux variations between our observations and others, mostly
ascribed to flaring activity. 
Interestingly, during the HRC-I observations, the brown dwarf
No.~84/Mayrit~433123 (S\,Ori~25) underwent an X-ray brightening by a factor five
with respect to the EPIC/{\em XMM-Newton} epoch. 
Besides, we revisited old {\em ROSAT} data and found new flaring activity
in the $\sigma$~Orionis star No.~37/Mayrit~102101~AB. 
To facilitate further studies, we also compiled the {\em ROSAT} sources
presented by Wolk (1996).
From this compilation, we noticed that he tabulated X-ray emission from the
brown dwarf Mayrit~433123, but he was not able to classify it as one of the
first discovered substellar objects. 

The X-ray luminosity function that we presented here ranges from spectral type
O9.5V, which corresponds to a mass of about 18\,$M_\odot$, to M6.5, below the
hydrogen burning mass limit at 0.07\,$M_\odot$.
We found a tendency of early-type stars in multiple systems or with spectral
peculiarities to display X-ray emission.
On the other side of the luminosity function, the two detected brown dwarfs and
the least massive young star candidate are among the $\sigma$ Orionis members
with the highest values of $L_X/L_J$ luminosity ratios.
We found X-ray emission from only two stars in the spectral type interval from
early A to intermediate-late G.

We noticed that most of the $\sigma$~Orionis T~Tauri stars with the largest
infrared excesses have not been detected in X-ray surveys in the area, which
supports the scenario of a lower frequency and intensity of X-ray emission of
classical (accreting) T~Tauri stars than weak-line (non-accreting) T~Tauri
stars.
The only very red ($J-K_{\rm s} >$ 1.5\,mag) young star detected  with
HRC-I/{\em Chandra} was No.~45/Mayrit~609206, which is a classical
T~Tauri star with a strong H$\alpha$ emission for its spectral type
(K7.0), photometric variability, and a spectral energy distribution typical of
Class~II objects. 

Finally, we investigated the spatial distribution of the X-ray cluster members,
which is strongly affected by the degradation of the sensitivity towards the
borders of the HRC-I detector.
While roughly all the X-ray sources with count rates $CR >$ 0.1\,ks$^{-1}$ at
less than 4\,arcmin to the cluster centre were detected, the estimated numbers
of missed X-ray cluster members in the 4--10 and 10--20\,arcmin annuli are 30
and 100, respectively.
Since the core of $\sigma$~Orionis extends up to 20\,arcmin from the
centre, defined by the Trapezium-like $\sigma$~Ori system, additional de-centred
pointings with HRC-I/{\em Chandra}, EPIC/{\em XMM-Newton}, or the future Wide
Field Imager + Hard X-ray Imager instruments onbard the ESA-NASA-JAXA space
mission {\em International X-ray Observatory} are necessary to investigate the
full X-ray luminosity function of the cluster. 
To conclude, a few shallow pointings around the cluster centre will probably be
more efficient to detect and characterise new X-ray young brown dwarfs in 
$\sigma$~Orionis than a single deep pointing centred on the Trapezium-like
system.

\begin{acknowledgements}

We are indebt to the anonymous referee for his/her quick, polite, greatly
valuable report.
JAC is an {\em investigador Ram\'on y Cajal} at the CAB, 
JFAC is a researcher of the Consejo Nacional de Investigaciones  
Cient\'{\i}ficas y Tecnol\'ogicas (CONICET) at the UNComa, and
JLS is an AstroCAM post-doctoral fellow at the UCM. 
This research made use of the SIMBAD, operated at Centre de Donn\'ees
astronomiques de Strasbourg, France, and NASA's Astrophysics Data System.
PWDetect has been developed by scientists at Osservatorio Astronomico di
Palermo.
Financial support was provided by the Universidad Complutense de Madrid, the
Comunidad Aut\'onoma de Madrid, the Spanish Ministerio de Ciencia e
Innovaci\'on, the Secretar\'{\i}a de Ciencia y Tecnolog\'{\i}a de la Universidad
Central de C\'ordoba, and the Argentinian CONICET under grants 
AyA2008-06423-C03-03, 		% (WSO) 
AyA2008-00695,			% (estrellax II)
PRICIT S-2009/ESP-1496, and	% (AstroMadrid) 
PICT 2007-02177.		% (SECyT)

\end{acknowledgements}

\appendix

\section{HRC-I/{\em Chandra} compared to other X-ray space missions} 
\label{section.comparison}

\subsection{IPC/{\em Einstein}}
\label{section.ipceinstein}

We identified the eight 2E sources detected at less or about 15\,arcmin to
the cluster centre with the Imaging Proportional Counter (IPC) onboard {\em
Einstein} (Harris et~al. 1994).
Given the large {\em Einstein} position errors of 30--50\,arcsec tabulated in
the 2E catalogue, the origin of each X-ray source can be a combination of
several bright sources (e.g., 2E~1470 = $\sigma$~Ori~AB + D + E + IRS1~AB). 
Besides, there was a ninth 2E source at about 16\,arcsec to the cluster centre,
2E~1483, which we associated to our HRC-I source No.~99.
The {\em Einstein} Two-Sigma catalogue (Moran et~al. 1996) only provided the
marginal detection of three additional {\em ROSAT} sources (Mayrit~528005~AB,
Mayrit~653170, and Mayrit~306125~AB) and five possible spurious X-ray detections
and, hence, we did not use it.

\subsection{HRI/{\em ROSAT}}
\label{section.hrirosat}

We recovered in our HRC-I observations all except one of the 24 sources (23
young stars and the galaxy 2E~1456) detetced by Caballero et~al. (2009) with the
High Resolution Imager (HRI) onboard {\em ROSAT}.
The exception was the flaring star Mayrit~969077 (2E~1487), the most separated 
X-ray source to the cluster centre in the HRI observation, which fell out
of the HRC-I field of view.
Of the 23 sources, eight were reported to vary by Caballero et~al. (2009).
In their variability study, the authors imposed a minimum number of associated
X-ray events of $N$ = 20.  
In the present work, we revisited their HRI/{\em ROSAT} dataset and applied the
same methodology as in Caballero et~al. (2009) to eight X-ray sources with 5 $<
N <$ 20 not investigatd by them. 
The results of this analysis are summarised in
Table~\ref{table.hrirosatourxrayparameters}. 
The eight X-ray sources correspond to the active field star No.~31
([W96]~4771--1056; Section~\ref{subsection.fieldstars}) and seven young
$\sigma$~Orionis stars, of which two are variable according to the robust
Poisson-$\chi^2$ criterion in Caballero et~al. (2009). 
The two (highly) variable stars are No.~4 (Mayrit~348349, Haro~5--13), the
strong H$\alpha$ emitter that showed the flare decay during the HRC-I
observations, and No.~37 (Mayrit~102101~AB, [W96]~rJ053851--20130236), an
M3-type, accreting, double-lined spectroscopic binary (Wolk 1996; Sacco et~al.
2008; Caballero et~al. 2008). 
In both cases, flaring activity was responsible for the large variation in count
rates (of up to a factor ten). 

%__________________________________________________ 
   \begin{table}
      \caption[]{X-ray parameters of faint HRI/{\em ROSAT} sources in
      $\sigma$~Orionis not listed by Caballero et~al. (2009)$^{a}$.}    
         \label{table.hrirosatourxrayparameters}
     $$ 
         \begin{tabular}{l l ccccc}
            \hline
            \hline
            \noalign{\smallskip}
No.	& Name			& $N$	& $\overline{CR}$	& $\sigma_{CR}$	& $\overline{\delta CR}$	& $\chi^2$	\\
	& 			& 	& [ks$^{-1}$]		& [ks$^{-1}$]	& [ks$^{-1}$]			&		\\
            \noalign{\smallskip}
            \hline
            \noalign{\smallskip}
4	& Mayrit 348349		& 11	& 11.46			& 10.03		& 1.95				& 114.6		\\ % 
16	& Mayrit 97212		& 11	& 3.66			& 0.95		& 1.22				& 0.361		\\ % 
18	& Mayrit 403090		& 16	& 5.07			& 1.75		& 1.58				& 4.663		\\ % 
27	& Mayrit 489165		& 6	& 2.77			& 1.68		& 1.24				& 2.176		\\ % 
29	& Mayrit 92149 AB	& 6	& 3.24			& 1.01		& 0.77				& 1.540		\\ % 
31	& [W96] 4771--1056	& 6	& 3.76			& 1.00		& 1.09				& 1.126		\\ % 
32	& Mayrit 374056		& 13	& 5.61			& 1.88		& 1.88				& 4.617		\\ % 
37	& Mayrit 102101~AB	& 7	& 7.77			& 4.86		& 1.77				& 45.91		\\ % 
          \noalign{\smallskip}
            \hline
         \end{tabular}
     $$ 
\begin{list}{}{}
\item[$^{a}$] Number of associated X-ray events, mean and standard deviation of 
	the net count rate, mean of the error on the count rate, and normalized
	double-weighted $\chi^2$ of the X-ray series (see Table~1 in Caballero
	et~al. 2009).  
\end{list}
   \end{table}
%

%__________________________________________________ 
   \begin{table*}
      \caption[]{Optical/near-infrared counterparts of EPIC/{\em XMM-Newton} 
      sources in Franciosini et~al. (2006) not listed in 
      Tables~\ref{table.beyond} or~\ref{table.xraydetections}.}   
         \label{table.xmmnewton}
     $$ 
         \begin{tabular}{l cc c ll}
            \hline
            \hline
            \noalign{\smallskip}
NX		& $\alpha$ 	& $\delta$ 	& CR	    	& Name  		& Class			\\
(Fr06)		& (J2000)	& (J2000)	& [ks$^{-1}$]   & 			& 			\\
            \noalign{\smallskip}
            \hline
            \noalign{\smallskip}
4		& 05 38 00.56	& --02 45 09.7	&  9.7$\pm$1.0  & Mayrit 861230		& Possible young star	\\ % 
6		& 05 38 06.50	& --02 28 49.4	& 0.67$\pm$0.18 & Mayrit 717307		& Young star		\\ % 
7		& 05 38 06.74 	& --02 30 22.8 	& 19.0$\pm$0.9  & Mayrit 662301		& Young star		\\ % Bright X-ray stars absent in our work: NX~7 (Mayrit~662301, Kiso A-0904~67).
11		& 05 38 13.20	& --02 26 08.8	&  1.6$\pm$0.3  & Mayrit 757321		& Young star		\\ % 
14		& 05 38 15.53	& --02 42 05.1	&  2.7$\pm$0.4  & [FPS2006] NX 14	& Possible galaxy	\\ % NX 41 in LC2008, blue i-Ks, unknown
17		& 05 38 17.78	& --02 40 50.1	& 0.97$\pm$0.17 & Mayrit 498234		& Young star		\\ % 
25		& 05 38 23.32	& --02 44 14.2	& 0.85$\pm$0.17 & Mayrit 589213		& Young star		\\ % 
26		& 05 38 23.55	& --02 41 31.8	& 0.96$\pm$0.19 & Mayrit 459224		& Young star		\\ % 
33		& 05 38 27.51	& --02 35 04.2 	&  1.5$\pm$0.4  & Mayrit 265282		& Young star		\\ % In the innermost 5\,arcmin: NX~33 (Kiso A-0976 329, S\,Ori~J053827.5--023504).
42		& 05 38 30.98	& --02 34 03.8	& 0.64$\pm$0.13 & [FPS2006] NX 42	& Possible field star	\\ % Prob. fore. star?
47		& 05 38 32.68	& --02 31 15.6	& 0.57$\pm$0.13 & [KJN2005] 3.01 325	& Field star		\\ % Non-member in Sacco et~al. 2008
50		& 05 38 33.02	& --02 39 27.9	& 0.53$\pm$0.11 & [FPS2006] NX 50	& Possible field star?	\\ % New candidate member?
67		& 05 38 39.13	& --02 28 00.4	&  1.3$\pm$0.2  & Mayrit 487350		& Young brown dwarf	\\ % Interesting: [SE2004]~70 + S\,Ori~68.
82		& 05 38 45.98	& --02 45 23.2	& 0.73$\pm$0.18 & Mayrit 563178		& Possible young star	\\ % 
89		& 05 38 47.66	& --02 30 37.4	&  2.2$\pm$0.5  & Mayrit 326008		& Young star		\\ % 
97$^{a}$	& 05 38 49.93	& --02 41 22.8	& 1.08$\pm$0.15 & Mayrit 332167		& Young star		\\ % 
100		& 05 38 50.39	& --02 26 47.7	&  1.0$\pm$0.2  & Mayrit 559009		& Young star		\\ % 
101		& 05 38 51.01	& --02 27 45.7	& 0.70$\pm$0.17 & [KJN2005] 1.02 156 AB	& Field star		\\ % SB, non-member in Sacco et~al. 2008 
111		& 05 38 54.92	& --02 28 58.3	& 0.69$\pm$0.16 & Mayrit 449020		& Young star		\\ % 
114		& 05 38 56.99	& --02 31 25.6	& 0.65$\pm$0.13 & SO110979		& Field star		\\ % 
116		& 05 38 58.84	& --02 34 13.2	& 0.63$\pm$0.13 & [D90] 3		& Radiogalaxy		\\ % 
118		& 05 38 59.23	& --02 33 51.4	& 0.25$\pm$0.07 & Mayrit 252059		& Young star		\\ % In the innermost 5\,arcmin: NX~118 (Haro 5-18, SWW~227).
147		& 05 39 12.32	& --02 30 06.4	&  2.1$\pm$0.4  & Mayrit 544049 	& Possible young star	\\ % 
149		& 05 39 14.47	& --02 28 33.4 	& 10.9$\pm$0.6  & Mayrit 631045		& Possible young star	\\ % Bright X-ray stars absent in our work: NX~149 (Mayrit~631045, S\,Ori~J053914.5--022834).
157		& 05 39 18.97	& --02 30 55.6	&  3.0$\pm$0.4  & Mayrit 596059 	& Young star?		\\ % 
160		& 05 39 20.97	& --02 30 33.5	&  5.8$\pm$0.5  & Mayrit 633059		& Young star		\\ % Moderately bright: NX~160 (Mayrit~633059, S\,Ori~3).
167		& 05 39 26.73	& --02 26 17.4	&  2.9$\pm$0.6  & Mayrit 856047 	& Young star		\\ % 
169		& 05 39 50.56	& --02 38 27.0	&  1.8$\pm$0.3  & [SWW2004] 222 AB	& Field star		\\ % % NX 169; SB and no member in Sa08; [SWW2004] J053930.567-023826.96
174		& 05 39 39.83	& --02 33 16.0	& 19.8$\pm$1.0  & Mayrit 841079		& Young star		\\ % Bright X-ray stars absent in our work: NX~174 (Mayrit~841079, V603~Ori, source of Herbig-Haro object HH~445).
175		& 05 39 39.99 	& --02 43 09.7	&  1.9$\pm$0.5  & Mayrit 931117		& Young star		\\ % 
         \noalign{\smallskip}
            \hline
         \end{tabular}
     $$ 
\begin{list}{}{}
\item[$^{a}$] Star NX~97 (\object{Mayrit~332167}, [SWW2004]~200) has colours,
lithium, radial velocity, and H$\alpha$ emission consistent with membership in
cluster (Sacco et~al. 2008).
Its Mayrit number is firstly given here.
\end{list}
   \end{table*}

For completeness, in Table~\ref{table.roundnround} we present a reappraisal of
the X-ray sources near $\sigma$~Ori in the novel work by Wolk (1996).
Years later, his work is acknowledged as a cornerstone in the study of the
$\sigma$~Orionis cluster.
In the table, we list the names and spectral types\footnote{Symbols ``$\times$''
and ``...'' in the spectral type column in Table~\ref{table.roundnround}
indicate that the stars were spectroscopically investigated by Wolk (1996) but
their spectral types were not given and that the stars were not
spectroscopically investigated, respectively.} of the optical counterparts and
coordinates and count rates of the X-ray sources from Wolk (1996), coordinates
of the near-infrared counterparts from 2MASS, identification number in our work,
and recommended name. 

Among the 58 X-ray sources listed in Table~\ref{table.roundnround}, there are
three double and one triple detections (including $\sigma$~Ori~AB), between two
and four probable spurious detections (marked with ellipses and question marks),
two galaxies (No.~9/2E~1456 and No.~62), and three field active stars
(No.~31/[W96]~4771--1056, No.~51/[SWW2004]~166, and
``R053930--0238''/[SWW2004]~222~AB).  
The remaining sources correspond to young objects in the $\sigma$~Orionis
cluster. 
A few sources were not detected in the {\em Chandra} images, mostly due to the
difference in sizes of fields of view.

Interestingly, Wolk (1996) tabulated at 0.81$\pm$0.19\,ks$^{-1}$ the count rate
of the X-ray source ``R053908--0239'' (No.~84/Mayrit~433123), which is the young
{\em brown dwarf} S\,Ori~25 (see Section~\ref{section.browndwarfs}).
This was the first report of X-ray emission from a substellar object.
Unfortunately, Wolk (1996) did not collect optical or near-infrared photometry
of the object and could not classify it.

\subsection{ACIS-S/{\em Chandra}}
\label{section.acischandra}

Of the 42 X-ray sources detected with ACIS-S/{\em Chandra} by Skinner
et~al. (2008), 40 were HRC-I/{\em Chandra} X-ray sources with significance of
detection larger than 5.4 (Table~\ref{table.xraydetections}).
One of the other two sources, the X-ray galaxy [SSC2008]~40 (CXO~40), was
recovered with our new 10-spurious search (Section~\ref{section.beyond}). 
We did not detect [SSC2008]~39 (CXO~39, [FPS2006]~NX~116).
It is probably related to the nearby radio source [D90]~3 (Drake 1990; Caballero
2009), which was also tabulated in the National Radio Astronomy Observatory Very
Large Array Sky Survey (NVSS at 1465\,MHz; Condon et~al. 1998). 
The measured angular separation, $\rho \sim$ 1.4\,arcsec, is consistent
with the large NVSS mean error in declination of more than 6\,arcsec.
[D90]~3 might be a radio-galaxy with variable X-ray emission.

\subsection{EPIC/{\em XMM-Newton}}
\label{section.epicxmmnewton}

In Tables~\ref{table.beyond} and~\ref{table.xraydetections}, there are 87 X-ray
sources in common between our observations with HRC-I/{\em Chandra} and the ones
with EPIC/{\em XMM-Newton} by Franciosini et~al. (2006).
However, the authors reported 175 detections. 
Only thirty of the 88 unidentified sources, listed in
Table~\ref{table.xmmnewton}, have optical and near-infrared counterparts. 
They are 18 $\sigma$~Orionis stars and brown dwarfs with signposts of youth,
four cluster member candidates, four field stars, two possible field stars, a
possible galaxy, and the radiogalaxy [D90]~3.
In the table, the uncertainty in the actual stellar counterpart of two X-ray
sources is indicated with a question mark. 
Following the criterion in L\'opez-Santiago \& Caballero (2008), we classified
the other 58 X-ray sources with no 2MASS counterpart as faint active galaxies
(the X-ray sources NX~46 and NX~123 had blue optical counterparts in the Guide
Star Catalog).  

We were able to identify 23 sources not detected by Franciosini et~al. (2006).
Of them, the authors provided EPIC count-rate upper levels for seven young
stars and candidates (marked with the symbol ``$<$'' in the NX column in
Table~\ref{table.xraydetections}).
The source No.~25 (Mayrit~3020~AB, $\sigma$~Ori~IRS1), at only 3\,arcsec from
$\sigma$~Ori~AB, was not resolved by EPIC.
Most of the remaining 15 new sources fell at angular separations to the
pointing centre larger than $\rho \sim$ 15\,arcmin (e.g., the bright X-ray
galaxy No.~9/2E~1454 or the young star candidate No.~46/Mayrit~1093033) or
shorter than $\rho \sim$ 3\,arcmin (where the background level due to
$\sigma$~Ori~AB in the EPIC observations was high). 
Among the 23 sources not detected by Franciosini et~al. (2006), eight sources
were detected independently with ACIS-S by Skinner et~al. (2008).
There were also eight young stars with signposts of youth, including the
early-type stars No.~70/Mayrit~13084 ($\sigma$~Ori~D) and No.~74/Mayrit~182305
(HD~294272~A), six young star candidates, and two galaxies (No.~9/2E~1456 and
No.~64/UCM0536--0239, which was also detected by Skinner et~al. 2008). 
They all have low significances of detection in our HRC-I data.

\section{Notes on individual objecs}

\subsection{Notes to Table~\ref{table.nonmayrit}}
\label{section.notes.table1}

\begin{itemize}
\item[*] No.~25/Mayrit 3020 AB ($\sigma$~Ori~IRS1~AB) is a Class~II (or Class-I
proplyd?) binary star located at $\rho$ = 3.32$\pm$0.06\,arcsec, $\theta$ =
19.6$\pm$1.4\,deg, to $\sigma$~Ori~AB.  
In turn, it forms a binary system separated by $\rho \approx$ 0.24\,arcsec,
$\theta \approx$ 318\,deg (Bouy et~al. 2009; Hodapp et~al. 2009).  
Coordinates and $J$-band magnitude of Mayrit~3020~AB are from the unresolved
adaptive optics observations in Caballero (2006). 
The tabulated $H$- and $K_{\rm s}$-band magnitudes, from Bouy et~al. (2009), are
for the primary Mayrit~3020~A. 
The secondary Mayrit~3020~B has $H$ = 12.84$\pm$0.07\,mag and $K_{\rm s}$ =
12.65$\pm$0.07\,mag. 
\item[*] No.~31/{[W96]} 4771--1056 is a possible field star discovered by Wolk
(1996). 
He derived K1 spectral type and found H$\alpha$ in absorption.
The Li~{\sc i} $\lambda$6708\,{\AA} equivalent width was smaller than expected 
for an early K-type cluster member.
The star does not follow the spectro-photometric sequence of the cluster.
\item[*] No.~39/Mayrit~168291~A, No.~47/Mayrit~68229, and No.~57/Mayrit~492211
have lithium absorption, radial velocity, and H$\alpha$ emission consistent with
membership in $\sigma$~Orionis (Sacco et~al. 2008). 
Their Mayrit numbers are firstly given here.
No.~39/Mayrit~168291~A has a fainter visual companion, tentatively called 
Mayrit~168291~B, at about 3.5\,arcsec to the northeast.
Neither DENIS nor 2MASS resolved the system.
\item[*] No.~58/Mayrit~21023 is located at $\rho \approx$ 21\,arcsec, $\theta
\approx$ 23\,deg, to $\sigma$~Ori~AB. 
Their coordinates and $JHK_{\rm s}$ magnitudes are from Caballero (2007b).
Besides, the DENIS catalogue tabulates $i$ = 14.31$\pm$0.03\,mag, which seems
to be affected by the glare of the nearby $\sigma$~Ori system.
\end{itemize}

\subsection{Notes to Table~\ref{table.noncounterpart}}
\label{section.notes.table2}

\begin{itemize}
\item[*] No.~62: digitisations of the Palomar Optical Sky Survey show an
extended source (probably the X-ray host galaxy) in the background of a field
dwarf. 
The brown dwarf cluster member candidate {S\,Ori~43} (B\'ejar et~al.
1999) is also in a 6\,arcsec-radius cone search around the X-ray source.
\item[*] No.~93: it has a close, faint, blue, extended, USNO-B1 visual
companion.  
Although this source is probably of extragalactic nature, it does not seem to be
the origin of the X-ray source.
\item[*] No.~96: the 2MASS photometric quality flag of the infrared source close
to the X-ray source is EEA, an indication of binarity.
Public IRAC/{\em Spitzer} images resolve the 2MASS source into two point-like 
sources.
\item[*] No.~97: Mayrit~68191 might be its actual optical counterpart.
\item[*] No.~107: [BNL2005]~1.02~156 may be its actual optical counterpart,
which could correspond to the X-ray source [FPS2006]~NX~101  
(Table~\ref{table.xmmnewton}). 
\end{itemize}

\section{Long tables}

\begin{longtable}{l cc ccc cc ll}
\caption{\label{table.xraydetections} HRC-I/{\em Chandra} X-ray detections with
significance larger than 5.1.}\\ 
\hline
\hline
            \noalign{\smallskip}
No.	& $\alpha$ 	& $\delta$ 	& $\Delta \alpha$, $\Delta \delta$	& $S$		& Offaxis	& CR		& Flux				& NX	& CXO	\\
	& (J2000)	& (J2000)	& [arcsec] 				& ($\sigma$)	& [arcmin] 	& [ks$^{-1}$]   & [10$^{-17}$\,W\,m$^{-2}$]	& (Fr06)& (Sk08)\\
            \noalign{\smallskip}
\hline
            \noalign{\smallskip}
\endfirsthead
\caption{HRC-I/{\em Chandra} X-ray detections with
significance larger than 5.1 (cont.).}\\
\hline
\hline
            \noalign{\smallskip}
No.	& $\alpha$ 	& $\delta$ 	& $\Delta \alpha$, $\Delta \delta$	& $S$		& Offaxis	& CR		& Flux				& NX	& CXO	\\
	& (J2000)	& (J2000)	& [arcsec] 				& ($\sigma$)	& [arcmin] 	& [ks$^{-1}$]   & [10$^{-17}$\,W\,m$^{-2}$]	& (Fr06)& (Sk08)\\
            \noalign{\smallskip}
\hline
            \noalign{\smallskip}
\endhead
\hline
\endfoot
1	& 05 38 44.76	& --02 36 00.1	& 0.04	& 423   &  0.26		&286.6  $\pm$ 1.8       &  261.1        & 80	& 19	\\ % 2E 1470	
2	& 05 38 38.49	& --02 34 55.1  & 0.11  & 170	&  2.05 	& 47.7  $\pm$ 0.8       &   44.7        & 65	& 10	\\ % (new Fl~47, rot. mod.)	
3	& 05 38 47.19	& --02 35 40.3  & 0.11  & 145	&  0.43 	& 29.5  $\pm$ 0.6       &   26.3        & 84	& 23	\\	
4	& 05 38 40.28	& --02 30 18.7  & 0.40  & 142	&  5.74 	& 48.5  $\pm$ 0.9       &   43.6        & 69	& ...	\\	
5	& 05 38 53.37	& --02 33 23.1  & 0.22  & 131	&  3.15 	& 27.7  $\pm$ 0.6       &   24.0        & 109	& 37	\\	
6	& 05 38 44.22	& --02 32 33.7  & 0.22  & 130	&  3.35 	& 28.3  $\pm$ 0.6       &   25.5        & 79	& ...	\\	
7	& 05 37 53.09	& --02 33 34.2  & 2.02  & 123	& 13.3  	& 91.3  $\pm$ 1.4       &  106.3        & 2	& ...	\\ % 2E 1454	   
8	& 05 39 36.48	& --02 42 17.5  & 2.02  & 112	& 14.2  	&100.8  $\pm$ 1.7       &  108.8        & 172	& ...	\\ % 2E 1484  
9	& 05 37 56.34	& --02 45 11.9  & 2.12  & 105	& 15.4  	&101.8  $\pm$ 1.9       &  119.2        & ...	& ...	\\ % 2E 1456	   
10	& 05 39 01.48	& --02 38 56.3  & 0.47  & 99.8  &  4.98 	& 20.9  $\pm$ 0.5       &   18.5        & 125	& 42	\\ % Table 2 in Fr06 
11	& 05 38 43.55	& --02 33 25.4  & 0.11  & 97.7  &  2.53 	& 21.1  $\pm$ 0.7       &   21.0        & 76	& ...	\\	
12	& 05 38 38.22	& --02 36 38.2  & 0.11  & 93.5  &  2.01 	& 14.3  $\pm$ 0.4       &   13.3        & 64	& 9	\\ % (new Fl~46, rot. mod.?)	
13	& 05 38 32.84	& --02 35 39.1  & 0.18  & 92.2  &  3.22 	& 18.6  $\pm$ 0.5       &   16.4        & 49	& 4	\\	
14	& 05 37 54.40	& --02 39 29.6  & 1.66  & 92.0  & 13.3  	& 63.6  $\pm$ 1.4       &   74.3        & 3	& ...	\\ % 2E 1455	   
15	& 05 38 44.24	& --02 40 19.3  & 0.36  & 86.2  &  4.45 	& 15.0  $\pm$ 0.4       &   13.5        & 78	& 18	\\	
16	& 05 38 41.29	& --02 37 22.4  & 0.14  & 79.8  &  1.85 	& 10.1  $\pm$ 0.3       &    9.3        & 70	& 13	\\	
17	& 05 38 49.18	& --02 38 22.0  & 0.14  & 69.6  &  2.62 	&  9.5  $\pm$ 0.4       &    9.5        & 93	& 29	\\ % (new Fl~66, faint flare)	
18	& 05 39 11.62	& --02 36 02.7  & 0.72  & 68.1  &  6.47 	& 15.5  $\pm$ 0.5       &   14.4        & 145	& ...	\\	
19	& 05 38 07.87	& --02 31 31.3  & 1.87  & 66.4  & 10.4  	& 27.8  $\pm$ 0.8       &   33.5        & 8	& ...	\\ % 2E 1459	    
20	& 05 38 35.87	& --02 30 43.4  & 0.54  & 65.5  &  5.72 	& 13.1  $\pm$ 0.5       &   12.2        & 62	& ...	\\ % 2E 1468	   
21	& 05 38 48.00	& --02 27 14.1  & 1.37  & 65.4  &  8.68 	& 21.9  $\pm$ 0.6       &   21.4        & 90	& ...	\\	
22	& 05 38 35.45	& --02 31 51.7  & 0.43  & 52.4  &  4.77 	&  8.5  $\pm$ 0.3       &    7.6        & 60	& ...	\\	
23	& 05 38 48.69	& --02 36 16.1  & 0.11  & 52.0  &  0.83 	&  4.6  $\pm$ 0.2       &    4.5        & 92	& 27	\\ 
24	& 05 38 35.22	& --02 34 38.0  & 0.18  & 50.9  &  2.91 	&  6.9  $\pm$ 0.5       &    7.1        & 58	& 8	\\	
25	& 05 38 44.84   & --02 35 57.1  & 0.11  & 48.8  &  0.22 	&  4.1  $\pm$ 0.2       &    4.7        & (80)  & 20	\\ %	DOUBLE detection     
26	& 05 38 47.46  	& --02 35 25.3  & 0.18  & 48.3  &  0.64 	&  3.9  $\pm$ 0.2       &    3.5        & 87	& 24	\\ % (new Fl~80, rot. mod.)	
27	& 05 38 53.18   & --02 43 52.6  & 1.15  & 46.2  &  8.20 	& 12.8  $\pm$ 0.5       &   13.3        & 106	& ...	\\ % (new Fl~99, faint flare)	
28	& 05 38 35.87   & --02 43 50.4  & 1.15  & 45.4  &  8.32 	& 12.2  $\pm$ 0.5       &   14.0        & 61	& ...	\\	
29	& 05 38 47.90   & --02 37 19.2  & 0.18  & 43.1  &  1.53 	&  3.07 $\pm$ 0.19      &    2.91       & 88	& 25	\\	
30	& 05 39 07.58   & --02 32 39.0  & 0.90  & 41.9  &  6.35 	&  7.4  $\pm$ 0.4       &    7.5        & 138	& ...	\\	
31	& 05 39 00.53   & --02 39 38.8  & 0.65  & 41.1  &  5.27 	&  5.9  $\pm$ 0.3       &    5.9        & 122	& 41	\\	
32	& 05 39 05.39   & --02 32 30.2  & 0.90  & 38.6  &  5.97 	&  6.3  $\pm$ 0.3       &    6.3        & 132	& ...	\\	
33	& 05 39 18.05   & --02 29 29.1  & 2.30  & 37.6  & 10.3  	& 15.2  $\pm$ 0.7       &   20.3        & 156	& ...	\\	
34	& 05 38 34.23   & --02 34 16.0  & 0.36  & 36.7  &  3.30 	&  3.3  $\pm$ 0.2       &    2.9        & 55	& ...	\\	
35	& 05 38 54.09   & --02 49 29.8  & 3.17  & 34.4  & 13.8  	& 23.9  $\pm$ 1.2       &   41.3        & 110	& ...	\\	 
36	& 05 38 31.57   & --02 35 14.9  & 0.40  & 33.8  &  3.59 	&  2.64 $\pm$ 0.19      &    2.68       & 44	& 3	\\	 
37	& 05 38 51.45   & --02 36 20.3  & 0.14  & 33.2  &  1.50 	&  2.08 $\pm$ 0.16      &    2.19       & 102	& 33	\\	 
38	& 05 38 29.12   & --02 36 02.5  & 0.40  & 33.1  &  4.14 	&  3.1  $\pm$ 0.2       &    3.2        & 39	& 1	\\	 
39	& 05 38 34.32   & --02 35 00.1  & 0.22  & 32.7  &  2.98 	&  3.0  $\pm$ 0.3       &    3.5        & 56	& 7	\\ % Table 2 in Fr06
40	& 05 38 52.00   & --02 46 43.4  & 2.41  & 32.2  & 11.0  	& 12.8  $\pm$ 0.7       &   23.0        & 104	& ...	\\	 
41	& 05 38 45.36   & --02 41 58.8  & 0.97  & 32.0  &  6.09 	&  4.5  $\pm$ 0.3       &    5.0        & 81	& 21	\\	 
42	& 05 38 26.43   & --02 34 28.3  & 0.76  & 28.6  &  5.02 	&  3.2  $\pm$ 0.2       &    3.1        & 31	& ...	\\	 
43	& 05 38 50.04   & --02 37 35.4  & 0.22  & 26.3  &  2.01 	&  1.38 $\pm$ 0.13      &    1.45       & 98	& 31	\\	 
44	& 05 39 32.41   & --02 39 44.1  & 2.70  & 23.9  & 12.3  	& 11.3  $\pm$ 0.8       &   20.9        & 170	& ...	\\ % 2E 1482	  
45	& 05 38 27.32   & --02 45 09.0  & 2.77  & 23.2  & 10.3  	&  8.1  $\pm$ 0.6       &   15.5        & 32	& ...	\\	 
46	& 05 39 24.39   & --02 20 45.2  & 3.74  & 23.0  & 18.0  	& 32    $\pm$ 3         &   62	        & ...	& ...	\\	 
47	& 05 38 41.36   & --02 36 44.3  & 0.18  & 20.8  &  1.37 	&  0.96 $\pm$ 0.12      &    0.92       & 71	& 14	\\ % (new Fl~54, rot. mod.?)	 
48	& 05 38 33.35   & --02 36 17.4  & 0.47  & 19.9  &  3.11 	&  1.20 $\pm$ 0.13      &    1.19       & 52	& 5	\\ %	[SWW2004] J053833.280-023617.81, [SSC2008] 5	   
49	& 05 38 51.75   & --02 36 03.2  & 0.25  & 18.8  &  1.52 	&  0.87 $\pm$ 0.10      &    0.80       & 103	& 34	\\ %	Fr06: Table 1	      
50	& 05 39 02.74   & --02 29 55.8  & 1.22  & 17.6  &  7.32 	&  3.5  $\pm$ 0.4       &    3.3        & 126	& ...	\\	 
51	& 05 38 53.06   & --02 38 53.4  & 0.36  & 17.2  &  3.52 	&  0.91 $\pm$ 0.11      &    0.98       & 105	& 36	\\ %	[SWW2004] J053853.040-023853.53, [OJV2006] 6	    
52	& 05 38 48.29   & --02 36 40.9  & 0.22  & 17.2  &  1.02 	&  0.74 $\pm$ 0.10      &    0.65       & 91	& 26	\\ %	Fr06: Table 1	   
53	& 05 39 14.99   & --02 31 37.4  & 1.69  & 16.4  &  8.47 	&  4.1  $\pm$ 0.4       &    6.0        & 150	& ...	\\	 
54	& 05 38 49.23   & --02 41 24.8  & 1.26  & 16.2  &  5.59 	&  1.9  $\pm$ 0.2       &    2.6        & 94	& 28	\\	 
55	& 05 38 59.05   & --02 47 12.7  & 3.24  & 15.2  &  11.8 	&  7.0  $\pm$ 0.7       &   14.9        & 117	& ...	\\	 
56      & 05 38 38.72   & --02 30 21.2  & 1.33  & 14.7  &  5.81 	&  1.7  $\pm$ 0.2       &    1.6        & 66	& ...	\\     
57	& 05 38 27.74 	& --02 43 00.9  & 1.76  & 14.4  &  8.42 	&  3.0  $\pm$ 0.3       &    4.7        & 34	& ...	\\ %	[SWW2004] J053827.639-024301.36      
58	& 05 38 45.31 	& --02 35 41.1  & 0.29  & 14.0  &  0.23 	&  0.53 $\pm$ 0.08      &    0.69       & ...	& ...	\\	 
59	& 05 38 42.28 	& --02 37 14.6  & 0.22  & 14.0  &  1.60 	&  0.52 $\pm$ 0.08      &    0.54       & $<$	& 15	\\ %	SWW 48, Fr06:<63.91	
60	& 05 38 34.05 	& --02 36 37.2  & 0.40  & 14.0  &  3.00 	&  0.67 $\pm$ 0.10      &    0.79       & 54	& 6	\\	 
61	& 05 38 43.02 	& --02 36 14.3  & 0.22  & 13.7  &  0.75 	&  0.46 $\pm$ 0.08      &    0.51       & ...	& 16	\\ 
62      & 05 38 13.85	& --02 35 00.1  & 2.56  & 13.3  &  7.99 	&  3.1  $\pm$ 0.4       &    7.0        & $<$	&  ...  \\ %	Non-member star in Sa08 and S Ori 43, Fr06:<3.71	 
63	& 05 38 59.65 	& --02 45 08.2  & 3.46  & 13.2  &  9.88 	&  4.5  $\pm$ 0.5       &   11.0        & 119	& ...	\\ % Table 2 in Fr06	 
64	& 05 38 41.23 	& --02 37 37.6  & 0.29  & 13.0  &  2.06 	&  0.51 $\pm$ 0.08      &    0.56       & ...	& 12	\\	 
65	& 05 38 43.31 	& --02 32 00.9  & 0.58  & 12.6  &  3.92 	&  0.92 $\pm$ 0.16      &    0.91       & 75	& ...	\\	 
66	& 05 39 22.81 	& --02 33 34.0  & 2.59  & 11.8  &  9.55 	&  3.7  $\pm$ 0.5       &    6.9        & 161	& ...	\\	 
67	& 05 38 46.85	& --02 36 43.4  & 0.22  & 11.7  &  0.88 	&  0.31 $\pm$ 0.06      &    0.38       & $<$	& 22	\\ %	B 3.01-67, Fr06:<0.53, Sk08: Low-significance wavdetect detection (2<{significance}<3)
68      & 05 38 58.27	& --02 38 51.3  & 0.79  & 11.7  &  4.32 	&  0.86 $\pm$ 0.13      &    0.70       & 115	& 38	\\ %	[SSC2008] 38	  
69	& 05 38 29.87	& --02 23 38.4  & 3.96  & 9.72  & 12.9  	&  5.4  $\pm$ 0.9       &   14.0        & 41	& ...	\\ % Table 2 in Fr06	 
70	& 05 38 45.63	& --02 35 58.7  & 0.25  & 9.60  &  0.09 	&  0.33 $\pm$ 0.09      &    0.29       & ...	& ...	\\	 
71	& 05 38 08.26	& --02 35 56.0  & 2.95  & 8.97  &  9.35 	&  2.1  $\pm$ 0.3       &    5.0        & 9	& ...	\\	 
72	& 05 38 33.60	& --02 44 13.5  & 2.20  & 8.87  &  8.86 	&  1.7  $\pm$ 0.3       &    3.3        & 53	& ...	\\	 
73	& 05 39 05.17	& --02 33 00.7  & 0.68  & 8.54  &  5.65 	&  0.48 $\pm$ 0.13      &    0.66       & 131	&  ...  \\	 
74	& 05 38 34.79	& --02 34 15.8  & 0.65  & 8.48  &  3.17 	&  0.45 $\pm$ 0.12      &    0.39       & ...	& ...	\\	 
75	& 05 39 25.20	& --02 38 22.4  & 3.02  & 8.44  & 10.2  	&  2.6  $\pm$ 0.5       &    7.3        & 165	& ...	\\ 
76      & 05 38 51.88	& --02 33 32.6  & 0.36  & 8.42  &  2.81 	&  0.30 $\pm$ 0.09      &    0.26       & ...	& 35	\\	
77	& 05 37 51.59	& --02 35 26.7  & 4.32  & 8.39  & 13.5  	&  3.1  $\pm$ 0.6       &   10.9        & 1	& ...	\\	 
78	& 05 38 50.80	& --02 36 26.6  & 0.36  & 8.31  &  1.39 	&  0.28 $\pm$ 0.09      &    0.33       & $<$	& ...	\\ %	K 8, Fr06:<4.98       
79	& 05 38 31.41	& --02 36 33.7  & 0.65  & 8.20  &  3.63 	&  0.45 $\pm$ 0.12      &    0.47       & $<$	& 2	\\ %	SWW 50, Fr06:<0.31	      
80	& 05 39 11.60	& --02 31 05.3  & 3.17  & 7.94  &  8.06 	&  2.0  $\pm$ 0.4       &    5.4        & 144	& ...	\\	 
81	& 05 39 01.17	& --02 36 39.1  & 0.90  & 7.76  &  3.94 	&  0.55 $\pm$ 0.14      &    0.68       & 124	& ...	\\	 
82	& 05 38 18.35	& --02 35 38.3  & 1.69  & 7.65  &  6.84 	&  1.0  $\pm$ 0.3       &    2.4        & 19	& ...	\\ 
83      & 05 39 01.05	& --05 33 38.5  & 0.68  & 7.62  &  4.44 	&  0.55 $\pm$ 0.14      &    0.53       & ...	& ...	\\	
84	& 05 39 08.98	& --02 39 58.6  & 1.73  & 7.34  &  7.10 	&  1.0  $\pm$ 0.3       &    1.8        & 140	& ...	\\	 
85	& 05 38 39.73	& --02 40 19.7  & 0.68  & 7.20  &  4.68 	&  0.38 $\pm$ 0.11      &    0.51       & 68	& 11	\\	 
86 	& 05 38 36.87	& --02 36 43.2  & 0.40  & 7.12  &  2.36 	&  0.23 $\pm$ 0.07      &    0.15       & $<$	& ...	\\ %	SWW 4, Fr06:<0.39	   
87 	& 05 38 20.23	& --02 38 02.8  & 1.80  & 6.95  &  6.72 	&  0.9  $\pm$ 0.2       &    1.7        & 20	& ...	\\	 
88 	& 05 39 45.79	& --02 40 35.8  & 6.37  & 6.94  & 15.7  	&  5.6  $\pm$ 1.3       &   27.5        & ...	&  ...  \\	 
89 	& 05 39 07.54	& --02 28 22.3  & 4.68  & 6.89  &  9.29 	&  2.7  $\pm$ 0.7       &    9.7        & 137	& ...	\\	 
90 	& 05 39 16.94	& --02 25 43.6  & 4.68  & 6.88  & 12.8  	&  3.0  $\pm$ 0.7       &   11.1        & 154	& ... 	\\	 
91 	& 05 38 41.59	& --02 28 20.0  & 1.80  & 6.85  &  7.63 	&  0.9  $\pm$ 0.2       &    1.6        & 72	& ...	\\	 
92 	& 05 38 49.70	& --02 34 52.8  & 0.32  & 6.70  &  1.42 	&  0.17 $\pm$ 0.07      &    0.20       & ...	& 30	\\	 
93 	& 05 38 42.14	& --02 43 16.1  & 2.45  & 6.65  &  7.43 	&  1.2  $\pm$ 0.3       &    3.0        & 73	& ...	\\	 
94 	& 05 38 01.45	& --02 25 53.5  & 6.73  & 6.41  & 14.9  	&  4.1  $\pm$ 1.0       &   29.2        & ...	& ...	\\	 
95 	& 05 39 24.42	& --02 34 03.1  & 3.53  & 6.21  &  9.84 	&  1.7  $\pm$ 0.5       &    5.7        & 164	& ...	\\ %	[SWW2004] J053924.358-023401.25      
96 	& 05 39 06.66	& --02 38 05.0  & 1.40  & 5.92  &  5.67 	&  0.57 $\pm$ 0.16      &    1.1        & ...	& ...	\\	  
97 	& 05 38 43.87	& --02 37 06.1  & 0.25  & 5.92  &  1.29 	&  0.12 $\pm$ 0.05      &    0.18       & $<$	& 17	\\ %	SWW 15, Fr06:<0.51     
98 	& 05 39 33.42	& --02 20 37.1  & 7.02  & 5.70  & 19.4  	&  4.8  $\pm$ 1.3       &   43.1        & ...	& ...	\\ %	[SWW2004] J053933.792-022039.74    
99 	& 05 39 37.50	& --02 26 58.5  & 7.16  & 5.62  & 15.7  	&  3.9  $\pm$ 1.0       &   25.6        & 173	& ... 	\\ % 2E 1483?, [SWW2004] J053937.293-022656.70   
100	& 05 38 26.16	& --02 29 10.4  & 2.66  & 5.51  &  8.30 	&  1.0  $\pm$ 0.3       &    2.5        & 30	& ...	\\     
101	& 05 38 50.16	& --02 36 54.0  & 0.32  & 5.42  &  1.50 	&  0.13 $\pm$ 0.05      &    0.13       & 99	& 32	\\ % Sk08: Low-significance wavdetect detection (2<{significance}<3). Probable XMM-Newton counterpart is source NX 99 in Table B.1 of FPS06. High value of <E> suggests possible extragalactic background source.	 
102	& 05 38 41.55	& --02 39 09.1  & 0.14  & 5.41  &  3.42 	&  0.08 $\pm$ 0.06      &    0.01       & ...	& ...	\\	
103	& 05 39 16.91	& --02 41 18.1  & 2.66  & 5.39  &  9.49 	&  1.0  $\pm$ 0.3       &    2.7        & 153	& ...	\\ %	Fr06: Table 1	 
104	& 05 38 37.36	& --02 42 50.0  & 1.48  & 5.35  &  7.23 	&  0.49 $\pm$ 0.15      &   0.89        & 63	& ...	\\     
105	& 05 38 35.58	& --02 34 04.5  & 0.18  & 5.35  &  3.11 	&  0.08 $\pm$ 0.05      &   0.05        & ...	& ...	\\	
106	& 05 39 11.14	& --02 36 48.1  & 1.44  & 5.22  &  6.42 	&  0.49 $\pm$ 0.15      &   0.68        & 143	& ...	\\     
107	& 05 38 28.43	& --02 32 30.7  & 1.51  & 5.11  &  5.48 	&  0.47 $\pm$ 0.15      &   0.91        & 35	& ...	\\	 
          \noalign{\smallskip}
\end{longtable}

\begin{longtable}{l l l l l}
\caption{\label{table.xraycounterparts} Optical/near-infrared counterparts of
X-ray sources in Table~\ref{table.xraydetections}.}\\ 
\hline
\hline
            \noalign{\smallskip}
No.	& 2MASS  		& Name			& Alternative name	& Class	\\
	& designation		&			&			&	\\
            \noalign{\smallskip}
\hline
            \noalign{\smallskip}
\endfirsthead
\caption{Optical/near-infrared counterparts of X-ray sources
in Table~\ref{table.xraydetections} (cont.).}\\
\hline
\hline
            \noalign{\smallskip}
No.	& 2MASS  		& Name			& Alternative name	& Class	\\
	& designation		&			&			&	\\
            \noalign{\smallskip}
\hline
            \noalign{\smallskip}
\endhead
\hline
\endfoot
1	& 05384476--0236001 	& Mayrit AB        	& $\sigma$~Ori~AB + ``F''	& Young star		\\      
2	& 05383848--0234550    	& Mayrit 114305 AB	& [W96] 4771--1147 AB 		& Young star		\\      
3	& 05384719--0235405    	& Mayrit 42062 AB  	& $\sigma$~Ori~E + ``Eb''	& Young star		\\      
4	& 05384027--0230185    	& Mayrit 348349    	& Haro 5--13			& Young star		\\      
5	& 05385337--0233229    	& Mayrit 203039    	& [W96] 4771--1049		& Young star		\\      
6	& 05384424--0232336    	& Mayrit 207358    	& [W96] 4771--1055		& Young star		\\      
7	& 05375303--0233344    	& Mayrit 789281    	& 2E 1454			& Young star		\\ % [LC2008] 23, [W96] 4771--0775      
8	& 05393654--0242171 	& Mayrit 863116 AB 	& RX J0539.6--0242 AB		& Young star		\\ % RX J0539.6-0242 AB, 2E 1484 AB      
9	& 05375630--0245130    	& 2E 1456	   	& 2E 0535.4--0246 		& Galaxy		\\      
10	& 05390149--0238564    	& Mayrit 306125 AB 	& HD 37525 AB			& Young star		\\      
11	& 05384355--0233253    	& Mayrit 156353    	& [SWW2004] 36			& Young star		\\ % [SWW2004] J053843.449-023325.33    
12	& 05383822--0236384 	& Mayrit 105249    	& [W96] rJ053838--0236		& Young star		\\      
13	& 05383284--0235392    	& Mayrit 180277    	& [W96] rJ053832--0235b		& Young star		\\       
14	& 05375440--0239298    	& Mayrit 783254    	& 2E 1455			& Young star		\\   
15	& 05384423--0240197 	& Mayrit 260182    	& [W96] 4771--1051		& Young star		\\      
16	& 05384129--0237225 	& Mayrit 97212     	& [W96] rJ053841--0237		& Young star		\\      
17	& 05384917--0238222    	& Mayrit 157155    	& [W96] rJ053849--0238		& Young star		\\      
18	& 05391163--0236028    	& Mayrit 403090    	& [W96] 4771--1038		& Young star		\\      
19	& 05380784--0231314    	& Mayrit 615296    	& 2E 1459			& Young star		\\ % [LC2008] 37       
20	& 05383587--0230433    	& Mayrit 344337 AB    	& [W96] 4771--1097		& Young star		\\ % 2E 1458     
21	& 05384803--0227141    	& Mayrit 528005 AB 	& [W96] 4771--0899 AB		& Young star		\\      
22	& 05383546--0231516 	& Mayrit 285331    	& [W96] rJ053835--0231		& Young star		\\        
23	& 05384868--0236162    	& Mayrit 61105     	& [SWW2004] 35			& Young star		\\ 
25  &{\em 05384484--0235571}    & Mayrit 3020 AB	& $\sigma$~Ori~IRS1~AB 		& Young star		\\ % DOUBLE detection	 
26	& 05384746--0235252    	& Mayrit 53049     	& [SWW2004] 78			& Young star		\\ % Class II     
27	& 05385317--0243528    	& Mayrit 489165    	& [SWW2004] 47			& Young star		\\      
28	& 05383587--0243512    	& Mayrit 489196    	& TY Ori			& Young star		\\ % Halpha     
29	& 05384791--0237192 	& Mayrit 92149 AB     	& [W96] R053847--0237 AB	& Young star		\\      
30	& 05390760--0232391    	& Mayrit 397060    	& V507 Ori			& Young star		\\ % Halpha      
31	& 05390052--0239390    	& [W96] 4771--1056 	& [SSC2008] 41			& Possible field star	\\ % [W96] 4771--1056, Mayrit 322123 (?), Wolk (1996): K1, Li I = 0.107, Ha = +0.520   
32	& 05390540--0232303    	& Mayrit 374056    	& [W96] 4771--1075		& Young star		\\      
33	& 05391807--0229284    	& Mayrit 634052    	& [W96] 4771--0598		& Young star		\\ % Debris disc     
34	& 05383422--0234160    	& Mayrit 189303    	& HD 294272 B			& Young star		\\      
35	& 05385410--0249297    	& Mayrit 822170    	& RX J0538.9--0249		& Young star		\\	
36	& 05383157--0235148    	& Mayrit 203283    	& [W96] rJ053831--0235		& Young star		\\	
37	& 05385145--0236205    	& Mayrit 102101 AB    	& [W96] rJ053851--0236		& Young star		\\ % Halpha	
38	& 05382911--0236026    	& Mayrit 234269    	& [SWW2004] 177			& Young star		\\	
39	& 05383431--0235000 	& Mayrit 168291 AB	& [W96] rJ053834--0234  	& Young star		\\ % [HHM2007] 592, [GNR2008] 9, [SSC2008] 7
40	& 05385200--0246436    	& Mayrit 653170    	& RU Ori			& Young star		\\	
41	& 05384537--0241594    	& Mayrit 359179    	& V595 Ori			& Young star		\\ % Halpha	
42	& 05382639--0234286    	& SO210038         	& 2MASS J05382639--0234286	& Field star		\\	
43	& 05385003--0237354 	& Mayrit 124140    	& [W96] pJ053850--0237		& Young star candidate	\\	
44	& 05393256--0239440    	& Mayrit 750107    	& [W96] rJ053932--0239		& Young star		\\	
45	& 05382725--0245096    	& Mayrit 609206    	& V505 Ori			& Young star		\\	
46	& 05392456--0220441    	& Mayrit 1093033   	& [HHM2007] 1030		& Young star candidate	\\ % No. 46	
47	& 05384135--0236444	& Mayrit 68229     	& [W96] pJ053841--0236		& Young star		\\	
48	& 05383335--0236176    	& Mayrit 172264    	& [SWW2004] 130			& Young star candidate	\\ % [SWW2004] J053833.280-023617.81, [SSC2008] 5   
49	& 05385173--0236033 	& Mayrit 105092    	& [FPS2006] NX 103		& Young star		\\	
50	& 05390276--0229558 	& Mayrit 453037 AB   	& [W96] R053902--0229 AB	& Young star		\\	
51	& 05385306--0238536 	& [SWW2004] 166    	& [OJV2006] 6			& Field star		\\ % [SWW2004] J053853.040-023853.53, [OJV2006] 6, [SFR2008] S11
52	& 05384828--0236409    	& Mayrit 67128     	& [FPS2006] NX 91		& Young star		\\	
53	& 05391506--0231376    	& Mayrit 524060    	& HD 37564			& Young star		\\	
54	& 05384921--0241251    	& Mayrit 332168 AB    	& [SWW2004] 205			& Young star		\\	
55	& 05385911--0247133    	& Mayrit 707162    	& [W96] rJ053859--0247 AB	& Young star		\\	
57	& 05382774--0243009    	& Mayrit 492211    	& [SWW2004] 87			& Young star		\\ % [SWW2004] J053827.639-024301.36	 
58  &{\em 05384531--0235413}    & Mayrit 21023     	& [BHM2009] 14			& Young star candidate	\\ % SigOri-MAD-14	
59	& 05384227--0237147    	& Mayrit 83207     	& [W96] pJ053842--0237		& Young star		\\ % Class II	
60	& 05383405--0236375    	& Mayrit 165257    	& [W96] rJ053833--0236		& Young star		\\	
61	& 05384301--0236145    	& Mayrit 30241     	& [HHM2007] 687			& Young star candidate	\\ 
63	& 05385955--0245080    	& Mayrit 591158    	& [W96] 4771--0026		& Young star		\\ % SO420039 in GH08, primordial lithium abundance	
64	& 05384123--0237377    	& UCM0536--0239     	& [HHM2007] 668			& Galaxy		\\	
65	& 05384333--0232008    	& Mayrit 240355 AB    	& [SWW2004] 144			& Young star		\\ % Class II	
66	& 05392286--0233330    	& Mayrit 590076    	& [W96] rJ053923--0233		& Young star		\\	
67	& 05384684--0236435    	& Mayrit 53144     	& [BNL2005] 3.01--67		& Young star		\\ % low g
69	& 05382993--0223381 	& [W96] rJ053829--0223	& SO120731			& Field star		\\	  
70	& 05384561--0235588    	& Mayrit 13084     	& $\sigma$~Ori~D		& Young star		\\ 	
71	& 05380826--0235562 	& Mayrit 547270 AB    	& Kiso A--0976 316		& Young star		\\ % Halpha	
72	& 05383368--0244141    	& Mayrit 521199    	& TX Ori			& Young star		\\ % Halpha, Class II, Ca II, Si	
73	& 05390524--0233005    	& Mayrit 355060    	& [SWW2004] 175			& Young star		\\	
74	& 05383479--0234158    	& Mayrit 182305    	& HD 294272 A			& Young star		\\	
75	& 05392519--0238220    	& Mayrit 622103    	& BG Ori			& Young star		\\ 
77	& 05375161--0235257    	& Mayrit 797272    	& [W96] rJ053751--0235		& Young star		\\ % alt. [SWW2004] 125 	
78	& 05385077--0236267    	& Mayrit 94106     	& [KJN2005] 8			& Young star		\\	
79	& 05383141--0236338    	& Mayrit 203260    	& Haro 5--11			& Young star		\\	
80	& 05391151--0231065    	& Mayrit 497054    	& V509 Ori			& Young star		\\	
81	& 05390115--0236388    	& Mayrit 249099    	& [KJN2005] 9			& Young star		\\	
82	& 05381834--0235385    	& Mayrit 396273    	& S\,Ori J053818.2--023539	& Young brown dwarf	\\ % low g 
84	& 05390894--0239579    	& Mayrit 433123    	& S\,Ori 25			& Young brown dwarf	\\	
85	& 05383972--0240197    	& Mayrit 270196    	& [HHM2007] 655			& Young star candidate	\\	
86	& 05383687--0236432    	& Mayrit 126250 AB    	& [SWW2004] 154	AB		& Young star candidate	\\	
87	& 05382021--0238016 	& Mayrit 387252    	& S\,Ori J053820.1--023802	& Young star		\\	
88	& 05394619--0240320 	& Mayrit 960106    	& V1147 Ori			& Young star		\\ % HRC-I border
89	& 05390759--0228234    	& Mayrit 571037    	& [W96] rJ053907--0228		& Young star		\\	
90	& 05391717--0225433    	& Mayrit 785038    	& Kiso A--0904 80		& Young star		\\ % Halpha	
92	& 05384970--0234526    	& Mayrit 100048    	& [HHM2007] 754			& Young star		\\ % Class II	
94	& 05380167--0225527 	& Mayrit 887313    	& [SE2004] 53			& Young star candidate	\\	
95	& 05392435--0234013 	& Mayrit 605079    	& [SWW2004] 127			& Young star candidate	\\ % Lithium-depleted, [SWW2004] J053924.358-023401.25	 
98	& 05393378--0220398 	& Mayrit 1178039   	& [SWW2004] 138			& Young star candidate	\\ % HRC-I border, [SWW2004] J053933.792-022039.74	  
99	& 05393729--0226567    	& Mayrit 957055    	& [SWW2004] 163			& Young star candidate	\\ % 2E 1483?, [SWW2004] J053937.293-022656.70	
103	& 05391699--0241171    	& Mayrit 578123    	& [FPS2006] NX 153		& Young star candidate	\\	
          \noalign{\smallskip}
\end{longtable}

\begin{longtable}{l cccc cc ll}
\caption{\label{table.roundnround} A reappraisal to the X-ray sources near $\sigma$~Ori in Wolk (1996).}\\ 
\hline
\hline
            \noalign{\smallskip}
Source	&$\alpha_{\rm Wo96}$&$\delta_{\rm Wo96}$& CR   			& Sp. 	& $\alpha_{\rm 2MASS}$	& $\delta_{\rm 2MASS}$	& No.	& Name  	\\
identification	& (J2000)	& (J2000)	& [ks$^{-1}$]           & type	& (J2000)	& (J2000)	&	&			\\
            \noalign{\smallskip}
\hline
            \noalign{\smallskip}
\endfirsthead
\caption{A reappraisal to the X-ray sources near $\sigma$~Ori in Wolk (1996) (cont.).}\\
\hline
\hline
            \noalign{\smallskip}
Source	&$\alpha_{\rm Wo96}$&$\delta_{\rm Wo96}$& CR   			& Sp. 	& $\alpha_{\rm 2MASS}$	& $\delta_{\rm 2MASS}$	& No.	& Name  	\\
identification	& (J2000)	& (J2000)	& [ks$^{-1}$]           & type	& (J2000)	& (J2000)	&	&			\\
            \noalign{\smallskip}
\hline
            \noalign{\smallskip}
\endhead
\hline
\endfoot
R053751--0235	& 05 37 51.5	& --02 35 25	& 1.49 $\pm$ 0.66       & M0	& 05 37 51.61	& --02 35 25.7	& 77	& Mayrit 797272 	\\ % R053751--0235   
4771 0775	& 05 37 52.1	& --02 40 40	& 1.20 $\pm$ 0.59       & K0	& ...		& ...		& ...	& ...			\\ % R053752--0240   % 1RXH J053752.2-024049	     
4771 0921	& 05 37 52.7	& --02 33 33	& 13.4 $\pm$ 1.13       & K0	& 05 37 53.03	& --02 33 34.4	& 7	& Mayrit 789281 	\\ % R053752--0233   
R053754--0239	& 05 37 53.9	& --02 39 27	& 18.2 $\pm$ 1.57       & ...	& 05 37 54.40	& --02 39 29.8	& 14	& Mayrit 783254 	\\ % R053754--0239   
R053755--0245	& 05 37 55.4	& --02 45 07	& 22.4 $\pm$ 1.67       & ...	& 05 37 56.30	& --02 45 13.0	& 9	& 2E 1456		\\ % R053755--0245   
		& 05 37 55.7	& --02 45 16	& 22.4 $\pm$ 1.56       &	& 		& 		&	&			\\ % 		     
		& 05 37 56.0	& --02 45 12	& 23.5 $\pm$ 1.67       & 	& 		& 		&	&			\\ % 		     
4771 0947	& 05 38 06.8	& --02 30 30	& 1.01 $\pm$ 0.47      &$\times$& 05 38 06.74 	& --02 30 22.8	& ...	& Mayrit 662301 	\\ % R053806--0230   % NX 7
4771 0854	& 05 38 07.5	& --02 31 27	& 6.31 $\pm$ 0.85       & ...	& 05 38 07.84	& --02 31 31.4	& 19	& Mayrit 615296 	\\ % R053807--0231   
R053808--0235	& 05 38 08.0	& --02 35 50	& 1.14 $\pm$ 0.63       & K5	& 05 38 08.26	& --02 35 56.2	& 71	& Mayrit 547270 AB 	\\ % R053808--0235   
R053813--0234	& 05 38 13.9	& --02 34 57	& 1.80 $\pm$ 0.42       & ...	& ...		& ...		& 62	& ...			\\ % R053813--0234   % galaxy; no member in Sa08
R053814--0236	& 05 38 14.6	& --02 36 37	& 0.60 $\pm$ 0.14       & ...	& ...		& ...		& ...	& ...			\\ % R053814--0236   % galaxy; 2XMM J053814.1-023646	    
R053820--0237	& 05 38 20.3	& --02 37 47	& 2.01 $\pm$ 0.47       & M0	& 05 38 20.21	& --02 38 01.6	& 87	& Mayrit 387252 	\\ % R053820--0237   % ???; [W96] rJ053820-0237 ("the ghost star")
R053827--0242	& 05 38 27.7	& --02 42 58	& 2.25 $\pm$ 0.62       & ...	& 05 38 27.74	& --02 43 00.9	& 57	& Mayrit 492211 	\\ % R053827--0242   
R053828--0236	& 05 38 28.8	& --02 36 00	& 0.95 $\pm$ 0.61       & K5	& 05 38 29.11	& --02 36 02.6	& 34	& Mayrit 234269 	\\ % R053828--0236   
R053831--0235	& 05 38 31.4	& --02 35 04	& 0.85 $\pm$ 0.42       & K7	& 05 38 31.57	& --02 35 14.8	& 36	& Mayrit 203283 	\\ % R053831--0235   
R053832--0235	& 05 38 32.7	& --02 35 35	& 3.20 $\pm$ 0.78       & ...	& 05 38 32.84	& --02 35 39.2	& 13	& Mayrit 180277 	\\ % R053832--0235   
R053833--0236	& 05 38 33.7	& --02 36 25	& 0.36 $\pm$ 0.69       & K5	& 05 38 34.05	& --02 36 37.5	& 60	& Mayrit 165257 	\\ % R053833--0236   
R053834--0234	& 05 38 34.2	& --02 34 16	& 1.60 $\pm$ 0.73       & K0	& 05 38 34.22	& --02 34 16.0	& 34	& Mayrit 189303 	\\ % R053834--0234   
		& 05 38 34.1	& --02 34 18	& 0.92 $\pm$ 0.64       &	& 		& 		&	&			\\ % 		     
R053835--0231	& 05 38 35.2	& --02 31 48	& 8.23 $\pm$ 0.95       & K5	& 05 38 35.46	& --02 31 51.6	& 22	& Mayrit 285331 	\\ % R053835--0231   
4771 1097	& 05 38 35.7	& --02 30 39	& 3.52 $\pm$ 0.76       & K5	& 05 38 35.87	& --02 30 43.3	& 20	& Mayrit 344337 AB 	\\ % R053835--0230   
R053838--0236	& 05 38 38.1	& --02 36 38	& 3.44 $\pm$ 0.76       & K5	& 05 38 38.22	& --02 36 38.4	& 12	& Mayrit 105249 	\\ % R053838--0236   
4771 1147	& 05 38 38.3	& --02 34 51	& 12.1 $\pm$ 1.07       & K0	& 05 38 38.48	& --02 34 55.0	& 2	& Mayrit 114305 	\\ % R053838--0234   
R053840--0230	& 05 38 39.8	& --02 30 17	& 2.51 $\pm$ 0.75       & M0	& 05 38 40.27	& --02 30 18.5	& 4	& Mayrit 348349 	\\ % R053839--0230   
R053841--0237	& 05 38 41.2	& --02 37 20	& 3.03 $\pm$ 0.75       & K3	& 05 38 41.29	& --02 37 22.5	& 16	& Mayrit 97212  	\\ % R053841--0237   
R053843--0233	& 05 38 43.3	& --02 33 20	& 1.23 $\pm$ 0.73       & ...	& 05 38 43.55	& --02 33 25.3	& 11	& Mayrit 156353 	\\ % R053843--0233   
4771 1051	& 05 38 44.0	& --02 40 18	& 6.10 $\pm$ 0.84       & K5	& 05 38 44.23	& --02 40 19.7	& 15	& Mayrit 260182 	\\ % R053844--0240   
4771 1055	& 05 38 44.1	& --02 32 30	& 7.49 $\pm$ 0.92       & ...	& 05 38 44.24	& --02 32 33.6	& 6	& Mayrit 207358 	\\ % R053844--0232   
R053845--0236	& 05 38 44.6	& --02 35 58	& 120  $\pm$ 2.99   & {\em O9.5}& 05 38 44.76	& --02 36 00.1	& 1	& Mayrit AB		\\ % R053844--0235   
		& 05 38 45.4	& --02 36 00	& 130  $\pm$ 3.76       &	& 		& 		&	&			\\ % R053845--0236   
R053845--0241	& 05 38 45.3	& --02 41 58	& 1.59 $\pm$ 0.63       & ...	& 05 38 45.37	& --02 41 59.4	& 41	& Mayrit 359179 	\\ % R053845--0241   
R053847--0235	& 05 38 47.0	& --02 35 35	& 8.33 $\pm$ 1.03     & {\em B2}& 05 38 47.19	& --02 35 40.5	& 3	& Mayrit 42062 AB	\\ % R053847--0235   
R053847--0237	& 05 38 47.6	& --02 37 18	& 2.34 $\pm$ 0.80       & K5	& 05 38 47.91	& --02 37 19.2	& 29	& Mayrit 92149 AB	\\ % R053847--0237   
4771 0899	& 05 38 47.6	& --02 27 09	& 4.28 $\pm$ 0.93       & K3	& 05 38 48.03	& --02 27 14.1	& 21	& Mayrit 528005 AB	\\ % R053847--0227   
R053849--0238	& 05 38 49.0	& --02 38 21	& 6.40 $\pm$ 0.87       & K5	& 05 38 49.17	& --02 38 22.2	& 17	& Mayrit 157155 	\\ % R053849--0238   
R053851--0236	& 05 38 51.3	& --02 36 15	& 2.65 $\pm$ 0.50       & K5	& 05 38 51.45	& --02 36 20.5	& 37	& Mayrit 102101 AB 	\\ % R053851--0236   
4771 0080	& 05 38 51.8	& --02 46 41	& 5.39 $\pm$ 0.98       & K7	& 05 38 52.00	& --02 46 43.6	& 40	& Mayrit 653170 	\\ % R053851--0246   
R053852--0238	& 05 38 52.7	& --02 38 56	& 1.91 $\pm$ 0.62       & ...	& 05 38 53.06	& --02 38 53.6	& 51	& [SWW2004] 166 	\\ % R053852--0238   
R053853--0243	& 05 38 52.9	& --02 43 47	& 2.88 $\pm$ 0.83      &$\times$& 05 38 53.17	& --02 43 52.8	& 27	& Mayrit 489165 	\\ % R053852--0243   
4771 1049	& 05 38 53.2	& --02 33 20	& 12.0 $\pm$ 1.08       & K5	& 05 38 53.37	& --02 33 22.9	& 5	& Mayrit 203039 	\\ % R053853--0233   
4771 0119	& 05 38 54.2	& --02 49 27	& 4.03 $\pm$ 0.90       & ...	& 05 38 54.10	& --02 49 29.7	& 35	& Mayrit 822170 	\\ % R053854--0249   
4771 0226	& 05 38 59.1	& --02 44 59	& 3.65 $\pm$ 0.85       & ...	& 05 38 59.55	& --02 45 08.0	& 63	& Mayrit 591158 	\\ % R053859--0244   
4771 1056	& 05 39 00.4	& --02 39 35	& 0.65 $\pm$ 0.82       & K1	& 05 39 00.52 	& --02 39 39.0	& 31	& [W96] 4771--1056	\\ % R053900--0239   
R053901--0238	& 05 39 01.3	& --02 38 54	& 5.96 $\pm$ 0.90     & {\em B5}& 05 39 01.49	& --02 38 56.4	& 10	& Mayrit 306125 AB	\\ % R053901--0238   
R053901--0240	& 05 39 01.6	& --02 40 57	& 0.97 $\pm$ 0.22       & ...	& ...		& ...		& ...	& [FPS2006] NX 121 (?) 	\\ % R053901--0240   % ???; galaxy; no optical/nIR counterpart
R053902--0229	& 05 39 02.6	& --02 29 47	& 2.61 $\pm$ 0.61       & K7	& 05 39 02.76	& --02 29 55.8	& 50	& Mayrit 453037 AB	\\ % R053902--0229   
4771 1075	& 05 39 05.3	& --02 32 24	& 2.75 $\pm$ 0.73       & K5	& 05 39 05.40	& --02 32 30.3	& 32	& Mayrit 374056 	\\ % R053905--0232   
4771 1092	& 05 39 07.4	& --02 32 33	& 3.20 $\pm$ 0.75       & K3	& 05 39 07.60	& --02 32 39.1	& 30	& Mayrit 397060 	\\ % R053907--0232   
R053908--0239	& 05 39 08.7	& --02 39 54	& 0.81 $\pm$ 0.19       & ...	& 05 39 08.94	& --02 39 57.9	& 84	& Mayrit 433123 	\\ % R053908--0239   % !!! S Ori 25
4771 0901	& 05 39 11.3	& --02 30 59	& 1.89 $\pm$ 0.44       & K5	& 05 39 11.51	& --02 31 06.5	& 80	& Mayrit 497054 	\\ % R053911--0230   
4771 1038	& 05 39 11.6	& --02 35 58	& 1.67 $\pm$ 0.69       & K5	& 05 39 11.63	& --02 36 02.8	& 18	& Mayrit 403090 	\\ % R053911--0235   
R053914--0228	& 05 39 14.8	& --02 28 27	& 1.48 $\pm$ 0.62       & ...	& 05 39 14.47	& --02 28 33.4	& ...	& Mayrit 631045 	\\ % R053914--0228   % NX 149
R053916--0233	& 05 39 16.8	& --02 33 03	& 0.77 $\pm$ 0.18       & ...	& ...		& ...		& ...	& [FPS2006] NX 155 (?)	\\ % R053916--0233   % NX 155; galaxy; 2XMM J053917.5-023313
4771 0598	& 05 39 18.2	& --02 29 27	& 3.02 $\pm$ 0.78       & ...	& 05 39 18.07	& --02 29 28.4	& 33	& Mayrit 634052 	\\ % R053918--0229   
4771 0910	& 05 39 19.5	& --02 30 36	& 2.95 $\pm$ 0.68       & K3	& 05 39 18.83	& --02 30 53.1	& ...	& Mayrit 596059		\\ % R053919--0230   % Mayrit 596059 (NX~157...) and Mayrit 633059 (NX~160)
R053922--0233	& 05 39 23.1	& --02 33 33	& 5.46 $\pm$ 1.27       & K7	& 05 39 22.86	& --02 33 33.0	& 66	& Mayrit 590076 	\\ % R053923--0233   
R053930--0238	& 05 39 30.4	& --02 38 20	& 1.85 $\pm$ 0.43       & ...	& 05 39 30.56	& --02 38 27.0	& ...	& [SWW2004] 222 AB	\\ % R053930--0238   % NX 169; SB and no member in Sa08; [SWW2004] J053930.567-023826.96
R053932--0239	& 05 39 32.7	& --02 39 40	& 3.35 $\pm$ 0.74      &$\times$& 05 39 32.56	& --02 39 44.0	& 44	& Mayrit 750107 	\\ % R053932--0239   
R053936--0242	& 05 39 36.4	& --02 42 17	& 16.9 $\pm$ 1.42       & ...	& 05 39 36.54	& --02 42 17.1	& 8	& Mayrit 863116 AB	\\ % R053936--0242   
		& 05 39 36.8	& --02 42 19	& 14.5 $\pm$ 1.26       &	& 		& 		&	&			\\ % R053937--0242   
R053947--0226	& 05 39 47.6	& --02 26 05	& 2.54 $\pm$ 0.68       & ...	& 05 39 47.42	& --02 26 16.2	& ...	& Mayrit 1106058	\\ % R053947--0226   % 2E 1486, no fov 
R053947--0232	& 05 39 47.4	& --02 32 22	& 0.50 $\pm$ 0.96       & ...	& 05 39 47.84	& --02 32 24.9	& ...	& Mayrit 969077 	\\ % R053947--0232   % no fov
          \noalign{\smallskip}
\end{longtable}

\end{document}